\documentclass[12pt]{article}
\usepackage[usenames, dvipsnames]{xcolor}
\usepackage{jcapmod}

\usepackage{tocloft}
\usepackage[normalem]{ulem}
\usepackage[]{todonotes}
\usepackage{verbatim}
\usepackage{mathrsfs}
\usepackage{cleveref}
\usepackage[margin=.8in]{geometry}

\usepackage{multirow}

\usepackage{caption}
\usepackage{longtable}
\usepackage{afterpage}
\usepackage{tikz}
\usepackage{setspace,caption}
\usepackage{lipsum}
\usetikzlibrary{matrix,arrows,calc}
\usepackage{bbm}
\usepackage{slashed}
\usepackage{graphicx}
\usepackage{subcaption}
\usepackage{psfrag}
\usepackage{tensor}
\usepackage{fouridx}
\usepackage{enumitem}
\usepackage{dynkin-diagrams}

\usetikzlibrary{arrows,shapes,snakes,automata,backgrounds,petri}

\usepackage{bm}
\usepackage{mdframed}
\usepackage{multirow}
\usepackage{soul}
\usepackage{bbold}
\usepackage{multicol}
\usepackage{tikz-cd}
\usepackage{rotating}

\usetikzlibrary{arrows}

  \usepackage{mdframed}
\tikzset{
commutative diagrams/.cd,
arrow style=tikz,
diagrams={>=latex}}

\usepackage{amsmath,amsthm}
\usepackage{amssymb}
\usepackage{mathtools}

\usepackage{feynmf}
\usepackage{marvosym}

\usepackage{import}

\usepackage[utf8]{inputenc}
\usepackage{amsmath}
\usepackage{tikz}
\usepackage{extpfeil}
\usepackage{amsfonts}
\usepackage{amssymb}
\usepackage{bm}
\usepackage{graphicx}
\usepackage{array}
\usepackage{lipsum}
\usepackage{float}
\usepackage{multirow}
\usepackage{hhline}
\usepackage{tabularx}
\usepackage{graphicx}
\usepackage{afterpage}
\usepackage{longtable}
\usepackage{epstopdf}
\usepackage{ytableau}
\usetikzlibrary{decorations.markings}

\usepackage[titletoc]{appendix}
\makeatletter
\newtheorem*{rep@theorem}{\rep@title}
\newcommand{\newreptheorem}[2]{%
\newenvironment{rep#1}[1]{%
 \def\rep@title{#2 \ref{##1}}%
 \begin{rep@theorem}}%
 {\end{rep@theorem}}}
\makeatother
\newtheorem{lemma}{Lemma}[section]
\newreptheorem{lemma}{Lemma}

\newreptheorem{conj}{Conjecture}

\theoremstyle{definition}
\newtheorem{defn}[lemma]{Definition}

\theoremstyle{remark}
\newtheorem*{rem*}{Remark}

\newcommand{\jht}[1]{{}}

\definecolor{cobalt}{RGB}{44, 98, 120}
\definecolor{celadon}{rgb}{0.67, 0.88, 0.69}
\definecolor{dm}{cmyk}{.20, 0, .30, 0}
\definecolor{burgundy}{rgb}{0.5, 0.0, 0.13}
\definecolor{plotBlue}{RGB}{94, 130, 181}

\newcommand*\xoverline[2][0.75]{
    \sbox{\myboxA}{$\m@th#2$}
    \setbox\myboxB\null
    \ht\myboxB=\ht\myboxA
    \dp\myboxB=\dp\myboxA
    \wd\myboxB=#1\wd\myboxA
    \sbox\myboxB{$\m@th\overline{\copy\myboxB}$}
    \setlength\mylenA{\the\wd\myboxA}
    \addtolength\mylenA{-\the\wd\myboxB}
    \ifdim\wd\myboxB<\wd\myboxA
       \rlap{\hskip 0.5\mylenA\usebox\myboxB}{\usebox\myboxA}%
    \else
        \hskip -0.5\mylenA\rlap{\usebox\myboxA}{\hskip 0.5\mylenA\usebox\myboxB}%
    \fi}
\makeatother

\newcommand{\be}{\begin{equation}}
\newcommand{\ee}{\end{equation}}
\newcommand{\ba}{\begin{aligned}}
\newcommand{\ea}{\end{aligned}}

\def\mb{\mathbb}

\def\mc{\mathcal}
\def\bp{\begin{pmatrix}}
\def\ep{\end{pmatrix}}

\def\br{\breve}
\def\mk{\mathfrak}

\begin{document}

\newcommand{\main}{.}
\begin{titlepage}

\setcounter{page}{1} \baselineskip=15.5pt \thispagestyle{empty}

\bigskip\

\vspace{2cm}
\begin{center}
{\Huge
A Tale of Bulk and Branes:\\
\vspace{0.5cm}
Symmetry TFT of 6D SCFTs from IIB/F-theory
}
\end{center}

\vspace{1cm}

\begin{center}
Jiahua Tian$^{1}$,\ Yi-Nan Wang$^{2,3}$

\vspace{1 cm}

\emph{$^1$School of Physics and Electronic Science, East China Normal University, \\Shanghai, China, 200241}\\
\vspace{.3cm}
\emph{$^2$School of Physics, Peking University, \\Beijing 100871, China}\\
\vspace{.3cm}
\emph{$^3$Center for High Energy Physics, Peking University, \\
Beijing 100871, China}

\end{center}

\vspace{1cm}
\noindent

\begin{abstract}

We study the 7D Symmetry Topological Field Theory (SymTFT) associated to a 6D SCFT from the IIB/F-theory geometric engineering approach. The 6D (2,0) or (1,0) SCFT is constructed from IIB on a non-compact complex surface possibly with 7-branes. We derive the general form of 7D SymTFT actions from the compactification of IIB action on the boundary link of the base manifold of an elliptic Calabi-Yau threefold, for both the cases with or without flavor 7-branes intersecting the boundary link. Along the way we find new terms in the SymTFT action from the worldvolume action of flavor 7-branes involving the flavor center symmetries. We crosscheck the results against those obtained from either holographic constructions or the dual M-theory picture. Our construction potentially leads to a classification of the 7D SymTFTs which parallels the known geometric classification of the 6D SCFTs.

\end{abstract}

\end{titlepage}

\clearpage

\tableofcontents

\newpage

\section{Introduction and summary}

The Symmetry Topological Field Theory (SymTFT) provides a unified way to describe a class of quantum field theories living in flat spacetime $\mb{R}^{d-1,1}$ with different global forms~\cite{Witten:AdSCFT&TFT,Kong:2020cie,Gaiotto:2020iye,Sakura:SymTFT_String,Apruzzi:6DSymTFT,Moradi:2022lqp,Hubner:GenSymm_Ftheory,Freed:2022qnc,Sakura:NoninvertibleHolography,Kaidi:2022cpf,Sakura:3DSymTFT,Freed:2022iao,Kaidi:2023maf,Chen:2023qnv,Bhardwaj:Generalized_Charge_II,Cao:2023rrb,Apruzzi:2023uma,Bhardwaj:2023idu,Bhardwaj:2023fca,Baume:2023kkf,Wen:2023otf,Lan:2023uuq,Bhardwaj:2023bbf,Brennan:2024fgj,Antinucci:2024zjp,Bonetti:2024cjk,Apruzzi:2024htg,DelZotto:2024tae,Bhardwaj:2024qrf,Wen:2024udn,Franco:2024mxa,Putrov:2024uor,Bhardwaj:2024kvy,Bhardwaj:2024ydc,Copetti:2024onh,Antinucci:2024bcm,Antinucci:2024ltv,Cvetic:2024dzu,Bhardwaj:2024xcx,GarciaEtxebarria:2024jfv,Bhardwaj:2024igy,Argurio:2024ewp,Chen:2024fno}. More precisely, a SymTFT $\mathcal{T}_{sym}$ is a topological field theory living in $\mb{R}^{d-1,1}\times [0,1]$, which has two boundaries: the topological boundary at one end of the interval $[0,1]$ and the physical (dynamical) boundary at the other end. 
With a choice of a proper set of topological boundary conditions of the fields in $\mathcal{T}_{sym}$, the generalized symmetries~\cite{Gaiotto_2015, McGreevy:2022oyu, Sakura:ICTP_Symmetries, Brennan:2023mmt,Bhardwaj:2023kri,Luo:2023ive} hence the global form of QFT$_d$ obtained after collapsing $\mb{R}^{d-1,1}\times [0,1]$ to $\mb{R}^{d-1,1}$ are fixed. On the other hand, the local dynamics of this QFT$_d$ are encoded in the dynamical boundary conditions of the fields in the corresponding $\mathcal{T}_{sym}$~\cite{Witten:AdSCFT&TFT, Sakura:SymTFT_String, Sakura:3DSymTFT, Kaidi:2022cpf}.

Interestingly, when a superconformal field theory (SCFT) living in flat space-time is geometrically engineered by putting superstring/M-theory on a non-compact space $X$, the action of the corresponding $\mathcal{T}_{sym}$ living in one-higher dimension can typically be derived by a dimensional reduction of the kinetic and the topological terms of the 10D/11D SUGRA action on the boundary link space $L:=\partial X$ of $X$~\cite{Sakura:SymTFT_String,Sakura:3DSymTFT}.

On the other hand, in the F-theory~\cite{Vafa:Ftheory,Morrison:1996na,Morrison:1996pp,Weigand:Ftheory} setup which is IIB superstring with 7-brane profile, the action of $\mathcal{T}_{sym}$ should include not only the dimensional reduction of the bulk IIB SUGRA action, but also the contribution from the 7-branes. For 6D SCFTs geometrically engineered by F-theory on non-compact elliptic Calabi-Yau threefolds $X_3$~\cite{Heckman:6DBaseClassification,Heckman:AtomicClassification}, their higher-form symmetries and the global form of flavor symmetries have been worked out~\cite{Dierigl:2020myk,Bhardwaj:Higher_form_5D6D,Apruzzi:2020zot,Apruzzi:2021mlh,Heckman:2022suy,Hubner:GenSymm_Ftheory}. While the SymTFT action of the higher-form symmetries have been studied either by exploiting the topological data of the tensor branch of the corresponding 6D SCFT~\cite{Apruzzi:2020zot, Apruzzi:6DSymTFT} or by looking at the dual M-theory description~\cite{Hubner:GenSymm_Ftheory}, that of the 0-form (flavor) symmetries, which are inevitably part of the whole set of the global symmetries of the 6D SCFT in the presence of flavor 7-branes, is not yet fully investigated.

In this work, we derive the full SymTFT action of a 6D SCFT directly from IIB/F-theory compactification, including the background gauge fields for non-abelian 0-form symmetries and their center (flavor center). Given a 6D SCFT obtained from F-theory  on an elliptic CY3 $E\hookrightarrow X_3\rightarrow B_2$, the key observation we make in this work is that the action of $\mathcal{T}_{sym}$ can be obtained from dimensionally reducing the combination of the IIB SUGRA action and the flavor 7-brane action on the boundary link $L$ of the non-compact base variety $B_2$. Since the forms of both the IIB SUGRA action and the 7-brane action are fixed, the action of $\mathcal{T}_{sym}$ is completely determined by the topological data of $L$. In particular, we propose that the $SL(2,\mathbb{Z})$-neutral field strength $F_5$ is reduced on $H^*(L,\mathbb{Z})$ while the $SL(2,\mathbb{Z})$-doublet B-field and its field strength doublet $(H_3,F_3)$ are reduced on the twisted cohomology groups $H^*(L,(\mathbb{Z}\oplus\mathbb{Z})_\rho)$ where the twist $\rho$ is determined by the monodromies generated by the 7-branes on $B_2$. Due to the appearance of the combination $F_2 - B_2$ in the 7-brane action, the worldvolume field strength $F_2$ must be reduced in the same way as the doublet B-field for the consistency of the dimensional reduction. The above ansatz of reduction can be lifted in terms of differential cohomology groups of $L$ which enables us to conveniently capture the non-trivial contributions to the action of $\mathcal{T}_{sym}$ coming from the reduction of various fields on the torsional classes of $L$ with untwisted or twisted coefficients. In principle, one can achieve the computation of $\mathcal{T}_{sym}$ purely from the boundary geometry, without studying the details of the tensor branch of the theory.

For some detailed coefficients in $\mathcal{T}_{sym}$, we have also compared them with the 5D KK theory of the 6D SCFT on $S^1$, which can be alternatively constructed from M-theory on the resolved elliptic CY3. Following the results of \cite{Sakura:SymTFT_String,Hubner:GenSymm_Ftheory}, these coefficients can be computed via triple intersection number calculations.

As a working example, we show the case of the 6D (1,0) SCFT with the tensor branch
\be
\overset{\mk{so}(2n+8)}{(-4)}-[\mk{sp}(2n)]\,,
\ee
whose 7D SymTFT action is derived to be~\footnote{We do not distinguish $d$ and $\delta$ in the paper since the background gauge fields can be chosen as either $U(1)$-valued or $\mb{Z}_n$ valued.}
\be
\frac{S_{\mathcal{T}_{sym}}}{2\pi}=\int_{\mc{M}^7}\frac{1}{8}a_3da_3+\frac{1}{2} f_2df_4+\frac{1}{2}\j_1 d\Upsilon_5+\frac{1}{2}a_3 f_2 f_2-\frac{1}{4}\j_1 f_2 w(F_{2,\mathfrak{sp}(2n)})^2-\frac{1}{4}a_3 w(F_{2,\mk{sp}(2n)})^2\,.
\ee
Here $a_3\in H^3(\mc{M}^7,\mb{Z}_4)$ is a $\mb{Z}_4$ valued 3-form background gauge field for a $\mb{Z}_2$ 2-form symmetry after choosing a polarization.  $(f_2,f_4)$ are the background gauge fields for the dual pair of 1-form/3-form $\mb{Z}_2$ symmetries. $(\j_1,\Upsilon_5)$ are the background gauge fields for a speculated dual pair of 0-form/4-form $\mb{Z}_2$ symmetries. In fact, $\j_1$ corresponds to the potential flavor center symmetry $\mb{Z}_2\subset Sp(2n)$. $F_{2,\mathfrak{sp}(2n)}$ is the field strength for the $\mathfrak{sp}(2n)$ non-abelian flavor symmetry and $w(F_{2,\mathfrak{sp}(2n)})$ is the second (generalized) Stiefel-Whitney class describing the obstruction of lifting an $Sp(2n)/\mb{Z}_2$ bundle to an $Sp(2n)$ bundle. 

Another example is the 6d (1,0) $(A_{n-1},A_{n-1})$ conformal matter~\cite{DelZotto:2014hpa} with the tensor branch
\be
[\mk{su}(n)]-\overset{\mk{su}(n)}{(-2)}-\overset{\mk{su}(n)}{(-2)}-[\mk{su}(n)]\,.
\ee
The SymTFT action reads
\be
\frac{S_{\mathcal{T}_{sym}}}{2\pi}=\int_{\mathcal{M}^7} \frac{1}{6} a_3 d a_3+\frac{1}{n}\j_{1} d \Upsilon_{5}-\frac{n-1}{3n}a_3(w(F_{2,\mk{su}(n)_1})^2-w(F_{2,\mk{su}(n)_2})^2)\,.
\ee
$a_3$ is the background gauge field for a potential $\mb{Z}_3$ 2-form symmetry. $(\j_{1},\Upsilon_5)$ are the background gauge fields of the dual $\mb{Z}_n$ 0-form/4-form symmetries, where $\j_1$ corresponds to the $\mb{Z}_n$ flavor center. 
$F_{2,\mk{su}(n)_i}$ denotes the field strength for the $i$-th $\mk{su}(n)$ flavor symmetry from left to the right and $w$ stands for the second Stiefel-Whitney class describing the obstruction of lifting a $PSU(n)$ bundle to an $SU(n)$ bundle.

The structure of this paper is outlined as follows. We first review in Section~\ref{sec:IIB_action_lift} certain aspects of IIB SUGRA action and its topological lift that will be useful for the subsequent discussions. We give a detailed discussion of 6D $\mathcal{N} = (2,0)$ theories in Section~\ref{sec:SymTFT_20} and of 6D $\mathcal{N} = (1,0)$ theories without and with flavor branes in Section~\ref{sec:6D10_noflav} and Section~\ref{sec:6D10_flav}, respectively. We calculate $H^*(L,\mathbb{Z})$ and $H^*(L,(\mathbb{Z}\oplus\mathbb{Z})_\rho)$ in several representative cases. We will see in those cases that as $H^*(L,\mathbb{Z})$ is determined by $\pi_1(L)$, $H^*(L,(\mathbb{Z}\oplus\mathbb{Z})_\rho)$ can be calculated by analyzing the monodromies of both the gauge and the flavor 7-branes living in $B_2$. We will compare the results with those obtained from the dual M-theory construction in Section~\ref{sec:Mtheory}. We present more non-trivial examples in Section~\ref{sec:examples}. Finally in appendix \ref{sec:CY3SymTFT} we discuss some aspects of relative homology in the elliptic CY3 cases and in appendix \ref{app:cohomo_twistcoeff} we summarize knowledge of twisted cohomology.

\section{IIB action and its topological lift}\label{sec:IIB_action_lift}

The action of SymTFT $\mathcal{T}_{sym}$ generally takes the form:
\begin{equation}
\label{SymTFT-action}
    \frac{S_{\mathcal{T}_{sym}}}{2\pi} = \int_{M^{d+1}} \sum_{r=0}^{p-1} \frac{1}{l_r}a_{r+1}\cup \delta b_{d-r-1} + \mathcal{A}(a_1,\cdots,a_p)
\end{equation}
with the BF-couplings $a_{r+1}\cup \delta b_{d-r-1}$ and the \emph{twist} $\mathcal{A}(a_1,\cdots,a_p)$~\cite{Bhardwaj:Generalized_Charge_I, Bhardwaj:Generalized_Charge_II} and the gauge fields $a_{r+1}\in H^{r+1}(M^{d+1},\mb{Z}_{l_r})$, $b_{d-r-1}\in H^{d-r-1}(M^{d-r-1},\mb{Z}_{l_r})$. Note that we use the normalization that the flux of $a_{r+1}$ and $b_{d-r-1}$ are quantized as $\int_{M^{r+2}}\delta a_{r+1}\in\mb{Z}$, $\int_{M^{d-r}}\delta b_{d-r-1}\in\mb{Z}$ for any cycles $M^{r+2}$, $M^{d-r}$. It was shown in~\cite{Sakura:SymTFT_String} that $\mathcal{T}_{sym}$ can be obtained from string/M-theory compactification on the boundary link of a non-compact space $M^{D-d}$ for $D = 10$ or $11$ and in particular $S_{\mathcal{T}_{sym}}$ can be obtained from the reduction of 10D or 11D SUGRA with a set of suitable rules of reduction of the various fields in the SUGRA action. In particular, it was shown explicitly in~\cite{Sakura:SymTFT_String, Sakura:3DSymTFT} that $\mathcal{A}(a_1,\cdots,a_p) \subset S_{\mathcal{T}_{sym}}$ can be obtained from properly reducing the topological couplings of the 11D SUGRA and in~\cite{Apruzzi:2023uma} that each $\int\frac{1}{l_r}a_{r+1}\delta b_{d-r-1}\subset S_{\mathcal{T}_{sym}}$ can be obtained from reducing the BF-couplings in an auxiliary $(D+1)$-dimensional action.

In this work we focus on the top-down construction of $\mathcal{T}_{sym}$ associated to a 6D SCFT $\mathcal{T}_{6D}$ obtained from IIB compactification on a two complex-dimensional non-compact variety $B_2$. As the base geometry is classified in~\cite{Heckman:6DBaseClassification, Heckman:AtomicClassification}, we will devote this section to the discussion of the auxiliary 11D action which is the topological lift of the IIB SUGRA action, of which we will later perform a dimensional reduction on the boundary link $L:=\partial B_2$ to obtain $S_{\mathcal{T}_{sym}}$.

The familiar 10D IIB SUGRA action is given by~\cite{Polchinski:String_vol2}:
\begin{equation}\label{eq:IIB_action}
    S_{\text{IIB}} = \frac{1}{2\kappa_0^2} \int d^{10}x \sqrt{-g} \left( R - \frac{\partial_\mu \tau \partial^\mu\overline{\tau}}{ 2 \tau_2^2} - \mathcal{M}_{ij}F_3^i\cdot F_3^j - \frac{1}{4}|\widetilde{F}_5|^2 \right) -  \frac{\epsilon_{ij}}{8\kappa_0^2} \int C_4\wedge F_3^i\wedge F_3^j
\end{equation}
where
\begin{equation}\label{eq:defs_fields}
    F_3^i = dB_2^i = \begin{pmatrix}
        dB_2 \\
        dC_2
    \end{pmatrix},\  \widetilde{F}_5 = dC_4 + \frac{1}{2} \epsilon_{ij} B_2^i\wedge F_3^j,\ \mathcal{M} = \frac{1}{\tau_2} \begin{pmatrix}
        |\tau|^2 & -\tau_1 \\
        -\tau_1 & 1
    \end{pmatrix}\,.
\end{equation}
Nevertheless, the above action is not convenient for a top-down construction of $S_{\mathcal{T}_{sym}}$. Instead, we rewrite $S_{\text{IIB}}$ as the following auxiliary \emph{topological IIB action} in an 11D space with a proper normalization~\cite{Belov:2006xj, Apruzzi:2023uma}~\footnote{We would often omit the wedge, cup and star products throughout the paper.} (see also \cite{Lawrie:2023tdz,Yu:2023nyn} for similar discussions):
\begin{equation}\label{eq:topIIB}
    \frac{S_{\text{top-IIB}}}{2\pi} := \int_{M^{11}} I_{11} = \int_{M^{11}} \frac{1}{2}\widetilde{F}_5 d\widetilde{F}_5 + \widetilde{F}_1d\widetilde{F}_9 + H_3d\widetilde{H}_7 - \widetilde{F}_3d\widetilde{F}_7 + H_3 \widetilde{F}_1 \widetilde{F}_7 - H_3\widetilde{F}_3\widetilde{F}_5\,.
\end{equation}
The various fields in~(\ref{eq:topIIB}) satisfy the following set of Bianchi identities \emph{on-shell}:
\begin{equation}\label{eq:Bianchi}
    dH_3 = 0,\ d\widetilde{H}_7 = \widetilde{F}_3\widetilde{F}_5 - \widetilde{F}_1\widetilde{F}_7,\ d\widetilde{F}_p = H_3\widetilde{F}_{p-2}\,.
\end{equation}
Using~(\ref{eq:Bianchi}) it is not hard to check that $d I_{11} = 0$.

\begin{figure}[h]
    \centering
    \includegraphics[width=0.7\textwidth]{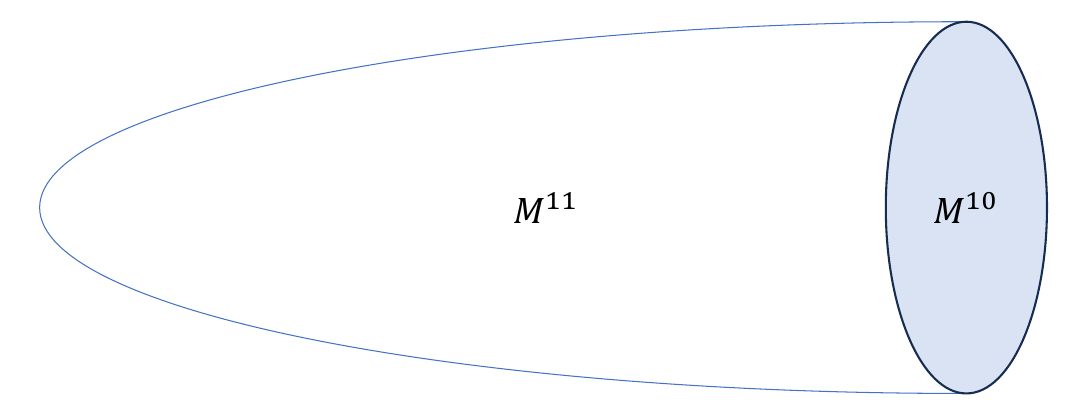}
    \caption{$M^{10}$ as the boundary of $M^{11}$.}
    \label{fig:M11_M10}
\end{figure}

To see the equivalence between $S_{\text{top-IIB}}$ and $S_{\text{IIB}}$ in the physical 10D spacetime $M^{10}$, we take a closer look at the relation between $M^{11}$ and $M^{10}$ which is illustrated in Figure~\ref{fig:M11_M10}.
The physical 10D spacetime $M^{10}$ is the boundary of $M^{11}$ where $S_{\text{top-IIB}}$ lives. Note that though the physical spacetime $M^{10}$ must be equipped with a metric, the auxiliary $M^{11}$ does not have to be so since after all $S_{\text{top-IIB}}$ is defined without metric.

To recover $S_{\text{IIB}}$ from $S_{\text{top-IIB}}$, we plug in~(\ref{eq:topIIB}) the solutions of~(\ref{eq:Bianchi}) and apply Stokes' theorem on $M^{11}$. For example, from~(\ref{eq:Bianchi}) we have~\cite{Bergshoeff:2001pv}
\begin{equation}
    \widetilde{F}_3 = F_3 - H_3C_{0},\ \widetilde{F}_5 = F_5 - H_3C_2
\end{equation}
with closed $F_3$ and $F_5$, from which we have:
\begin{equation}
    H_3 \widetilde{F}_3 \widetilde{F}_5 = H_3 (F_3 - H_3C_0)(F_5 - H_3C_2) = H_3 F_3 F_5
\end{equation}
since $H_3H_3 \equiv 0$. Therefore, by Stokes' theorem we have
\begin{equation}
    \frac{S_{\text{top-IIB}}}{2\pi} \supset \int_{M^{11}} - H_3 \widetilde{F}_3 \widetilde{F}_5 = \int_{M^{11}} d (- H_3 F_3 C_4) = \int_{M^{10}} - H_3F_3C_4
\end{equation}
which correctly reproduces the IIB SUGRA topological coupling in~(\ref{eq:IIB_action}) up to an overall constant.

The kinetic term in~(\ref{eq:IIB_action}) can also be reproduced in essentially the same manner. For that it is convenient to introduce sources for $F_p$ for $p\geq 5$ which will be dropped later, in other words we temporarily assume $dF_p$ is not identically 0 for $p\geq 0$. In this case we have:
\begin{equation}
    \begin{split}
        H_3d\widetilde{H}_7 - \widetilde{F}_3d\widetilde{F}_7 \supset H_3 d H_7 - F_3 d F_7 = d (H_3 H_7 - F_3 F_7)\,.
    \end{split}
\end{equation}
Applying the Hodge relation $H_7 = \tau_2^2 * H_3$ and $F_7 = - * F_3$ the above term becomes $\tau_2^2 H_3\cdot H_3 + F_3\cdot F_3$ on $M^{10}$ which is proportional to the kinetic term $\mathcal{M}_{11} H_3\cdot H_3 + \mathcal{M}_{22} F_3\cdot F_3$ in~(\ref{eq:topIIB}).

In general, one can lift the 10/11D action to a topological action in one higher dimension which after dimensional reduction on a boundary link reproduces both the BF-couplings and the twist in $S_{\mathcal{T}_{sym}}$. The validity of this method can also be seen from matching the resulting equations of motion as in~\cite{Apruzzi:2023uma}. It has to be kept in mind that the relations involving Hodge star on $M^{10}$ have to be implemented by hand.

For our purpose it will be more convenient to further rewrite $S_{\text{top-IIB}}$ in terms of differential cohomology of $L$, which captures the torsional structure of the spacetime in a more natural way~\footnote{For an introduction of differential cohomology and its applications in physics, in particular in our context, see~\cite{HopkinsSinger, Hsieh:2020jpj, Sakura:SymTFT_String}.}. For this let us recall that the \emph{bordism formula} says that given a $(d+1)$-dimensional space $M^{d+1}$ with boundary $M^d := \partial M^{d+1}$ and a differential cohomology class $a\in\breve{H}^{d+1}(M^d,\mathbb{Z})$, we have
\begin{equation}
    \int_{M^d} \breve{a}|_{M^d} \equiv \int_{M^{d+1}} R(\breve{a}) \mod \mathbb{Z}
\end{equation}
where $R$ is defined via the short exact sequence:
\begin{equation}
    0 \rightarrow H^{d}(M^{d+1}, \mathbb{R}/\mathbb{Z}) \xrightarrow{i} \breve{H}^{d+1}(M^{d+1}, \mathbb{Z}) \xrightarrow{R} \Omega^{d+1}(M^{d+1}) \rightarrow 0\,.
\end{equation}
Here $\Omega^{d+1}(M^{d+1})$ is the group of closed $(d+1)$-forms. In our setup we have $d = 10$ and $R(\breve{a}) = I_{11}$ which is indeed closed, therefore we need to find the pre-image of $R$ which will be the desired differential cohomological lift $\breve{I}_{11}$ of $I_{11}$ the integration of which (more precisely its restriction to $M^{10}$) over $M^{10}$ gives the \emph{secondary invariant} of $M^{10}$. In other words, we will focus on the reduction of
\begin{equation}\label{eq:topIIB_DC_S}
    \frac{S}{2\pi} = \int_{M^{10}} \breve{I}_{11}
\end{equation}
on any consistent boundary link. From~(\ref{eq:topIIB}) it is natural to see that
\begin{equation}\label{eq:topIIB_DC}
    \breve{I}_{11} = \frac{1}{2} \breve{F}_5 \delta \breve{F}_5 + \breve{F}_1 \delta \breve{F}_9 + \breve{H}_3 \delta \breve{H}_7 - \breve{F}_3 \delta \breve{F}_7 + \breve{H}_3 \breve{F}_1 \breve{F}_7 - \breve{H}_3 \breve{F}_3 \breve{F}_5\,.
\end{equation}

It is subtle that $\breve{I}_{11}$ is written in terms of $\breve{F}_{p}$'s as if each $\breve{F}_{p}$ is the pre-image of $R$ of $\widetilde{F}_{p}$ for various $p$-form field strengths, whereas $\widetilde{F}_{p}$ is non-closed. Generally one can write each $\widetilde{F}_p$ as a sum of non-closed and closed parts as follows:
\begin{equation}
    \widetilde{F}_p = \widetilde{F}_p^{(nc)} + \widetilde{F}_p^{(c)}.
\end{equation}
The pre-image of $R$ of the closed part $\widetilde{F}_p^{(c)}$ will be our $\breve{F}_p$ in~(\ref{eq:topIIB_DC}). The non-closed part $\widetilde{F}_p^{(nc)}$ plays the role of background flux as in~\cite{Apruzzi:2023uma} which always integrates to integer on cycles of $\partial B$ due to flux quantization when performing string compactification on a non-compact variety $B$. Since the reduction of $\widetilde{F}_p^{(nc)}$ leads to integral shifts which is inessential to our purpose, we will simply ignore their contribution and focus only on $\widetilde{F}_p^{(c)}$'s and their differential cohomological lifts $\breve{F}_p$. Note that $\breve{I}_{11}$ will play the same role in our setup as the M-theory differential cohomological action $\breve{I}_{12}$ played in~\cite{Sakura:SymTFT_String, Sakura:3DSymTFT}.

A missing important property of~(\ref{eq:topIIB_DC}) is its manifest $SL(2,\mb{Z})$-invariance, which will play an essential role in subsequent discussions. Physically, to ensure that the field after decomposition in spacetime is well-defined under an $SL(2,\mathbb{Z})$ monodromy $\rho$, $(\breve{H}_3, \breve{F}_3)$ should be decomposed on the twisted differential cohomology group $\breve{H}^*(L,(\mathbb{Z}\oplus \mathbb{Z})_\rho)$ where $\rho$ is a twist represented by an $SL(2,\mathbb{Z})$ matrix to be determined by the 7-brane profile~\cite{Douglas:Cremmer-Scherk, Katz:DimRed_B-field, Heckman:Brane_GenSymm}. To capture the monodromy action on $(\breve{H}_3, \breve{F}_3)$ in a convenient way, we introduce the notation $\breve{F}_3^\rho$ to be used throughout this work following the notation in~\cite{Heckman:Brane_GenSymm}. The superscript $\rho$ in $\breve{F}_3^\rho$ means that it is defined to be compatible with the monodromy $\rho$, or more precisely we have $\breve{F}_3^\rho\in \breve{H}^3(M^{10}, (\mathbb{Z}\oplus\mathbb{Z})_\rho)$. Then, by K\"unneth formula we must consider the decomposition:
\begin{equation}
    \breve{H}^3(M^{10}, (\mathbb{Z}\oplus\mathbb{Z})_\rho) = \breve{H}^{3-p}(\mathcal{M}^{7}, (\mathbb{Z}\oplus\mathbb{Z})_\rho)\times \breve{H}^{p}(L, (\mathbb{Z}\oplus\mathbb{Z})_\rho)
\end{equation}
given $M^{10} = \mathcal{M}^7\times L$. Concretely, in general we have~\cite{Heckman:Brane_GenSymm}
\begin{equation}\label{eq:def_F3rho}
    \breve{F}_3^\rho = \begin{pmatrix}
        \breve{H}_3 \\
        \breve{F}_3
    \end{pmatrix} / \text{Im}(\rho - 1) = \sum_{p=0}^3 \breve{f}_{p}^\rho \breve{u}_{3-p}\,,
\end{equation}
where $\breve{f}_{p}^\rho$ is a $p$-form field in spacetime $\mathcal{M}^7$ that is well-defined in the presence of $\rho$-monodromy.

After packaging $(\breve{H}_3, \breve{F}_3)$ into $\breve{F}_3^\rho$, we need to study the pairing between $\breve{H}_3$ and $\breve{F}_3$ which in particular appears in $\breve{H}_3 \breve{F}_3 \breve{F}_5$ in~(\ref{eq:topIIB_DC}). For this we use the Dirac pairing on the lattice $\mathbb{Z}^2/\text{Im}(\rho - 1)$ which descends from the Dirac pairing on the lattice $\mathbb{Z}^2$. In the simplest case where there are no 7-branes so that $\rho = 1$ and $\breve{F}_3^\rho = (\breve{H}_3, \breve{F}_3
)$~\footnote{We will explain in more detail and make use of this point in section~\ref{sec:SymTFT_20}.}, we have
\begin{equation}
    \breve{H}_3 \breve{F}_3 = \frac{1}{2} \breve{F}_3^\rho \breve{F}_3^\rho
\end{equation}
where on the right hand side the ordinary Dirac pairing is applied. Motivated by this we rewrite $\breve{H}_3 \breve{F}_3 \breve{F}_5$ as $\frac{1}{2} \breve{F}_3^\rho \breve{F}_3^\rho \breve{F}_5$ for arbitrary $\rho$.

It is less obvious how to rewrite the BF terms $\breve{H}_3 \delta \breve{H}_7 - \breve{F}_3 \delta \breve{F}_7$ in a manifestly $SL(2,\mathbb{Z})$-invariant fashion. For this it is illuminating to recall that the B-field $pB_2 + qC_2$ couples electrically to a $(p,q)$-string while the dual field $rB_6 + sC_6$ couples electrically to an $(s,r)$ 5-brane. Therefore a meaningful Dirac pairing should combine $(B_2, C_2)$ with $(C_6, B_6)$. We define $\breve{F}_7^\rho = (\breve{F}_7, \breve{H}_7)/\text{Im}(\rho-1)$. Thus in the simplest case where $\rho = 1$ we have
\begin{equation}
    \breve{H}_3 \delta \breve{H}_7 - \breve{F}_3 \delta \breve{F}_7 = \breve{F}_3^\rho \delta \breve{F}_7^\rho
\end{equation}
with the ordinary Dirac pairing between $\breve{F}_3^\rho = (\breve{H}_3, \breve{F}_3)$ and $\breve{F}_7^\rho = (\breve{F}_7, \breve{H}_7)$. Motivated by this we rewrite $\breve{H}_3 \delta \breve{H}_7 - \breve{F}_3 \delta \breve{F}_7$ as $\breve{F}_3^\rho \delta \breve{F}_7^\rho$ for arbitrary $\rho$.

In summary, the starting point of our top-down construction of $S_{\mathcal{T}_{sym}}$ from IIB is
\begin{equation}\label{eq:IIB_starting_point}
    \frac{S_{\br{\text{IIB}}}}{2\pi} = \int_{M^{10}} \breve{I}_{11} := \int_{M^{10}} \frac{1}{2} \breve{F}_5 \delta \breve{F}_5 + \breve{F}_3^\rho \delta \breve{F}_7^\rho - \frac{1}{2} \breve{F}_3^\rho \breve{F}_3^\rho \breve{F}_5 + \breve{F}_1 \delta \breve{F}_9 + \breve{H}_3 \breve{F}_1 \breve{F}_7\,.
\end{equation}
Note that though the field $\breve{F}_1$ and its dual $\breve{F}_9$ also transform non-trivially under monodromy while the couplings involving them have not been written in a manifestly $SL(2,\mathbb{Z})$-invariant fashion, we will show in subsequent sections that 
their dimensional reduction on $L$ do not contribute non-trivially to $S_{\mathcal{T}_{sym}}$.

\section{SymTFT of 6D $\mathcal{N}=(2,0)$ theory}\label{sec:SymTFT_20}

In this section we study $\mathcal{T}_{sym}$ of a 6D $\mathcal{N} = (2,0)$ theory $\mathcal{T}^{(2,0)}_{6D}$. The action of $\mathcal{T}_{sym}$ has been derived in~\cite{Apruzzi:6DSymTFT} relying on the topological data of the tensor branch of $\mathcal{T}^{(2,0)}_{6D}$. In contrast to the known approach, we will re-derive $\mathcal{T}_{sym}$ in a top-down manner by directly dimensionally reducing $S_{\breve{\text{IIB}}}$~(\ref{eq:IIB_starting_point}) on the boundary link of the base variety which in this case is a non-compact K3 surface. More precisely, since $\mathcal{T}^{(2,0)}_{6D}$ is obtained via IIB compactification on $B_2 := \mathbb{C}^2/\Gamma$ for a finite subgroup $\Gamma$ of $SU(2)$, one can obtain the corresponding $\mathcal{T}_{sym}$ by compactifying IIB on $L := \partial B_2 = S^3/\Gamma$ along similar lines as in~\cite{Sakura:SymTFT_String, Sakura:3DSymTFT}. Note that the computation of SymTFT action from IIB action was also done in \cite{Lawrie:2023tdz}.

In this case the 10D spacetime is $\mathbb{R}^{1,5}\times B_2 = \mathbb{R}^{1,5}\times \mathbb{R}_+\times L := \mathcal{M}^7\times L$ where $L = S^3/\Gamma$. It is known that in general $\breve{H}^p(L)$ is non-trivial for $p = 0,2,3$~\cite{Sakura:SymTFT_String, Heckman:Brane_GenSymm}. Clearly, $\breve{F}_5$ should be reduced along elements of $\breve{H}^p(L,\mathbb{Z})$. Hence we have
\begin{equation}\label{eq:F5_ansatz}
    \breve{F}_5 = \breve{a}_5\star \breve{1} + \breve{a}_3\star \breve{u}_2 + \breve{a}_2\star \breve{v}
\end{equation}
where $\breve{1} \in \breve{H}^0(L)$, $\breve{u}_2 \in \breve{H}^2(L)$ and $\breve{v} \in \breve{H}^3(L)$. In~(\ref{eq:F5_ansatz}), $\breve{a}_2$ is non-dynamical since $\text{vol}(L)$ is formally infinite and $\breve{a}_5$ vanishes due to the self-duality of $\breve{F}_5$~\cite{Heckman:Brane_GenSymm}. Therefore the only non-trivial term in $S_{\mathcal{T}_{sym}}$ that descends from dimensionally reducing $\frac{1}{2}\breve{F}_5d \breve{F}_5 \subset \breve{I}_{11}$ on $L$ is
\begin{equation}\label{eq:BF_2form_20}
    \frac{S_{\br{\text{IIB}}}}{2\pi} \supset \frac{1}{2}\int_{L}\breve{u}_2\star\breve{u}_2 \int_{\mathcal{M}^7} a_3d a_3\,,
\end{equation}
which is immediately recognized to be the BF coupling of the background 3-form field of the 2-form symmetry of $\mathcal{T}^{(2,0)}_{6D}$ where the integral $\frac{1}{2}\int_{L}\breve{u}_2\star\breve{u}_2$ is the \emph{spin Chern-Simons invariant} $\text{CS}[L]$ of $L$~\cite{Sakura:SymTFT_String, GarciaEtxebarria:IIB_flux_noncomm}.

As a concrete example, when $\Gamma = \mathbb{Z}_n$ we have $2\text{CS}[L] = \frac{1-n}{n} \equiv \frac{1}{n} \mod 1$~\cite{Sakura:SymTFT_String}, hence~(\ref{eq:BF_2form_20}) becomes
\begin{equation}\label{eq:BF_20}
    \frac{1}{2n} \int_{\mathcal{M}^7}a_3d a_3
\end{equation}
which after quantization leads to the conformal blocks of the $A_n$ type 6D $\mathcal{N} = (2,0)$ theory~\cite{Witten:AdSCFT&TFT, Witten:Geometric_Langlands_6D}. Traditionally, to obtain~(\ref{eq:BF_20}) via geometric engineering, one has to first calculate the intersection matrix of the tensor branch of the 6D SCFT then take its Smith normal form to read off the crucial $1/2n$ factor~\cite{Apruzzi:6DSymTFT, Morrison:HigherForm_5D}. The fact that we get the correct BF-term without making use of the tensor branch geometry reflects the fact that our approach depends only on the topology of $L$ rather than that of the compact part of $B_2$.

The tricky part is the reduction of the doublet field strength $(\breve{H}_3, \breve{F}_3)$. As discussed in section~\ref{sec:IIB_action_lift} we need to consider the decomposition of $\breve{F}_3^\rho = (\breve{H}_3, \breve{F}_3)$ on
\begin{equation}\label{eq:Z2_20}
    \breve{H}^*(L,(\mathbb{Z}\oplus \mathbb{Z})_\rho) = \breve{H}^*(L,\mathbb{Z}\oplus \mathbb{Z}) = \breve{H}^*(L,\mathbb{Z})\oplus \breve{H}^*(L,\mathbb{Z})
\end{equation}
where the splitting is due to the triviality of $\rho$ in the absence of 7-branes. Therefore, in this case both $\breve{F}_3^1 := \breve{H}_3$ and $\breve{F}_3^2 := \breve{F}_3$ are reduced on $\breve{H}^*(L,\mathbb{Z})$ in the same manner as follows:
\begin{equation}\label{eq:B_ansatz_20}
    \begin{split}
        \breve{F}_3^i &= \breve{f}_3^\rho\star\breve{1}^i + \breve{f}_1^\rho\star \breve{u}_2^i,\ \text{for }\breve{1}^1 = \breve{1}^2 = \breve{1}\ \text{and}\ \breve{u}_2^1 = \breve{u}_2^2 = \breve{u}_2\,.
    \end{split}
\end{equation}
Given~(\ref{eq:B_ansatz_20}), the decomposition of $\breve{F}_3^\rho\breve{F}_3^\rho = \epsilon_{ij}\breve{F}_3^i \breve{F}_3^j$ on $\breve{H}^*(L,(\mathbb{Z}\oplus \mathbb{Z})_\rho)$ is
\begin{equation}\label{eq:trivial_F5F3F3_20}
    \epsilon_{ij}\breve{F}_3^i\breve{F}_3^j = \epsilon_{ij}\left( \breve{f}_1\breve{f}_1 \breve{u}_2^i\breve{u}_2^j + \breve{f}_1\breve{f}_3 \breve{u}_2^i\breve{1}^j + \breve{f}_3\breve{f}_1 \breve{1}^i\breve{u}_2^j + \breve{f}_3\breve{f}_3 \breve{1}^i\breve{1}^j \right)\,.
\end{equation}
which vanishes due to antisymmetrization by $\epsilon_{ij}$. This implies that there is no non-vanishing contribution to $S_{\mathcal{T}_{sym}}$ from the dimensional reduction of $\int \breve{B}_2\breve{H}_3\breve{F}_3\breve{F}_3$.

We now turn to the reduction of $\int \breve{F}_3^\rho\delta \breve{F}_7^\rho = \int \breve{F}_3^1\delta \breve{F}_7^1 - \breve{F}_3^2\delta \breve{F}_7^2$ which may lead to extra BF-couplings in $S_{\mathcal{T}_{sym}}$. The only possibly non-vanishing contribution to $S_{\mathcal{T}_{sym}}$ must come from the following decomposition:
\begin{equation}
    \breve{F}_3^i = \breve{f}_1^\rho\star \breve{u}_2^i,\ \breve{F}_7^i = \breve{f}_5^\rho\star \breve{u}_2^i
\end{equation}
where $\breve{F}_7^1 = \breve{F}_7$ and $\breve{F}_7^2 = \breve{H}_7$. The above decomposition leads to the term:
\begin{equation}\label{eq:trivial_BF_20}
    \int_L \epsilon_{ij}\breve{u}_2^i\breve{u}_2^j \int_{\mathcal{M}^7} f_1df_5
\end{equation}
which again vanishes due to the antisymmetrization by $\epsilon_{ij}$. Therefore, in the $\mathcal{N} = (2,0)$ case there is no BF-coupling for any $1$-form field.

As we have already commented in section~\ref{sec:IIB_action_lift} the couplings in~(\ref{eq:IIB_starting_point}) involving $\breve{F}_1$ are tricky since it is also not $SL(2,\mathbb{Z})$-neutral. In the $\mathcal{N} = (2,0)$ the situation is greatly simplified due to the absence of monodromy thus the only meaningful decomposition for $\breve{F}_1$ is $\breve{F}_1 = \breve{d}_1 \star \breve{1}$. This makes the only potentially non-vanishing contribution from $\int \breve{F}_1\delta \breve{F}_9$ to be proportional to the volume of $L$ which is infinite, hence there is no need consider its contribution to $S_{\mathcal{T}_{sym}}$. Physically, $\breve{d}_1$ is the field strength of a $U(1)$ $(-1)$-form symmetry whose curvature $R(\breve{d}_1)$ is equal to $d(\Re \tau) \in H^1(L,\mathbb{Z})$~\cite{Heckman:Brane_GenSymm,Vandermeulen:2022edk,Aloni:2024jpb,Santilli:2024dyz}. Since we do not consider $(-1)$-form symmetry in this work, it is also justified from this more physical point of view that we shall ignore its contribution to $S_{\mathcal{T}_{sym}}$. For the same reason we will not consider the contribution to $S_{\mathcal{T}_{sym}}$ from $\int\breve{H}_3\breve{F}_1\breve{F}_7$. A more complete analysis including $(-1)$-form symmetries will be presented in future works.

In summary, the only term in (\ref{eq:IIB_starting_point}) that contributes non-trivially to $S_{\mathcal{T}_{sym}}$ after dimensional reduction on $L$ is $\int \frac{1}{2} \breve{F}_5\delta\breve{F}_5$ which results in the BF-coupling~(\ref{eq:BF_2form_20}). Therefore, for 6D $\mathcal{N}=(2,0)$ theory we have
\begin{equation}\label{eq:SymTFT_20}
    \frac{S_{\mathcal{T}_{sym}}}{2\pi} = \frac{1}{2} \int_L \breve{u}_2\star \breve{u}_2 \int_{\mathcal{M}^7}a_3da_3 = \text{CS}[L] \int_{\mathcal{M}^7}a_3da_3
\end{equation}
where $a_3$ is the 3-form background field of the 2-form symmetry of $\mathcal{T}_{6D}^{(2,0)}$. In this case, it is clear from our discussion that the derivation of $S_{\mathcal{T}_{sym}}$ does not rely on the data of the compact part of $B_2$ at all.

\section{SymTFT of 6D $\mathcal{N}=(1,0)$ theory}\label{sec:SymTFT_10}

In this section we calculate $S_{\mathcal{T}_{sym}}$ corresponding to a 6D $\mathcal{N}=(1,0)$ SCFT $\mathcal{T}_{6D}^{(1,0)}$ whose tensor branch is $x_1 - x_2 - \cdots - x_r$, where $x_i$ denotes the negative of the self-intersection number of each compact rational curve on the blown up base surface. In this case we have $B_2 = \mathbb{C}^2/\Gamma$ where $\Gamma:=\frac{1}{p}(1,q) \simeq \mathbb{Z}_p \subset U(2)$ acts on $\mathbb{C}^2$ in the following manner~\cite{ToricVarieties:CoxLittleSchenck, Heckman:6DBaseClassification, Heckman:AtomicClassification}:
\begin{equation}\label{eq:pq_action}
    (z_1, z_2) \mapsto (\omega z_1, \omega^q z_2),\ \omega = e^{\frac{2\pi i}{p}}\ \text{for}\ \frac{p}{q} = x_1 - \frac{1}{x_2 - \frac{1}{\cdots - \frac{1}{x_r}}}\,.
\end{equation}

Again, rather than looking into the details of the tensor branch as in~\cite{Apruzzi:6DSymTFT}, we dimensionally reduce IIB with 7-branes on $L := \partial B_2$ directly. While in all cases we have $\Gamma\sim \mathbb{Z}_p$, hence $H_1(L,\mathbb{Z}) \cong \mathbb{Z}_p$, it is a lot more difficult to determine the cohomology of $SL(2,\mathbb{Z})$-twisted $(\mathbb{Z}\oplus \mathbb{Z})$-bundle on $L$ since there is now non-trivial monodromy $\rho$ in contrast to the simple relation~(\ref{eq:Z2_20}). We emphasize that in this section we study the theory whose tensor branch is linear, i.e. the base is $A$-type. Our method is general and does apply to $D$ and $E$-type bases as has already been discussed in the previous section.

Technically, there are two rather different cases to study. One case is when there are no 7-branes intersecting $L$ while the other is when there are such 7-branes which are called \emph{flavor 7-branes} in literature~\cite{Cvetic:Cut&Glue}. When there are no flavor 7-branes, the boundary remains $L$ while the effect of the gauge 7-branes living in the compact part of $B_2$ is fully reflected by the $\rho$-twist on certain cohomology group of $L$. The same effect in a slightly different context is analyzed in~\cite{Heckman:Brane_GenSymm}. The primary examples of the cases without flavor 7-brane are the \emph{non-Higgsable clusters} (NHC)~\cite{Morrison:NHC, Heckman:6DBaseClassification, Heckman:AtomicClassification}. We will see in section~\ref{sec:6D10_noflav} that the cases without flavor 7-branes are only slightly more complicated than the $\mathcal{N} = (2,0)$ cases due to the absence of the intersection between the 7-brane locus and $L$. On the other hand, the analysis becomes much subtler when there are flavor 7-branes. In this case one has to take into consideration the topological action of the 7-branes in addition to $S_{\text{top-IIB}}$. We will discuss all these complications and present the calculation of $S_{\mathcal{T}_{sym}}$ with flavor 7-branes in section~\ref{sec:6D10_flav}.

\subsection{SymTFT of 6D $\mathcal{N}=(1,0)$ theory without flavor branes}\label{sec:6D10_noflav}

In this section we calculate $S_{\mathcal{T}_{sym}}$ associated to a 6D $\mathcal{N}=(1,0)$ theory without flavor 7-branes. In this case we assume $\Delta\cap L = \emptyset$ where $\Delta$ is the loci of 7-branes in $B_2$. In principle, whenever there are 7-branes in the system one has to work with F-theory rather than (perturbative) IIB theory~\cite{Vafa:Ftheory}. But since $\Delta\cap L = \emptyset$ there is no need to take into consideration the 7-brane action in addition to~(\ref{eq:IIB_starting_point}) as long as the dimensional reduction is performed only on $L$, while not on the ``bulk'' $B_2$. Nevertheless the existence of 7-branes in the compact part of $B_2$ significantly modifies both the topology of $L$ and the fields living on it hence leads to distinct physical consequences than those in section~\ref{sec:SymTFT_20}. We will find that the dimensional reduction of~(\ref{eq:IIB_starting_point}) on $L$ leads to the correct $S_{\mathcal{T}_{sym}}$ as long as the twist of certain cohomology group of $L$ due to the 7-brane monodromy is appropriately taken care of.

Though the boundary link $L$ has now become a quotient of $S^3$ by a finite subgroup of $U(2)$ rather than by a finite subgroup of $SU(2)$, the rules of reduction of various fields remain unmodified, i.e. we have (cf.~(\ref{eq:F5_ansatz}) and~(\ref{eq:B_ansatz_20}))
\begin{equation}\label{eq:ansatz_10_noflav}
    \breve{F}_5 = \sum_{n=0}^5\breve{a}^i_{5-n}\star \breve{u}_{ni}\ ,\ \breve{F}_3^\rho = \sum_{n=0}^3 \breve{f}^{\rho i}_{3-n}\star \breve{t}_{ni}\,, \ 
\end{equation}
where $\breve{u}_{ni}\in \breve{H}^n(L,\mathbb{Z})$ and $\breve{t}_{ni}\in \breve{H}^n(L,(\mathbb{Z}\oplus\mathbb{Z})_\rho)$. The key difference between the above rules of reduction and the one given by~(\ref{eq:B_ansatz_20}) is that there is no simple factorization of $\breve{H}^*(L,(\mathbb{Z}\oplus \mathbb{Z})_\rho)$ as in~(\ref{eq:Z2_20}). Rather, $H^*(L,(\mathbb{Z}\oplus \mathbb{Z})_\rho)$ is non-trivial hence $S_{\mathcal{T}_{sym}}$ must be more complicated than~(\ref{eq:SymTFT_20}).

A careful reader may note that we have not written down the decomposition of $\breve{F}_1$ in~(\ref{eq:ansatz_10_noflav}). For this we suppose that $\breve{F}_1 = \sum_{n=0}^1 \breve{g}^i_{1-n}\star\breve{s}_{ni}$ for $\breve{s}_{ni} \in \breve{H}^n(L, G_\rho)$ where $G_\rho$ is certain $\rho$-twisted sheaf. We immediately see that no matter what $G_\rho$ is, it is only possible to obtain the field strength $\breve{g}_{1}^i$ of a $(-1)$-form symmetry or the field strength $\breve{g}_{0}^i$ of a $(-2)$-form symmetry from such decomposition. As the discussion of the $(-1)$-form and $(-2)$-form symmetries is out of the scope of the current work, we will simply ignore the couplings obtained from reducing $\breve{F}_1$ as we have done in section~\ref{sec:SymTFT_20}. The same argument applies in the next section when we discuss the cases with flavor 7-branes as well.

To see why $\breve{F}_3^\rho$ should be reduced along $\breve{H}^*(L,(\mathbb{Z}\oplus \mathbb{Z})_\rho)$, one can insert a D3-brane to probe the effect of the gauge 7-branes on various fields. Since the D3-brane worldvolume field strength $F_2$ must be compatible with the $SL(2,\mathbb{Z})$ twist generated by the monodromy $\rho$ of the 7-brane~\cite{Heckman:Brane_GenSymm} and the bulk B-field couples linearly to $F_2$ through $F_2 - B_2$ in the D3-brane WZW action, the B-field must be twisted by $\rho$ in the same way as $F_2$ for consistency.

More precisely, we consider the ECY3 $E\hookrightarrow X_3\rightarrow B_2$ and insert a D3-brane probe at $p\in B_2$. The monodromy $\rho$ acts on the relative homology $H_2(X_3,E_p)$ in a way such that a $(p,q)$-string ending on the D3-brane becomes a $\rho(p,q)$-string after looping around the 7-brane where $\rho$ descends from an action on $H_2(X_3,E_p)$ to an action on the \emph{asymptotic charges} in $H_1(E_p)$ through the connecting homomorphism $\partial: H_2(X_3,E_p)\rightarrow H_1(E_p)$~\cite{Halverson:Strong_Coupling, Grassi:D3brane, Grassi:2018wfy, Grassi:2021ptc}. Since a $(p,q)$-string is a $(p,q)$-dyon as seen from the $U(1)$ gauge theory on the D3-brane worldvolume, for the consistency of this gauge theory before and after the looping, its field strength $F_2$ must be twisted accordingly such that its coupling to either a $(p,q)$-dyon or a $\rho(p,q)$-dyon leads to identical dynamics. Since the doublet $B$-field is twisted in the same way as $F_2$, the corresponding cohomology of $L$ on which the $B$-field is reduced must be twisted by $\rho$ in the same manner for consistency. Hence, by inserting a D3-brane probe, we see that one has to replace~(\ref{eq:Z2_20}) by $H^*(L,(\mathbb{Z}\oplus\mathbb{Z})_\rho)$ for non-trivial $\rho$ in the presence of 7-branes even when they do not intersect $L$ at all.

Having explained the physics behind~(\ref{eq:ansatz_10_noflav}), we now look for non-zero secondary invariants of $L$ given by products of $\breve{u}_{ni}$ and $\breve{t}_{ni}$ since each such non-vanishing secondary invariant corresponds to a non-trivial term in $S_{\mathcal{T}_{sym}}$. Recall that for all NHCs, $\breve{H}^n(L,(\mathbb{Z}\oplus\mathbb{Z})_\rho)$ is non-trivial only when $n=1,3$~\cite{Aharony:Sfold_N=3SCFT, Hsieh:2020jpj, Heckman:Brane_GenSymm}~\footnote{A sketch of the method to calculate the twisted cohomology group is given in Appendix~\ref{app:cohomo_twistcoeff}. Note that for the case of type $I_n$ singularity, which appears in the cases with flavor branes, there would be a free $\mb{Z}$ part in the twisted cohomology, nonetheless we do not expect them to contribute to new $U(1)$ global symmetries.}. This fact greatly simplifies the searching for secondary invariants and the only possibly non-zero secondary invariants are
\begin{equation}\label{eq:nonvanishing_terms}
    \text{CS}[L]_{ij} = \frac{1}{2}\int_{L}\breve{u}_{2i}\star \breve{u}_{2j},\ \text{CS}_t[L]_{ij} = \int_{L} \breve{t}_{1i}\star \breve{t}_{3j},\ L[u,t]_{kij} = \int_L \breve{u}_{2k}\star\breve{t}_{1i}\star\breve{t}_{1j}\,.
\end{equation}
In~(\ref{eq:nonvanishing_terms}) the pairing between $\breve{t}_{1i}$ and $\breve{t}_{1j}$ and that between $\breve{t}_{1i}$ and $\breve{t}_{3j}$ are given by the Dirac pairing descending from the Dirac pairing between $\breve{F}_3^\rho$'s and between $\breve{F}_3^\rho$ and $\breve{F}_7^\rho$ which has been discussed in section~\ref{sec:IIB_action_lift}.

In summary, given~(\ref{eq:nonvanishing_terms}), from~(\ref{eq:IIB_starting_point}) we arrive at the following $S_{\mathcal{T}_{sym}}$ associated to a $\mathcal{T}_{6D}^{(1,0)}$ without flavor branes:
\begin{equation}\label{10-SymTFT-p1}
    \frac{S_{\mathcal{T}_{sym}}}{2\pi} = \int_{\mathcal{M}^7} \text{CS}[L]_{ij} a_{3i}da_{3j} + \text{CS}_t[L]_{ij} f_{2i}^\rho df_{4j}^\rho + \frac{L[u,t]_{kij}}{2} a_{3k} f_{2i}^\rho f_{2j}^\rho\,.
\end{equation}
To determine the coefficients in~(\ref{10-SymTFT-p1}), we need to calculate the secondary invariants in~(\ref{eq:nonvanishing_terms}). As $\text{CS}[L]_{ij}$ has been calculated in section~\ref{sec:SymTFT_20} already, we will study $\text{CS}_t[L]_{ij}$ and $L[u,t]_{kij}$ in order.

\paragraph{Calculation of $\text{\normalfont CS}_t[L]_{ij}$}

To calculate $\text{CS}_t[L]_{ij}$ it is useful to consider a torus bundle construction over $B_2$ as in~\cite{Cvetic:HigherForm_Anomalies}. Equivalently we look at the F-theory geometry $E\hookrightarrow X_3\rightarrow B_2$ and the restriction of $X_3$ to $L$. The topology of $X_3$ can be studied either directly in F-theory as in~\cite{Hubner:GenSymm_Ftheory} or in the dual M-theory picture as in~\cite{Morrison:HigherForm_5D, Sakura:SymTFT_String}.
Here, similar to lifting $\breve{t}_1\in \breve{H}^1(L,(\mathbb{Z}\oplus\mathbb{Z})_\rho)$ to $\breve{t}_2\in \breve{H}^2(\partial X_3, \mathbb{Z})$ as in~\cite{Heckman:Brane_GenSymm}, we also lift $\breve{t}_3\in \breve{H}^3(L,(\mathbb{Z}\oplus\mathbb{Z})_\rho)$ to $\breve{t}_4\in \breve{H}^4(\partial X_3, \mathbb{Z})$. Therefore, one can write:
\begin{equation}
    \int_L \breve{t}_{1i} \breve{t}_{3j} = \int_{\partial X_3} \breve{t}_{2i}\breve{t}_{4j}\,.
\end{equation}
In this manner we avoid dealing with $SL(2,\mathbb{Z})$-twisted cohomology classes. The above linking pairing on $\partial X_3$ can then be calculated using the Smith normal form of the matrix $\mathcal{M}_4$ of the intersection pairing ($SNF(\mathcal{M}_4)$) of $X_3$~\cite{Morrison:HigherForm_5D, Sakura:SymTFT_String, Hubner:GenSymm_Ftheory}. We will crosscheck the results against those obtained via dual M-theory calculation in section~\ref{sec:Mtheory}.

\paragraph{Calculation of $L[u,t]_{kij}$} 

To calculate $L[u,t]_{kij}$ let us first look at the pairing $\int_L \breve{u}_{2i}\breve{u}_{2j}$, a sub-case of which ($i = j = 1$) has already been discussed in section~\ref{sec:SymTFT_20}. Recall that we have~\cite{GarciaEtxebarria:IIB_flux_noncomm, Sakura:SymTFT_String}
\begin{equation}
    2\text{CS}[L]_{ij} := \int_{L}\breve{u}_{2i}\breve{u}_{2j} = L(\text{PD}(u_{2i}), \text{PD}(u_{2j})) \equiv L(\gamma_i, \gamma_j) \mod 1
\end{equation}
which is the linking pairing between the Poincar\'e duals of $u_{2i}$ and $u_{2j}$. For any $\breve{v}_2\in \breve{H}^2(L)$ we have
\begin{equation}
     \int_{L}\breve{u}_{2i}\breve{v}_2 = L(\gamma_i, \gamma_v) = 2n\text{CS}[L]_{ii} \int_{\gamma_i} \breve{v}_2
\end{equation}
where $2n\text{CS}[L]_{ii}$ is the normalization factor. Therefore, given $\breve{u}_{2i}$ a linear function is defined to be
\begin{equation}
    L(\gamma_i, \bullet) := \int_L\breve{u}_{2i} \bullet.
\end{equation}
Following the notation of~\cite{Heckman:Brane_GenSymm}, we have
\begin{equation}
    L[u,t]_{kij} = \int_L \breve{u}_{2k}\breve{t}_{1i} \breve{t}_{1j} = L(\gamma_k, \breve{t}_{1i} \breve{t}_{1j}) = 2n\text{CS}[L]_{kk} N_k \int_{\gamma_k} \breve{t}_{1i} \breve{t}_{1j} := 2n\text{CS}[L]_{kk}N_k L_t^{\gamma_k}(\breve{t})
\end{equation}
where $N_k$ is the normalization factor that accounts for turning the pairing between $\breve{u}_2$ and $\breve{v}_2$ to the pairing between $\breve{u}_2$ and $\breve{t}_1\breve{t}_1$. It is stated in~\cite{Heckman:Brane_GenSymm} that the pairing $L_t^{\gamma_k}(\breve{t})$ can be geometrized as the linking pairing on the torus bundle $T^3_k$ over $\gamma_k$ where the fiber is the restriction of the S-duality torus fiber over $L$ to $\gamma_k$, i.e. we have
\begin{equation}
    L_t^{\gamma_k}(\breve{t}) = \int_{T^3_k} \breve{t}_2\breve{t}_2
\end{equation}
for $\breve{t}_2 \in \breve{H}^2(T^3_k)$~\footnote{We will discuss this torus fibration in greater detail in the next section around Figure~\ref{fig:K}.}. Such linking pairing on $T^3_k$ can then be computed using the methods in~\cite{Cvetic:HigherForm_Anomalies}. In summary, with non-trivial $L[u,t]_{kij}$ we have
\begin{equation}\label{eq:a3f2f2}
    \frac{S_{\mathcal{T}_{sym}}}{2\pi} \supset \int \frac{1}{2} \breve{F}_5\star \breve{F}_3^\rho\star \breve{F}_3^\rho = \frac{1}{2} \int_L \breve{u}_{2k}\breve{t}_{1i} \breve{t}_{1j} \int a_{3k}f^\rho_{2i}f^\rho_{2j} = \frac{L[u,t]_{kij}}{2} \int_{\mathcal{M}^7} a_{3k}f^\rho_{2i}f^\rho_{2j}.
\end{equation}
Since from~(\ref{eq:ansatz_10_noflav}) $a_{3k}$ is the 3-form background gauge field and $f_{2i}^\rho$ is the 2-form background field, the term~(\ref{eq:a3f2f2}) is 't Hooft anomaly between 2-form and 1-form symmetries of $\mathcal{T}_{6D}^{(1,0)}$. In section~\ref{sec:Mtheory} we will calculate these coefficients in M-theory analogous to \cite{Hubner:GenSymm_Ftheory}.

\subsection{SymTFT of 6D $\mathcal{N}=(1,0)$ theory with flavor branes}\label{sec:6D10_flav}

In this section we calculate $S_{\mathcal{T}_{sym}}$ associated to $\mathcal{T}_{6D}^{(1,0)}$ with flavor branes. Geometrically, we study IIB compactification on a non-compact variety $B_2$ with flavor 7-branes wrapping relative cycles of $\Delta\subset B_2$, which implies that $\Delta\cap L \neq \emptyset$.

The most crucial physical consequence of $\Delta\cap L \neq \emptyset$ is that we now have to take into consideration the action of the 7-branes rather than merely their monodromies. Physically, the introduction of flavor 7-branes leads to non-trivial (and in general non-abelian) 0-form symmetries. Though we lack a means to directly capture those continuous non-abelian 0-form symmetries via a SymTFT (in particular in the symmetry category language~\cite{Apruzzi:2023uma}), we do expect the discrete center of such symmetries to appear in a SymTFT as in~\cite{Heckman:TopDown_TopologicalDefects} in a slightly different context.

There can be many stacks of flavor branes intersecting $L$, for simplicity we assume there is only one stack of flavor 7-branes with worldvolume action:
\begin{equation}\label{eq:Dbrane_total_action}
    S_7 = S_{DBI} + S_{WZW} = \frac{1}{4g^2} \int_{\mathbb{R}^{1,5}\times \Delta} \text{tr} \mathcal{F}_2\wedge*\mathcal{F}_2  - 2\pi \int_{\mathbb{R}^{1,5}\times \Delta} i^*C \wedge \text{tr} e^{\mathcal{F}_2} \wedge \sqrt{\frac{\mathcal{A}_T}{\mathcal{A}_N}}
\end{equation}
where $\mathcal{F}_2 = \mathbf{F}_2 - i^*B_2$ with $\mathbf{F}_2 = F_2 + F_2^{U(1)}$ for traceless $F_2$ and $U(1)$ field strength $F_2^{U(1)}$. Since $\Delta$ is non-compact and affine, $\mathcal{A}_T$ and $\mathcal{A}_N$ are trivial. Therefore, $S_7$ simplifies to
\begin{equation}\label{eq:7_action}
    S_7 = \frac{1}{4g^2} \int_{\mathcal{M}^7\times \gamma_1} \text{tr} \mathcal{F}_2\wedge*\mathcal{F}_2  - 2\pi \int_{\mathcal{M}^7\times \gamma_1} C \wedge \text{tr} e^{\mathcal{F}_2}
\end{equation}
where $\mathcal{M}^7 = \mathbb{R}^{1,5}\times \mathbb{R}_+$ and $\Delta = \mathbb{R}_+\times \gamma_1$, and the pull-back $i^*$ of $C$ and $B_2$ to the 7-brane worldvolume is understood.

We need to determine the differential cohomological lift of~(\ref{eq:7_action}). For this let us define $J_2 := F_2^{U(1)} - B_2$ and expand~(\ref{eq:7_action}) in terms of a rank-$n$ $F_2$ and $J_2$:
\begin{equation}\label{eq:7_action_expansion}
    \begin{split}
        S_{n\text{D7}} &= \frac{1}{4g^2} \int_{\mathcal{M}^7\times \gamma_1} \left(\text{tr}F_2\wedge*F_2 + n J_2\wedge*J_2\right) \\
        &\quad - 2\pi \int_{\mathcal{M}^7\times \gamma_1} \frac{1}{24} C_0 \text{tr}(F_2 + J_2)^4 + \frac{1}{6} C_2\text{tr}(F_2 + J_2)^3 + \frac{1}{2} C_4 \text{tr}(F_2 + J_2)^2 + C_6 \text{tr}(F_2 + J_2) + C_8\,.
    \end{split}
\end{equation}
Due to the appearance of $C_6\text{tr}(J_2)$ in $S_{n\text{D7}}$, $J_2$ takes value in $C^2(\mc{M}^7\times\gamma^1,\text{Tor}(\ker(\varrho-1)))$ where $\varrho$ is the $SL(2,\mb{Z})$ monodromy generated by the flavor 7-branes.

As discussed in the previous sections, the fields in~(\ref{eq:7_action_expansion}) may experience monodromy around the $\gamma^1$ circle induced by other 7-branes, which will be denoted by $\rho$. Therefore it is convenient to rewrite~(\ref{eq:7_action_expansion}) into an $SL(2,\mathbb{Z})$-covariant form in order to perform the decomposition on $SL(2,\mathbb{Z})$-twisted cohomology classes.

By construction it is obvious that $J_2$ must be $SL(2,\mathbb{Z})$-twisted since $F_2^{U(1)} := dA_1^{U(1)}$ mixes with $B_2$ to preserve its gauge invariance $\delta B_2 = -d\Lambda_1^{U(1)}$ with the compensating shift $A_1^{U(1)}\rightarrow A_1^{U(1)} + \Lambda_1^{U(1)}$~\cite{Douglas:brane_within_branes}. Hence we will write it as $J_2^\rho$ as we did for $(H_3, F_3)$ in section~\ref{sec:IIB_action_lift}.

We need to determine if the non-abelian field strength $F_2$ is to be twisted by $\rho$ as well. For this let us insert $n$ coinciding D3-probes into the system with monodromy $\rho$ generated by the 7-branes. The endpoint of a $(p,q)$-string on the D3 worldvolume becomes $n$ $(p,q)$-dyons in $\mathbf{n}$ of $SU(n)$ and dyonic $(p,q)$-charge under $U(1)\subset U(n)$~\footnote{We thank James Halverson for helpful discussion on this point.}. Thus it is clear that the $(p,q)$-charge is related only to $U(1)\subset U(n)$ and independent from its traceless $SU(n)$ part. As the $SL(2,\mathbb{Z})$ monodromy acts only on the $(p,q)$-charges, we conclude that in general the traceless $F_2$ is not twisted. Hence in general for D3-brane worldvolume field strength we write:
\begin{equation}\label{eq:splitting_F2}
    \mathcal{F}_2 = \mathbf{F}_2 - B_2 = F_2 + F_2^\rho - B_2^\rho := F_2 + J_2^\rho
\end{equation}
to emphasize $J_2^\rho$ does experience the twist by monodromy. The same must hold on the worldvolume of multiple overlapping 7-branes as well since after a D3-brane can dissolve into a 7-brane~\cite{Douglas:brane_within_branes, Sen:tachyon_condensation, Witten:DbraneKtheory}.

Before dimensionally reduce $S_{n\text{D7}}$ on $\gamma_1$, we first note that since $F_2$ is not twisted, its dimensional reduction on $\gamma_1$ is equivalent to a simple $S^1$-reduction. Therefore its dimensional reduction does not lead to any non-trivial term in $S_{\mathcal{T}_{sym}}$. Hence we will drop $\text{tr}F_2\wedge*F_2$ in~(\ref{eq:7_action_expansion}) and dimensionally reduce the following action:
\begin{equation}\label{eq:S7_rho}
    \begin{split}
        S_{n\text{D7}} &= \int_{\mathcal{M}^7\times \gamma_1} \frac{n}{2} J^\rho_2\wedge * \frac{J^\rho_2}{2g^2} \\
        &\quad - 2\pi \left( \frac{1}{24} C_0 \text{tr}(F_2 + \mathbf{J}^\rho_2)^4 + \frac{1}{6} C_2^\rho\text{tr}(F_2 + \mathbf{J}^\rho_2)^3 + \frac{1}{2} C_4 \text{tr}(F_2 + \mathbf{J}^\rho_2)^2 + C_6^\rho \text{tr}(F_2 + \mathbf{J}^\rho_2) + C_8 \right)
    \end{split}
\end{equation}
where $\mathbf{J}^\rho_2 = J^{\rho}_2 \text{Id}_{n}$ where $\text{Id}_{n}$ is the $n\times n$ identity matrix. Note that we have also added superscript $\rho$ to $C_2$ and $C_6$ as they must transform as doublets under $\rho$ as well. Neither $C_0$ nor $C_8$ is $SL(2,\mathbb{Z})$-neutral but later we will show that they do not lead to any non-trivial term in $S_{\mathcal{T}_{sym}}$.

As the last preparation before dimensionally reducing~(\ref{eq:S7_rho}) on $\gamma_1$, we will rewrite it as certain secondary invariant of $\mathcal{M}^7\times \gamma_1$ in terms of differential cohomology classes on $\gamma_1$. After this we will do the dimensional reduction.

\paragraph{Differential cohomological lift of the 7-brane action}

Before rewriting $S_7$ as certain secondary invariant of $\mathcal{M}^7\times \gamma_1$, we first write down the ordinary topological lift of it (cf. the topological lift~(\ref{eq:topIIB}) of~(\ref{eq:IIB_action})).

It has been worked out in~\cite{Apruzzi:2023uma} that the topological lift of a source-free generalized free Maxwell theory $\frac{1}{2} f \wedge* \frac{f}{g^2}$ in one higher dimension is $F^{(1)} d F^{(2)}$ where $F^{(1)} = f$ and $F^{(2)} = g^{-2} * f$. Similarly, the topological lift of $\frac{n}{2} J^\rho_2 * \frac{J^\rho_2}{2g^2}$ will be $2\pi n J_2^\rho d \Upsilon_6^\rho$ where $\Upsilon_6^\rho = \frac{J^\rho_2}{4\pi g^2}$. Therefore, we can write down the topological lift of~(\ref{eq:S7_rho}) in one higher dimension as follows:
\begin{equation}
    \begin{split}
        \frac{S_{\text{top-}7}}{2\pi} &= \int_{\mathcal{M}^9} n J^\rho_2 d \Upsilon_6^\rho \\
        &\quad -  \left( \frac{1}{24} F_1 \text{tr}(F_2 + \mathbf{J}^\rho_2)^4 + \frac{1}{6} F_3^\rho\text{tr}(F_2 + \mathbf{J}^\rho_2)^3 + \frac{1}{2} F_5 \text{tr}(F_2 + \mathbf{J}^\rho_2)^2 + F_7^\rho \text{tr}(F_2 + \mathbf{J}^\rho_2) + F_9 \right)
    \end{split}
\end{equation}
where $\Upsilon_2^\rho := g^{-2} * J_2^\rho$ and $\partial \mathcal{M}^9 = \mathcal{M}^7\times \gamma_1$ (cf. the configuration illustrated in Figure~\ref{fig:M11_M10}).

Applying the method in~\cite{Heckman:Brane_GenSymm}, it is straightforward to write down the differential cohomological lift of $S_{\text{top-}7}$ as a secondary invariant of $\mathcal{M}^7\times \gamma_1$ as follows:
\begin{equation}\label{eq:S_breve7}
    \begin{split}
        \frac{S_{\breve{7}}}{2\pi} &= \int_{\mathcal{M}^7\times \gamma_1}n \breve{J}_2^\rho \delta \breve{\Upsilon}_6^\rho \\
        &\quad - \left( \frac{1}{24} \breve{F}_1 \text{tr}(F_2 + \breve{\mathbf{J}}^\rho_2)^4 + \frac{1}{6} \breve{F}_3^\rho\text{tr}(F_2 + \breve{\mathbf{J}}^\rho_2)^3 + \frac{1}{2} \breve{F}_5 \text{tr}(F_2 + \breve{\mathbf{J}}^\rho_2)^2 + \breve{F}_7^\rho \text{tr}(F_2 + \breve{\mathbf{J}}^\rho_2) + \breve{F}_9 \right).
    \end{split}
\end{equation}
The $n$ factor of the BF-term $n\breve{J}_2^\rho \delta \breve{\Upsilon}_6^\rho$ is the dimension of the fundamental representation of the algebra supported on $\gamma_1$. Since $\text{tr}((\mathbf{J}^\rho_2)^k) = n \text{tr}((J^\rho_2)^k)$ for any $k \geq 0$, this $n$ factor appears in all powers of $\mathbf{J}^\rho_2$ hence we will drop it in the following calculations by absorbing it back into $\breve{J}_2^\rho$.

\paragraph{Dimensional reduction of 7-brane action}

Having obtained $S_{\breve{7}}$ in~(\ref{eq:S_breve7}) we are ready to perform dimensional reduction of it on $\gamma_1$ to obtain an action on $\mathcal{M}^7$.

Since all the fields in~(\ref{eq:S_breve7}) are defined on the 7-brane worldvolume, we will have the following rules of reduction:
\begin{equation}\label{eq:ansatz_flavor7_1}
    \begin{split}
        \breve{F}^{\rho'}_3 &= \sum_{n=0}^3\breve{f}^{\rho'i}_{3-n}\star \breve{t}'_{ni}\ ,\ \breve{J}^{\rho'}_2 = \sum_{n=0}^2\breve{\j}^{\rho'i}_{2-n}\star \breve{t}'_{ni}\ ,\ 
        \breve{F}_5 = \sum_{n=0}^5\breve{a}^i_{5-n}\star \breve{u}_{ni}
    \end{split}
\end{equation}
where $\breve{t}'_{ni} \in \breve{H}^n(\gamma_1, (\mathbb{Z}\oplus \mathbb{Z})_{\rho'})$ and $\breve{u}_{ni} \in \breve{H}^n(\gamma_1, \mathbb{Z})$ and the summation over $i$ is made implicit. One should think of the decomposition of $\breve{J}_2^{\rho'}$ as a restriction from $U(1)$-valued form to Tor$(\ker(\rho'-1))$-valued form, where $\rho'$ is the monodromy along $\gamma_1$. 

We emphasize that $\rho'$ is collectively all the monodromies along but the ones around $\gamma_1$. In particular it does not include the monodromy generated by the stack of 7-branes on $\mathcal{M}^7\times \gamma_1$. We use $\rho'$ to distinguish it from $\rho$ which in later sections will be the monodromies including the one generated by the flavor 7-brane on $\gamma_1$. As having been discussed in section~\ref{sec:SymTFT_20} and~\ref{sec:6D10_noflav}, we ignore all the contributions from reducing $\breve{F}_1$, which is the reason why the corresponding rule of reduction is not written down in~(\ref{eq:ansatz_flavor7_1}).

The discussion can be naturally categorized into two cases with either a non-trivial $\rho'$ or a trivial $\rho'$. If $\rho'$ is non-trivial, the torsional part of $\breve{H}^n(L, (\mathbb{Z}\oplus \mathbb{Z})_{\rho'})$ is non-vanishing only for odd $n$ (see~\cite{Aharony:Sfold_N=3SCFT, Heckman:Brane_GenSymm} and Appendix~\ref{app:cohomo_twistcoeff}). Therefore, in dimensional reduction the only potentially non-vanishing secondary invariants of $\gamma_1$ are~\cite{Aharony:Sfold_N=3SCFT, Sakura:SymTFT_String, Heckman:Brane_GenSymm} (cf.~(\ref{eq:nonvanishing_terms})):
\begin{equation}\label{eq:nonvanishing_terms_flav}
    L_{\gamma_1}^i = \int_{\gamma_1} \breve{u}_{2i},\ L_{\gamma_1}^{ij} = \int_{\gamma_1}\breve{t}'_{1i}\star \breve{t}'_{1j}\,. 
\end{equation}
The above two secondary invariants of $\gamma_1$ will play important roles in the following discussions.

\bigskip

We define $S_{\breve{7}\text{WZW}} := \sum_n S_n = 2\pi\int_{\mathcal{M}^7\times \gamma_1} C_n\wedge \frac{1}{(4-n)!}\text{tr}\mathcal{F}_2^{4-n}$ and study each $S_n$ for $n = 0,2,4,6,8$ in order.

\noindent$\bullet$ $S_0$: Using~(\ref{eq:ansatz_flavor7_1}) we have:
\begin{equation}
    \frac{S_0}{2\pi} = \int_{\mathcal{M}^7\times \gamma_1} \breve{F}_1\star \frac{1}{24}\left( \text{tr}F_2^4 + 4\text{tr}F_2^3 \breve{J}_2^{\rho'} + 6\text{tr}F_2^2 (\breve{J}_2^{\rho'})^2 + (\breve{J}_2^{\rho'})^4 \right)\,.
\end{equation}
Given~(\ref{eq:nonvanishing_terms_flav}), the only potentially non-vanishing term is:
\begin{equation}\label{eq:vanishing_tt}
    L_{\gamma_1}^{ij} \int_{\mathcal{M}^7} F_1 \j^{\rho'}_{1i} \j^{\rho'}_{1j} \text{tr} F_2^2\,.
\end{equation}
Nevertheless, the above term must vanish since $L_{\gamma_1}^{ij}$ is symmetric while $\j^{\rho'}_{1i}\j^{\rho'}_{1j}$ is anti-symmetric under $i\leftrightarrow j$. Therefore there is no contribution to $S_{\mathcal{T}_{sym}}$ coming from the dimensional reduction of $S_0$.

\noindent$\bullet$ $S_2$: 
\begin{equation}
    \frac{S_2}{2\pi} 
    = \int_{\mathcal{M}^7\times \gamma_1} \breve{F}_3^{\rho'}\star \frac{1}{6} \left( \text{tr}F_2^3 + 3\text{tr}F_2^2 \breve{J}_2^{\rho'} + (\breve{J}_2^{\rho})^3 \right)\,.
\end{equation}
Due to vanishing $L_{\gamma_1}^{ij} f_{1i}f_{1j}$, the only potentially non-vanishing contribution to $S_{\mathcal{T}_{sym}}$ is:
\begin{equation}
    \frac{L_{\gamma_1}^{ij}}{2} \int_{\mathcal{M}^7} f^{\rho'}_{2i} \j^{\rho'}_{1j}\text{tr} F_2^2\,.
\end{equation}

\noindent$\bullet$ $S_4$: 
\begin{equation}
    \frac{S_4}{2\pi} 
    = \int_{\mathcal{M}^7\times \gamma_1} \breve{F}_5\star \frac{1}{2} \left( \text{tr}F_2^2 + (\breve{J}_2^{\rho'})^2 \right)\,.
\end{equation}
Again, due to vanishing $L_{\gamma_1}^{ij} f_{1i}f_{1j}$, the only potentially non-vanishing contribution to $S_{\mathcal{T}_{sym}}$ is:
\begin{equation}
    \frac{L_{\gamma_1}^i}{2} \int_{\mathcal{M}^7} a_{3i}\text{tr} F_2^2\,.
\end{equation}

\noindent$\bullet$ $S_6$: 
\begin{equation}
    \frac{S_6}{2\pi} 
    =  \int_{\mathcal{M}^7\times \gamma_1} \breve{F}_7^{\rho'}\star \breve{J}_2^{\rho'}\,.
\end{equation}
The potentially non-vanishing contribution to $S_{\mathcal{T}_{sym}}$ is:
\begin{equation}\label{eq:S6_reduction}
    L_{\gamma_1}^{ij} \int_{\mathcal{M}^7} f^{\rho'}_{6i}\j^{\rho'}_{1j}
\end{equation}
where $f^{\rho'}_{6i}$ is the background gauge field of a certain 5-form symmetry. Nonetheless, we will not consider the dual $(-1)$-form symmetry to this 5-form symmetry in this work, and we will omit the term (\ref{eq:S6_reduction}) in the later discussions.

\noindent$\bullet$ $S_8$: 
\begin{equation}
    \frac{S_8}{2\pi} = \int_{\mathcal{M}^7\times \gamma_1} \breve{F}_9\,.
\end{equation}
This the field dual to $\breve{F}_1$ and we will ignore it for the same reason we do not consider $\breve{F}_1$.

\bigskip

Recall that besides $S_{\breve{7}\text{WZW}}$ there is also the ``kinetic term'' $\int_{\mathcal{M}^7\times \gamma_1} \breve{J}_2^{\rho'} \delta \breve{\Upsilon}_6^{\rho'}$ which reduces to the following BF-coupling in lower dimension:
\begin{equation}\label{eq:reduction_DBI}
    L_{\gamma_1}^{ij} \int_{\mathcal{M}^7}  \j_{1i}^{\rho'} d \Upsilon_{5j}^{\rho'}
\end{equation}
where $\Upsilon_{5j}^{\rho'}$ is the dual of $\j_{1j}^{\rho'}$ coming from the reduction of $\Upsilon_6^{\rho'}$ on $\breve{t}'_{1i}$.

To summarize, the reduction of $S_{\breve{7}}/2\pi$ on $\gamma_1$ leads to the following action:
\begin{equation}\label{eq:symtft_from_7brane}
    \int_{\mathcal{M}^7} L_{\gamma_1}^{ij} \j_{1i}^{\rho'} d \Upsilon_{5j}^{\rho'} + \frac{L_{\gamma_1}^{ij}}{2} \j_{1j}^{\rho'} f_{2i}^{\rho'} \text{tr} F_2^2 + \frac{L_{\gamma_1}^i}{2} a_{3i} \text{tr} F_2^2\,.
\end{equation}
In the above expression, $\j_1^{\rho'}$ is the background 1-form field of certain 0-form symmetry. To see what this 0-form symmetry is, we look at specific examples where a single stack of 7-branes supports a $U(n) = (SU(n)\times U(1))/\mathbb{Z}_n$ symmetry or a $E_8\supset (G_F\times U(1)^m)/\mathbb{Z}_k$ symmetry where $T^n$ is the maximal torus of $G_F$ where $n+m = 8$ and $\mathbb{Z}_k$ is the center of $G_F$~\cite{Heckman:TopDown_TopologicalDefects}. As stated before, at the level of Lie algebra we have $\mathbf{F}_2 = F_2 + F_2^{\rho'}$, e.g. for $U(n)$ $F_2$ is an $\mathfrak{su}(n)$-valued 2-form and $F_2^{\rho'}$ is a $\mathfrak{u}(1)$-valued 2-form. It is clear from~(\ref{eq:reduction_DBI}) that the reduction of the ``kinetic term'' is proportional to $L_{\gamma_1}^{ij} = p/q$ with co-prime $p$ and $q$. In all the examples we have $q = n$ for $U(n)$ or $q = k$ for $(G_F\times U(1)^m)/\mathbb{Z}_k$, hence from~(\ref{eq:reduction_DBI}) we read off a $\mathbb{Z}_{n}$ or $\mathbb{Z}_{k}$ 0-form symmetry from reducing $\breve{J}^\rho_2$ (or equivalently $F_2^\rho$). Since the $\mathbb{Z}_n$ or $\mathbb{Z}_k$ acts diagonally on the traceless part and the $U(1)$ part, this discrete 0-form symmetry is identical to the center of the traceless part. We thus conclude that $\j_1^{\rho'}$ in~(\ref{eq:symtft_from_7brane}) is the center of the flavor symmetry of the 6D theory, which is also consistent with the fact that it comes from the reduction of the field strength of flavor center $J_2^\varrho$ in 8D. We also expect such arguments to apply to more general brane configurations other than the cases of $U(n)$ and $E_8$.

On the other hand, if $\rho'$ is trivial, all fields should be reduced on the elements of $H^*(\gamma_1,\mb{Z})=(\mb{Z},\mb{Z})$ rather than those of $H^*(\gamma_1,(\mb{Z}\oplus\mb{Z}))=(\mb{Z}\oplus\mb{Z},\mb{Z}\oplus\mb{Z})$, as we only expect a single 0-form symmetry field strength $\j_2$ from the reduction of flavor center field strength $J_2$ (as before we omit the field strength for a $(-1)$-form symmetry):
\be
\br{J}_2=\br{\j}_2\star \br{1}\,.
\ee
In this case, the SymTFT action coming from dimensional reducing~(\ref{eq:ansatz_flavor7_1}) contains only the following term:
\be
\frac{L_{\gamma_1}^i}{2}\int_{\mc{M}^7}a_{3i}(\text{tr}F_2^2+\j_2^2)\,.
\ee

\paragraph{Full SymTFT action}

Having obtained~(\ref{eq:symtft_from_7brane}), we need to further investigate its interplay with the ``bulk'' action $S_{\breve{\text{IIB}}}$ given by~(\ref{eq:IIB_starting_point}) and their dimensional reduction on the whole boundary $L$.

Before writing down the full SymTFT action we have to work out what the action to be dimensionally reduced is. For $\mathcal{N} = (2,0)$ it is clear that the action to be reduced is nothing but~(\ref{eq:IIB_starting_point}). For $\mathcal{N}=(1,0)$ without flavor branes, i.e. $\Delta\cap L = \emptyset$, we have argued in section~\ref{sec:6D10_noflav} that the action to be reduced is still~(\ref{eq:IIB_starting_point}) as long as the $\rho$-twist is taken care of. The case with flavor branes is subtler since $\Delta\cap L\neq \emptyset$, therefore it does not make sense to naively take the sum of $S_{\text{IIB}}$ given by~(\ref{eq:IIB_action}) and $S_7$ given by~(\ref{eq:Dbrane_total_action}) then dimensionally reduce it.

Fortunately, it is not our purpose to describe the full dynamics of the theory, rather we care only about the physics in deep IR where all dynamics are decoupled while only the data of symmetries are kept. In this limit the backreaction of the 7-brane on the geometry, in particular on the metric, becomes inessential and the whole system can be viewed as adding infinitely heavy 7-branes to $L$ consistently by hand. Hence, the data that remains important are topological, such as monodromy, and the topological action $S_{\text{top-IIB}} + S_{\text{top-}7}$ already captures all the data necessary for our purpose. Therefore, its differential cohomological lift $S_{\breve{\text{IIB}}} + S_{\breve{7}}$ will be the sought-after action whose dimensional reduction on $L$ leads to the desired $S_{\mathcal{T}_{sym}}$.

In summary, from~(\ref{10-SymTFT-p1}) and~(\ref{eq:symtft_from_7brane}), the SymTFT action in the presence of both the gauge and the flavor 7-branes, with either non-trivial or trivial $\rho'$, is:
\begin{equation}\label{10-SymTFT_pre}
    \begin{split}
        \frac{S_{\mathcal{T}_{sym}}}{2\pi} &= \int_{\mathcal{M}^7} \text{CS}[L]_{ij} a_{3i}da_{3j} + \text{CS}_t[L]_{ij} f_{2i}^\rho df_{4j}^\rho + \frac{L[u,t]_{kij}}{2} a_{3k} f_{2i}^\rho f_{2j}^\rho \\
        &\ + \sum_{\gamma_1,\mathrm{non-trivial}\ \rho'} \int_{\mathcal{M}^7} L_{\gamma_1}^{ij} \j_{1i}^{\rho'} d \Upsilon_{5j}^{\rho'} + \frac{L_{\gamma_1}^{ij}}{2} \j_{1j}^{\rho'} f_{2i}^{\rho'} \text{tr} F_{2,\gamma_1}^2 + \frac{L_{\gamma_1}^i}{2} a_{3i} \text{tr} F_{2,\gamma_1}^2\\
        &+ \sum_{\gamma_1,\mathrm{trivial}\ \rho'}\int_{\mc{M}^7}\frac{L_{\gamma_1}^i}{2} a_{3i} (\text{tr}F_2^2+\j_2^2)\,.
    \end{split}
\end{equation}
where the summation is over all irreducible components of $\Delta\cap L$ and $F_{2,\gamma_1}$ is the non-abelian 0-form flavor symmetry field strength labeled by the 1-cycle $\gamma_1$ that supports it. We also note that each $\rho'$ is associated to a $\gamma_1 \subset L$.

\paragraph{Matching the monodromies}

Clearly, the action~(\ref{10-SymTFT_pre}) cannot be the final form as there is a manifest mismatch between $\rho$ and each non-trivial $\rho'$ associated to an irreducible component of $\Delta\cap L$. More concretely, one has to work out the relation between the $\rho$-twisted fields in the first line of~(\ref{10-SymTFT_pre}) and the $\rho'$-twisted fields in its second line.

Before discussing the $\rho$-twisted fields and $\rho'$-twisted fields, we will first resolve a smaller problem on what the relation between the $a_3$ field in the first line and that in the second line of~(\ref{10-SymTFT_pre}) is, since the former is obtained from reducing $\breve{F}_5$ on $\breve{u}_2\in\breve{H}^2(L,\mathbb{Z})$ while the latter is obtained from reducing $\breve{F}_5$ on $\breve{u}_2\in\breve{H}^2(\gamma_1,\mathbb{Z})$. For this we simply make use of the pull-back $i^*: \breve{H}^2(L,\mathbb{Z})\rightarrow \breve{H}^2(\gamma_1,\mathbb{Z})$ induced by the embedding $i: \gamma_1 \rightarrow L$ to map each generator of $\breve{H}^2(L,\mathbb{Z})$ to the generator of $\breve{H}^2(\gamma_1,\mathbb{Z})$ for each $\gamma_1\in \text{Tor}H_1(L)$. Therefore the $a_{3}$ in the first line and in the second line of~(\ref{10-SymTFT_pre}) are indeed the same $a_{3}$ for each $\gamma_1$. Our notation is justified.

Next we study the relation between $f_{2i}^\rho$ and each $f_{2i}^{\rho'}$. For this we need to look at the relation between $\breve{H}^1(L,(\mathbb{Z}\oplus\mathbb{Z})_\rho) \simeq \mathbb{Z}^2/\text{Im}(\rho-1) := \Lambda_\rho$ and $\breve{H}^1(\gamma_1,(\mathbb{Z}\oplus\mathbb{Z})_{\rho'})\simeq \mathbb{Z}^2/\text{Im}(\rho'-1) := \Lambda^{\gamma_1}_{\rho'}$ for each $\rho'$ (see~\cite{Aharony:Sfold_N=3SCFT, Heckman:Brane_GenSymm} and Appendix~\ref{app:cohomo_twistcoeff} for a sketch of the calculation). Since $\rho$ is obtained from $\rho'$ by supplementing the monodromy along $\gamma_1$, naturally we have $\Lambda^{\gamma_1}_{\rho} \subseteq \Lambda^{\gamma_1}_{\rho'}$ and one can define the group homomorphism:
\begin{equation}\label{eq:lattice_homomorphism}
    \lambda_{\gamma_1}: \breve{t}'_1 \in \breve{H}^1(\gamma_1, (\mathbb{Z}\oplus \mathbb{Z})_{\rho'}) \mapsto \breve{t}_1 \in \breve{H}^1(\gamma_1, (\mathbb{Z}\oplus \mathbb{Z})_{\rho})
\end{equation}
for each $\gamma_1$ and the corresponding $\rho'$. It is not hard to see that $\Lambda_{\rho}^{\gamma_1} := \breve{H}^1(\gamma_1, (\mathbb{Z}\oplus \mathbb{Z})_{\rho})\cong \Lambda_{\rho}$ since both of them are given by the first cohomology group of the following chain complex (see appendix \ref{app:cohomo_twistcoeff} for details):
\begin{equation}
    \mathbb{Z}^2 \xrightarrow{1-\rho} \mathbb{Z}^2 \xrightarrow{1 + \rho + \cdots + \rho^{n-1}} \mathbb{Z}^2 \rightarrow \cdots.
\end{equation}
Physically, one has to require that the reduction of $\breve{F}_3$ makes sense for $\breve{t}_1$ in either $\Lambda_{\rho}$ or $\Lambda_{\rho'}$ to avoid apparent inconsistencies. Therefore, using each $\lambda_{\gamma_1}$ and the embedding $i$ we conclude that $\breve{F}_3^\rho$ must be reduced on $\breve{t}_1\in \Lambda_\rho$. In contrast to $\breve{F}_3^\rho$ which lives in $L$, each $\breve{J}_{2,\gamma_1}$ is localized on $\gamma_1$ which supports a stack of 7-branes. Therefore, a priori there is no need to match different $\j_{1i}^{\rho'}$'s. As the only term that involves both $\breve{J}_2$ and $\breve{F}^\rho_3$ is:
\begin{equation}
    S_2 \supset \int_{\mathcal{M}^7\times \gamma_1} \frac{1}{2} \breve{J}_2^{\rho'} \star \breve{F}_3^{\rho'} \text{tr}F_2^2,
\end{equation}
we will write down explicitly its dimensional reduction based on the discussion above as follows:
\begin{equation}
    \int_{\mathcal{M}^7\times \gamma_1} \frac{1}{2} \breve{J}_2^{\rho'} \star \breve{F}_3^{\rho'} \text{tr}F_2^2 = \frac{1}{2} \int_{\gamma_1} \breve{t}_{1i}^{\rho'}\breve{t}_{1j}^{\rho'} \int_{\mathcal{M}^7} \j_{1i}^{\rho'} f_{2j}^{\rho} \text{tr}F_2^2 = \frac{L_{\gamma_1}^{ij}}{2} \int_{\mathcal{M}^7} \j_{1i}^{\rho'} f_{2j}^{\rho} \text{tr}F_2^2
\end{equation}
where (cf.~(\ref{eq:ansatz_10_noflav})):
\begin{equation}\label{eq:ansatz_10_noflav_refined}
    \breve{J}_{2}^{\rho'} = \breve{\j}_{1i}^{\rho'} \star \breve{t}'_{1i},\ \breve{F}_3^{\rho'} \rightarrow \breve{F}_3^\rho = \breve{f}^\rho_{2i}\star \lambda_{\gamma_1} (\breve{t}'_{1i})
\end{equation}
for:
\begin{equation}
    \breve{t}'_{1i} \in \breve{H}^1(\gamma_1, (\mathbb{Z}\oplus \mathbb{Z})_{\rho'}),\ \breve{t}_{1i} = \lambda_{\gamma_1}(\breve{t}'_{1i}) \in \breve{H}^1(\gamma_1, (\mathbb{Z}\oplus \mathbb{Z})_{\rho}) \cong \breve{H}^1(L, (\mathbb{Z}\oplus \mathbb{Z})_{\rho})\,.
\end{equation}

In summary, we have the following SymTFT action:
\begin{equation}\label{10-SymTFT}
    \begin{split}
        \frac{S_{\mathcal{T}_{sym}}}{2\pi} &= \int_{\mathcal{M}^7} \text{CS}[L]_{ij} a_{3i}da_{3j} + \text{CS}_t[L]_{ij} f_{2i}^\rho df_{4j}^\rho + \frac{L[u,t]_{kij}}{2} a_{3k} f_{2i}^\rho f_{2j}^\rho \\
        &\ + \sum_{\gamma_1,\text{non-trivial\ }\rho'} \int_{\mathcal{M}^7} L_{\gamma_1}^{ij} \j_{1i}^{\rho'} d \Upsilon_{5j}^{\rho'} + \frac{L_{\gamma_1}^{ij}}{2} \j_{1j}^{\rho'} f_{2i}^{\rho} \text{tr} F_{2,\gamma_1}^2 + \frac{L_{\gamma_1}^i}{2} a_{3i} \text{tr} F_{2,\gamma_1}^2 \\
        &+ \sum_{\gamma_1,\text{trivial\ }\rho'}\int_{\mc{M}^7}\frac{L_{\gamma_1}^i}{2} a_{3i} (\text{tr} F_{2,\gamma_1}^2+\j_2^2)
    \end{split}
\end{equation}
where we have replaced different $\rho'$-twisted $f_{2i}^{\rho'}$ in~(\ref{10-SymTFT_pre}) by the $\rho$-twisted $f_{2i}^\rho$ obtained from reducing $\breve{F}_3^\rho$ on $\breve{t}_{1i}\in \Lambda$. In~(\ref{10-SymTFT}), $f_{2i}^\rho$, $a_{3i}$ and $f_{4i}^\rho$ correspond to the background gauge fields of 1-form, 2-form and 3-form symmetries respectively. There are also background gauge fields $\j_{1i}^{\rho'}$ for the center of each 0-form symmetry supported on $\gamma_1$. We have also included in $S_{\mathcal{T}_{sym}}$ the contribution from the flavor branes with trivial $\rho'$.

Furthermore, for the term tr$(F_{2,\gamma_1}^2)$, we can rewrite it as~\cite{Cordova:2019uob,Apruzzi:Fate_1form_6D,Heckman:2022suy}
\be
\ba
\label{F2-w2}
\text{tr}(F_{2,\gamma_1}^2)=-4\alpha_{\mathfrak{g}}w(F_{2,\gamma_1})\cup w(F_{2,\gamma_1})\cr
\alpha_{\mathfrak{g}}=\left\{\begin{array}{rl} \frac{n-1}{2n} & \mathfrak{g}=\mk{su}(n)\\
\frac{n}{4} & \mathfrak{g}=\mk{sp}(n)\\
\frac{2}{3} & \mathfrak{g}=\mk{e}_6\\
\frac{3}{4} & \mathfrak{g}=\mk{e}_7\\
\frac{2n+1}{8} & \mathfrak{g}=\mk{so}(4n+2)\\
\frac{1}{2} & \mathfrak{g}=\mk{so}(2n+1)\\
\end{array}
\right.\,.
\ea
\ee
While for $\mathfrak{g}=\mathfrak{so}(4n)$
\be
\ba
\text{tr}(F_{2,\gamma_1}^2)=-n\left(w^{(1)}(F_{2,\gamma_1})+w^{(2)}(F_{2,\gamma_1})\right)^2-2w^{(1)}(F_{2,\gamma_1})\cup w^{(2)}(F_{2,\gamma_1})\,.
\ea
\ee
Here $w^{(1,2)}(F_{2,\gamma_1})$ are the second Stiefel-Whitney classes parametrizing the obstruction of uplifting the corresponding gauge bundles (whose field strengths are $F_{2,\gamma_1}$ and $F_{2,\gamma_2}$ respectively) with non-simply-connected gauge group to a gauge bundle with simply-connected gauge group (see~\cite{Aharony:2013hda}). Note that the 4-form $\frac{1}{4}$tr$(F_{2,\gamma_1}^2)$ is integral over 4-cycles, by definition~\cite{Apruzzi:Fate_1form_6D}.

\paragraph{A physical argument for matching the monodromies} 

Before calculating $S_{\mathcal{T}_{sym}}$ in various examples, we give a physical argument on why the monodromies $\rho$ and $\rho'$'s have to match in the sense stated above.

\begin{figure}
    \centering
    \includegraphics[width=0.5\textwidth]{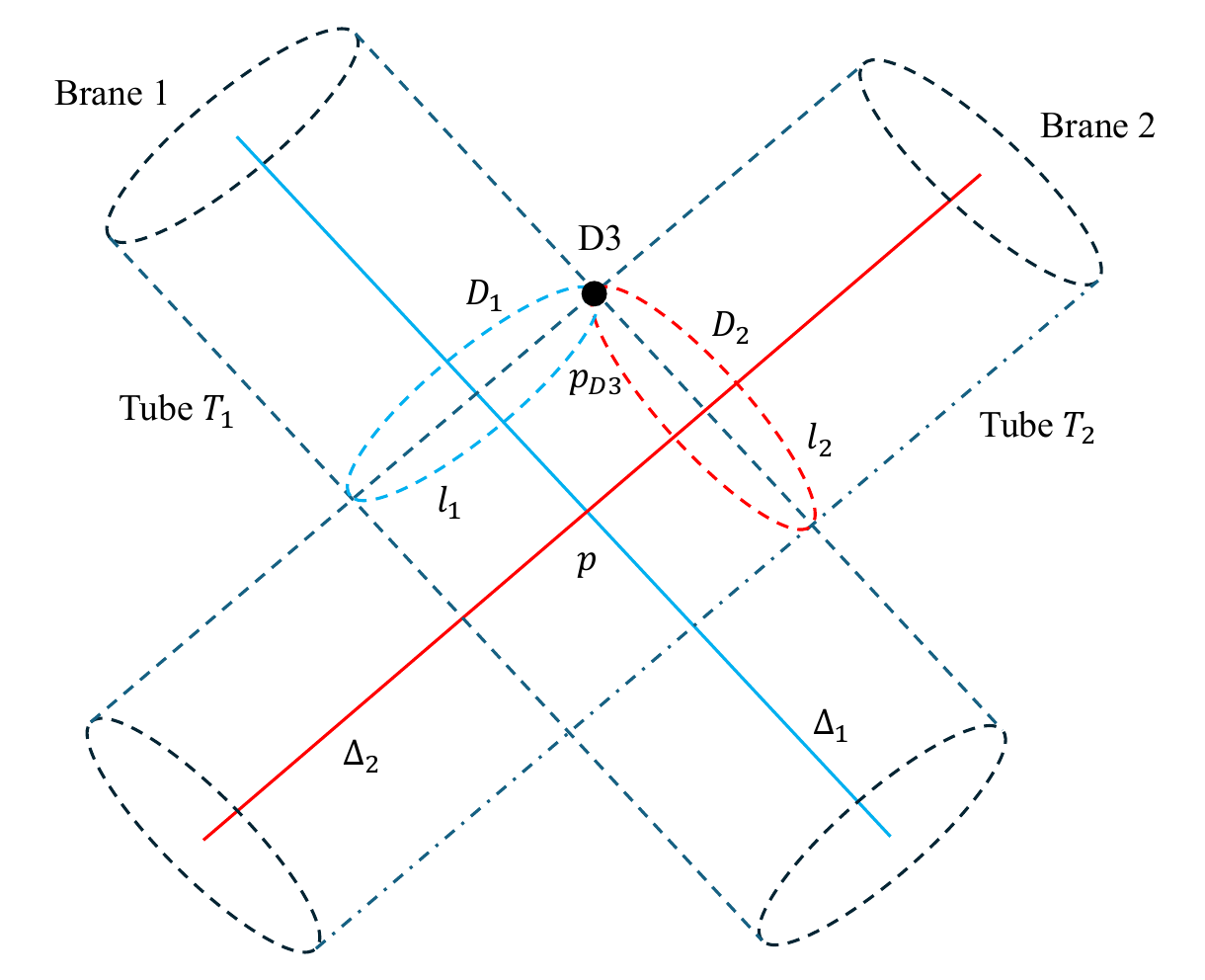}
    \caption{D3-brane probe in the background of intersecting stacks of 7-branes.}
    \label{fig:D3_12}
\end{figure}

Without loss of generality we study the configuration shown in Figure~\ref{fig:D3_12}.
The discriminant locus $\Delta_{1}$ intersects $\Delta_{2}$ at a point $p \in B_2$. We zoom in at the tubular neighbor $T_{1}$ of $\Delta_{1}$ and $T_{2}$ of $\Delta_{2}$ and insert a D3-brane probe at a point $p_{D3}$ near $p$ such that it experiences either the monodromy $\rho_1$ while looping $l_{1} = D_1 \cap T_1$ or the monodromy $\rho_2$ while looping $l_2 = D_2 \cap T_2$. Here $D_1$ and $D_2$ are disks intersecting transversally with $T_1$ and $T_2$, respectively.

We define $E\hookrightarrow X_{1,2}\rightarrow D_{1,2}$ to be the restriction of the elliptic CY3 $E\hookrightarrow X_3 \rightarrow B_2$ to $D_{1,2}$, respectively. The naive dyonic spectrum of the D3-brane worldvolume theory as seen from $T_{1,2}$ with flavor algebra $F_{1,2}$ can be calculated by counting the M2-branes in the dual M-theory picture wrapping cycles in the relative homology group $H_2(X_{1,2}, E_{p_{D3}})$~\cite{Banks:probing, DeWolfe:1998bi, Halverson:Strong_Coupling}. The monodromy acting on the string junctions becomes a homomorphism acting on the relative homology group as follows:
\begin{equation}
    \rho_2: H_2(X_1, E_{p_{D3}}) \rightarrow H_2(X_1, E_{p_{D3}}),\ \rho_1: H_2(X_2, E_{p_{D3}}) \rightarrow H_2(X_2, E_{p_{D3}})\,.
\end{equation}
Though in general $H_2(X_1, E_{p_{D3}}) \neq H_2(X_2, E_{p_{D3}})$, the uniqueness of the D3-brane worldvolume theory at $p = \Delta_1 \cap \Delta_2$ as a 4D $\mathcal{N} = 1$ SCFT requires that~\cite{Grassi:D3brane, Grassi:2021ptc}:
\begin{equation}\label{eq:homology_lattice_equal}
    \frac{H_2(X_1, E_{p_{D3}})}{\rho_2} \cong \frac{H_2(X_2, E_{p_{D3}})}{\rho_1}\,.
\end{equation}
This relation will play an important role in the later discussions.

In the naive 10D action~(\ref{10-SymTFT_pre}), there are different $\breve{F}_3^{\rho'}$ to be reduced on elements in different $\Lambda_{\rho'}^{\gamma_1} := H^1(\gamma_1,(\mathbb{Z}\oplus\mathbb{Z})_{\rho'})$ for each $\rho'$ associated to a $\gamma_1$. For simplicity and without loss of generality we focus on the case with two intersecting stacks of 7-branes on $\Delta_1\subset B_2$ and $\Delta_2\subset B_2$ where $\gamma_1^{(1)} = \Delta_1 \cap L$ and $\gamma_1^{(2)} = \Delta_2 \cap L$. The configuration of this system is exactly the same as illustrated in Figure~\ref{fig:D3_12}.

One crucial subtlety that is not depicted in Figure~\ref{fig:D3_12} is that $\gamma_1^{(1)}$ and $\gamma_1^{(2)}$ are not simply obtained from ``sliding'' $l_1$ and $l_2$ along the tubes $T_1$ and $T_2$ to $L$, respectively. Rather, $\gamma_1^{(1)}$ links to $l_1$ while $\gamma_1^{(2)}$ links to $l_2$ on $L$. Since $\Delta_1\cap \Delta_2\neq \emptyset$, $l_1$ links to $l_2$ on $L$. Therefore homologously we have $\gamma_1^{(1)} \cong l_2$ and $\gamma_1^{(2)} \cong l_1$, or equivalently, $\gamma_1^{(1)}$ links $\gamma_1^{(2)}$.

\begin{figure}
    \centering
    \includegraphics[width=0.5\textwidth]{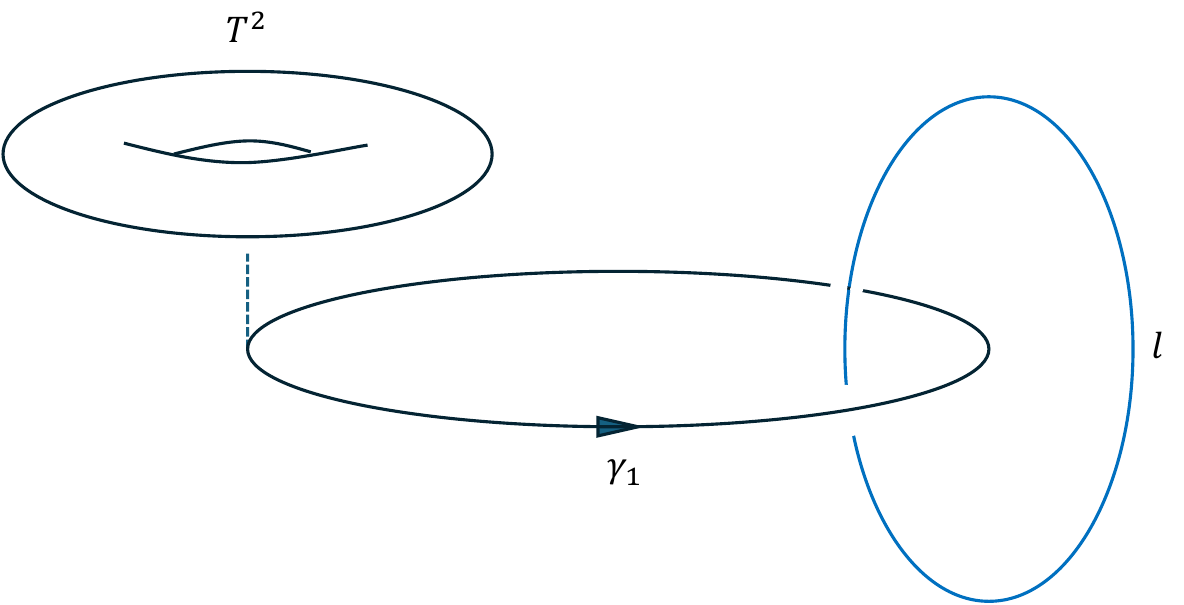}
    \caption{The geometrization of $H^1(\gamma_1, (\mathbb{Z}\oplus \mathbb{Z})_\rho)$ as an elliptic fibration $E\hookrightarrow K \rightarrow \gamma_1$ with monodromy $\rho$ acting on 1-cycles of $E$.}
    \label{fig:K}
\end{figure}

To geometrize $H^1(\gamma_1, (\mathbb{Z}\oplus \mathbb{Z})_\rho)$ we consider the elliptic fibration over $\gamma_1$ with $\rho$ acting on the 1-cycles of the $T^2$-fiber as illustrated in Figure~\ref{fig:K}~\cite{Cvetic:HigherForm_Anomalies, Heckman:Brane_GenSymm}.
The monodromy around $\gamma_1$ is generated by the 7-brane wrapping $l$ linking $\gamma_1$ and $H^1(\gamma_1, (\mathbb{Z}\oplus \mathbb{Z})_\rho) \cong \text{Tor}H_1(K, \mathbb{Z})$~\cite{Cvetic:HigherForm_Anomalies}. We have:
\begin{equation}
    H_1(K,\mathbb{Z}) \cong \mathbb{Z} \oplus (\mathbb{Z}^2/\text{Im}(\rho-1)) 
\end{equation}
where the non-torsional part $\mathbb{Z}$ is generated by the base $\gamma_1$ while the torsional part is isomorphic to $H^1(\gamma_1, (\mathbb{Z}\oplus \mathbb{Z})_\rho)$ as stated above. For the configuration in Figure~\ref{fig:D3_12}, since $\gamma_1^{(1)}$ links $\gamma_1^{(2)}$, the cohomology groups of our interest are $H^1(\gamma_1^{(1)}, (\mathbb{Z}\oplus \mathbb{Z})_{\rho_2})$ and $H^1(\gamma_1^{(2)}, (\mathbb{Z}\oplus \mathbb{Z})_{\rho_1})$ where $\rho_{2,1}$ is the monodromy along $\gamma_1^{(1,2)}$ generated by the 7-branes on $\gamma_1^{(2,1)}$, respectively. Since $\gamma_1^{(1)} \simeq l_2$ and $\gamma_1^{(2)} \simeq l_1$, we have:
\begin{equation}\label{eq:cohom_to_homo}
    \begin{split}
        &H^1(\gamma_1^{(1)}, (\mathbb{Z}\oplus \mathbb{Z})_{\rho_2}) \cong H^1(l_2, (\mathbb{Z}\oplus \mathbb{Z})_{\rho_2}) \cong \text{Tor}H_1(K_2,\mathbb{Z}),\\ &H^1(\gamma_1^{(2)}, (\mathbb{Z}\oplus \mathbb{Z})_{\rho_1}) \cong H^1(l_1, (\mathbb{Z}\oplus \mathbb{Z})_{\rho_1}) \cong \text{Tor}H_1(K_1,\mathbb{Z})
    \end{split}
\end{equation}
using the map to $K_{1,2}$. The base $l_{1,2}$ of $K_{1,2}$ can be viewed as the boundary of the disk $D_{1,2}$ that intersects transversally with $\gamma_1^{(2,1)}$. Using the elliptic fibration $E\hookrightarrow K_{1,2}\rightarrow l_{1,2}$ we further define the elliptic fibration $E\hookrightarrow X_{1,2}\rightarrow D_{1,2}$ for which we have:
\begin{equation}\label{eq:l_to_D}
    \text{Tor}H_1(K_{1,2},\mathbb{Z}) \cong H_2(X_{1,2}, E_p)
\end{equation}
with a reference point $p\in l_{1,2}$ since the non-torsion generator $l_{1,2}\subset K_{1,2}$ trivializes after extending the base from $l_{1,2}$ to $D_{1,2}$. We note that the elements in $H_2(X_{1,2}, E_p)$ are actually the \emph{Kodaira thimbles} defined in~\cite{Hubner:GenSymm_Ftheory}.

As we have argued previously, in order to have a uniquely defined 2-form field $f_2^\rho$ in $\mathcal{M}^7$ we have to enlarge each $\rho'$ to $\rho$, which is equivalent to finding the homomorphism~(\ref{eq:lattice_homomorphism}) from the $\Lambda_{\rho'}$ and $\Lambda_{\rho}$. For the configuration in Figure~\ref{fig:D3_12}, we enlarge $\rho_{1,2}$ to $\rho$ by supplementing $\rho_{2,1}$. Using~(\ref{eq:cohom_to_homo}) and~(\ref{eq:l_to_D}), it is easy to see that the corresponding lattice homomorphisms are given by the monodromy actions on the homology lattices as follows:
\begin{equation}
    \lambda_{\gamma_1}: \Lambda_{\rho_2}:= H_2(X_2, E_p) \rightarrow \frac{H_2(X_2, E_p)}{\rho_1},\ \lambda_{\gamma_2}: \Lambda_{\rho_1}:= H_2(X_1, E_p) \rightarrow \frac{H_2(X_1, E_p)}{\rho_2}\,.
\end{equation}
Because of~(\ref{eq:homology_lattice_equal}), it is clear that $\lambda_{\gamma_{1,2}}$ is the desired homomorphism from $\Lambda_{\rho_{1,2}}$ to a unique $\Lambda_{\rho}$.

In summary, we see that the enlargement from different $\rho'$ associated to each flavor brane locus $\gamma_1\subset L$ to $\rho$ generated by all the monodromies in the system is necessary for the uniqueness of the D3-brane worldvolume theory. Moreover, analyzing the dyonic spectrum of the worldvolume 4D $\mathcal{N} = 1$ SCFT via counting the string junctions naturally leads to the desired lattice homomorphisms which defines the unique cohomology group on which $\breve{F}_3^\rho$ is dimensionally reduced. Unlike the ``bulk field'' $\breve{F}_3^\rho$ propagating in $L$ which lives in the unique lattice $\Lambda_{\rho}$, each $J_2^{\rho'}$ is localized on the corresponding flavor 7-brane worldvolume $\mathcal{M}^7\times \gamma_1$. Hence a priori they do not have to live in $\Lambda_{\rho}$ as $\breve{F}_3^\rho$ does. In some sense, given the embedding $i:\gamma_1\hookrightarrow L$, the pull-back $i^*B_2^\rho$ on $\gamma_1$ ``forgets'' from $\rho$ the monodromy generated by the 7-branes on $\gamma_1$ leading to $B_2^{\rho'}$ on $\gamma_1$, which further combines linearly with $F_2^{\rho'}$ into $J_2^{\rho'}$.

\paragraph{Calculation of $L_{\gamma_1}^{ij}$}

The extra bit of topological data yet to be calculated is $L_{\gamma_1}^{ij}$. For this we again make use of the configuration in Figure~\ref{fig:K}. We consider the lift of $\int_{\gamma_1}\breve{t}_{1i} \breve{t}_{1j}$ in $K$, which turns $L_{\gamma_1}^{ij}$ into the linking pairing in $K$ with $\text{Tor} H_1(K) = \text{coker}(\rho - 1)$~\cite{Hubner:GenSymm_Ftheory, Heckman:Brane_GenSymm}. More precisely, we have:
\begin{equation}
    \int_{\gamma_1}\breve{t}_{1i} \breve{t}_{1j} \equiv \int_{K} \breve{u}_{2i} \breve{u}_{2j} \mod \mathbb{Z}\,.
\end{equation}
Therefore calculating $L_{\gamma_1}^{ij}$ is equivalent to calculating the CS invariant of $K$ which can be viewed as the boundary of $\mathbb{C}^2/\Gamma$ where $\Gamma$ is determined by $\rho$. This is the same calculation we have done in section~\ref{sec:SymTFT_20} and has been done in~\cite{Sakura:SymTFT_String} in the context of M-theory on non-compact K3.

\subsection{SymTFT coefficients from M-theory on elliptic CY3}\label{sec:Mtheory}

In order to fix the coefficients of the 7D SymTFT action~(\ref{10-SymTFT}) from topological terms, we consider the dimensional reduction of the 6D $\mc{N}=(1,0)$ theory on $S^1$, resulting in the 5D $\mc{N}=1$ KK theory. This 5D theory can be alternatively obtained as M-theory on the resolved elliptic CY3 $\widetilde{X}_3$ of the singular Weierstrass model $X_3$. For this 5D KK theory, we can employ the known methodology to compute its 6D SymTFT action~\cite{Sakura:SymTFT_String}.

Denote the generator of $H^1(S^1,\mb{Z})$ by $\theta$, we expand the background gauge fields for higher-form symmetries in (\ref{10-SymTFT}) as
\be
\ba
a_{3\alpha}&=\tilde{b}_{2\alpha}'\theta+\tilde{a}_{3\alpha}\,,\cr
f_{2i}&=\tilde{a}_{1i}\theta+\tilde{b}_{2i}\,,\cr
f_{4j}&=\tilde{c}_{3j}\theta+\tilde{b}_{4j}\,,\cr
\Upsilon_{5j}^{\rho'}&=\tilde\Upsilon_{4j}^{\rho'}\theta+\tilde{\Upsilon}'_{5j}\,.
\ea
\ee
The 5D background gauge fields are organized into dual pairs $(\tilde{b}'_2,\tilde{a}_3)$, $(\tilde{a}_1,\tilde{b}_4)$, $(\tilde{b}_2,\tilde{c}_3)$. For the 0-form symmetry field strengths $\mc{F}_{2,k}$ and the discrete 0-form symmetry $\j_{1j}^{\rho'}$, they could generate $(-1)$-form symmetries after the dimensional reduction on $S^1$. However these effects are trivialized after the $S^1$ circle goes to zero radius limit (as pointed out in \cite{Nardoni:2024sos}), and we would not consider these effects. After the dimensional reduction on $S^1$, we get the action
\be
\ba
\label{6d5d-reduction-gen}
\frac{S_{\mathcal{T}_{sym},5d}}{2\pi} &= 2\text{CS}[L]_{\alpha\beta}\int_{\mathcal{M}^6} \tilde{b}_{2\alpha}' d\tilde{a}_{3\beta} + \text{CS}_t[L]_{ij} \int_{\mathcal{M}^6} (\tilde{b}_{2i}d\tilde{c}_{3j}+\tilde{a}_{1i}d\tilde{b}_{4j}) +\cr
&\frac{1}{2}L[u,t]_{ij,\alpha} \int_{\mathcal{M}^6} (\tilde{b}_{2\alpha}' \tilde{b}_{2i} \tilde{b}_{2j}+2\tilde{a}_{3\alpha}\tilde{a}_{1i}\tilde{b}_{2j})\cr
    &
    + \sum_{\gamma_1,\text{non-trivial\ }\rho'} \int_{\mathcal{M}^6} L_{\gamma_1}^{ij} \tilde{\j}_{1i}^{\rho'} d \tilde{\Upsilon}_{4j}^{\rho'} + \frac{L_{\gamma_1}^{ij}}{2} \tilde{\j}_{1j}^{\rho'} \tilde{a}_{1i}^{\rho} \text{tr} \tilde{F}_{2,\gamma_1}^2 + \frac{L_{\gamma_1}^i}{2} \tilde{b}'_{2i} \text{tr} \tilde{F}_{2,\gamma_1}^2 \cr
    &+\sum_{\gamma_1,\text{trivial\ }\rho'}\int_{\mc{M}^6}\frac{L_{\gamma_1}^i}{2} \tilde{b}'_{2i} (\text{tr} \tilde{F}_{2,\gamma_1}^2+\j_2^2)\,.
\ea
\ee

Now we compare the result with the 5d computation from M-theory on elliptic CY3 \cite{Morrison:HigherForm_5D, Apruzzi:6DSymTFT,Hubner:GenSymm_Ftheory}. We discuss some detailed aspects of this computation in the appendix~\ref{sec:CY3SymTFT}. For the resolved elliptic CY3 $\widetilde{X}$, which has both compact divisors $S_i$ and compact curves $C_j$, one computes the Smith normal decomposition of the intersection matrix $q_{ij}=S_i\cdot C_j$:
\be
q=S\cdot \bp l_1 & 0 & 0 & \dots & 0\\
0 & l_2 & 0 & \dots & 0\\
\vdots & \vdots &\vdots & \ddots & \vdots\\
0 & 0 & l_k & \dots & 0\ep\cdot T\,.
\ee
The 1-form symmetry of the KK theory is given by
\be
\Gamma_{\rm 5d}=\bigoplus_{l_i>1} \widehat{\mb{Z}/l_i\mb{Z}}\,.
\ee
Each factor $\mb{Z}_{l_i}$ with $(l_i>1)$ corresponds to a torsional factor in the 5d 1-form symmetry. However, they have different origins in the 6D (1,0) theory, and the Smith normal decomposition is not unique. To correctly construct the compact divisor representatives $S^{(\alpha)}$ and  $S^{(i)}$ which correspond to $\tilde{b}_{2\alpha}'$ and $\tilde{b}_{2i}$ respectively, one should follow the following guidelines~\footnote{$S^{(\alpha)}$ and $S^{(i)}$ are rational linear combinations of compact divisors of $\widetilde{X}_3$, which are related to the compact divisor representatives $Z_i$ in \cite{Sakura:SymTFT_String} by a torsional degree factor $l_i$: $S^{(i)}=Z_i/l_i$, $S^{(\alpha)}=Z_\alpha/l_\alpha$.}:
\begin{itemize}
    \item For the representative $S^{(\alpha)}$ corresponding to $\tilde{b}_{2\alpha}'$ whose origin is 6D 3-form background gauge field leading to 2-form symmetry, it is expressed as a linear combination of $T^2$-fibrations over compact base curves in the tensor branch.
    \item For the representative $S^{(i)}$ corresponding to $\tilde{b}_{2i}$ whose origin is 6D 2-form background gauge field leading to 1-form symmetry, it is expressed as a linear combination of exceptional divisors (but not vertical divisor).
    \item Note that in either case we have $C_j\cdot S^{(i)}\in \mb{Z}$, $C_j\cdot S^{(\alpha)}\in \mb{Z}$ for all compact curves $C_j$.
\end{itemize}

Analogous to \cite{Sakura:SymTFT_String}, the coefficients $L[u,t]_{ij,\alpha}$ are then computed by
\be
L[u,t]_{ij,\alpha}=S^{(i)}\cdot S^{(j)}\cdot S^{(\alpha)}\,.
\ee
Note that in the computation there is no need to include the contributions from the 11D M-theory topological term $C_3\wedge X_8$, because it contributes non-trivially only to the self-intersection of a single $S^{(i)}$ or $S^{(\alpha)}$, which do not appear in our uplifted SymTFT action for 6D (1,0) theories.

The coefficients $L_{\gamma_1}^i=\int_{\gamma_1}\br{u}_{2i}$ are related to the 2-form symmetry and 0-form flavor symmetries in 6D. On $\widetilde{X}_3$, we denote the $k$-th non-compact exceptional divisor for the non-abelian flavor symmetry on $\gamma_1$ by $D_{\gamma_1}^{(k)}$, then we can compute the SymTFT coefficients from the triple intersection numbers
\be
\label{LSDD}
L_{\gamma_1}^{\alpha}\mc{C}^{\gamma_1}_{sym,jk}=S^{(\alpha)}\cdot D^{(j)}_{\gamma_1}\cdot D^{(k)}_{\gamma_1}\,.
\ee
The reasoning is that such coefficient should be consistent with the coefficients of $BFF$ term in (4.9) of \cite{Sakura:SymTFT_String}.

Here we have the symmetric Cartan matrix of the non-abelian flavor algebra $G_i$~\cite{Park:2011ji,Apruzzi:2019opn}
\be
\mc{C}^{\gamma_1}_{sym,jk}=-\frac{2\langle\alpha,\alpha\rangle_{\rm max}\langle\alpha_j,\alpha_k\rangle}{\langle\alpha_j,\alpha_j\rangle\langle\alpha_k,\alpha_k\rangle}\,,
\ee
where $\alpha_j$ is the $j$-th simple root of $G_i$, corresponding to the non-compact divisor $D^{(i)}_j$. $\langle\alpha,\alpha\rangle_{\rm max}$ is the square of length of the longest root of $G_i$. For simply-laced Lie algebra, $\mc{C}^i_{sym,jk}$ is equal to the Cartan matrix $\mc{C}^i_{jk}$, but they are different for non-simply laced Lie algebras. For instance for $C_n$ type Lie algebra, we have
\be
\mc{C}^i_{sym,jk}=\bp 
-4 & 2 & 0 & \dots & 0 & 0 \\
2 & -4 & 2 & \dots & 0 & 0\\
0 & 2 & -4 & \dots & 0 & 0\\
\vdots & \vdots & \vdots & \ddots & \vdots & \vdots\\
0 & 0 & 0 & \dots & -4 & 2\\
0 & 0 & 0 & \dots & 2 & -2
\ep\,.
\ee

Finally, for the coefficients CS$[L]_{\alpha\beta}$, CS$_t[L]_{ij}$ and $L_{\gamma_1}^{ij}$ in front of the BF terms in (\ref{10-SymTFT}), they are given by $\frac{1}{N}$ where $N$ is the torsional  degree of the $\mb{Z}_N$ gauge group in the BF term, which we would not elaborate further.

\section{Examples}\label{sec:examples}

In this section we calculate $S_{\mathcal{T}_{sym}}$ for several representative 6D $\mathcal{N} = (1,0)$ examples. In section~\ref{sec:SNNHC_egs} we calculate $S_{\mathcal{T}_{sym}}$ for the 6D SCFTs corresponding to single-node \emph{non-Higgsable clusters} (NHC, see~\cite{Morrison:NHC}). In section~\ref{sec:SO8on4_eg} we calculate $S_{\mathcal{T}_{sym}}$ for the 6D SCFT whose tensor branch is $\overset{\mk{so}(2n+8)}{(-4)}-[\mk{sp}(2n)]$. In section~\ref{sec:trivial1_nontri0_egs} we calculate $S_{\mathcal{T}_{sym}}$ for a 6D SCFT with non-trivial 0-form symmetry but trivial 1-form symmetry, which constitute a large subclass of all 6D SCFTs.

\subsection{Non-Higgsable Clusters}\label{sec:SNNHC_egs}

We first look at the simplest 6D $\mathcal{N} = (1,0)$ examples: single-node non-Higgsable clusters where there is only one $(-p)$-curve. In this case we have~\cite{Heckman:Brane_GenSymm}
\begin{equation}\label{eq:H_single_NHC}
    H^p(L, (\mathbb{Z}\oplus \mathbb{Z})_\rho) = (0, G, 0, G)
\end{equation}
where $G$ depends on the type of the node as shown in Table~\ref{tab:G_single_NHC}. We also have $\text{CS}[L] = \frac{1}{2p}$ and $\text{CS}_t[L]_{ij} = L_t^{\gamma_k}(\breve{t})$ in each case~\cite{Heckman:Brane_GenSymm}. For a single-node NHC in Table~\ref{tab:G_single_NHC} the $\rho$-twist is nothing but the $SL(2,\mathbb{Z})$ monodromy induced by the gauge 7-branes when looping around the torsional 1-cycle $\gamma_1 \subset L$.
\begin{table}[h]
    \centering
    \begin{tabular}{c|c|c|c|c|c}
         NHC & $\stackrel{\mathfrak{su}(3)}{3}$ & $\stackrel{\mathfrak{so}(8)}{4}$ & $\stackrel{\mathfrak{e}_6}{6}$ & $\stackrel{\mathfrak{e}_7}{8}$ & $\stackrel{\mathfrak{e}_8}{12}$ \\
        \hline
        \hline
        $(p,q)$ & $(3,1)$ & $(4,1)$ & $(6,1)$ & $(8,1)$ & $(12,1)$ \\
        \hline
        $\text{ord}(\rho)$ & 3 & 2 & 3 & 4 & 6 \\
        \hline
        $G$ & $\mathbb{Z}_3$ & $\mathbb{Z}_2\oplus \mathbb{Z}_2$ & $\mathbb{Z}_3$ & $\mathbb{Z}_2$ & $0$ \\
        \hline
        $L_t^{\gamma_k}(\breve{t})$ & $\frac{1}{3}$ & $\begin{pmatrix}
            0 & \frac{1}{2} \\
            \frac{1}{2} & 0
        \end{pmatrix}$ & $\frac{2}{3}$ & $\frac{1}{2}$ & 0
    \end{tabular}
    \caption{$G$ in (\ref{eq:H_single_NHC}) and $(p,q)$ (see~\ref{eq:pq_action}) of the single-node NHCs.}
    \label{tab:G_single_NHC}
\end{table}
Note that in Table~\ref{tab:G_single_NHC} the NHC $\stackrel{\mathfrak{f}_4}{5}$ and $\stackrel{\mathfrak{e}_7}{7}$ are excluded which will be discussed in a moment. Given~(\ref{eq:H_single_NHC}), (\ref{eq:ansatz_10_noflav}) becomes
\begin{equation}\label{eq:B_ansatz_10}
    \begin{split}
        \breve{F}_3^\rho = \breve{f}_2^\rho\star \breve{t}_1,\ \text{for}\ t_1^i\in H^1(L, (\mathbb{Z}\oplus \mathbb{Z})_\rho)=G
    \end{split}
\end{equation}
which is the rule of reduction in this case.

Plugging~(\ref{eq:B_ansatz_10}) for the single node NHC into the general expression~(\ref{10-SymTFT-p1}), we have:
\begin{equation}\label{eq:SymTFT_10}
    \frac{S_{\mathcal{T}_{sym}}}{2\pi} = \int_{\mathcal{M}^7} \frac{1}{2p} a_3da_3 + L_t^{\gamma_k}(\breve{t}) f_{2i}^\rho df_{4j}^\rho +  \frac{1}{2}L[u,t]_{ij} a_3 f_{2i}^\rho f_{2j}^\rho\,.
\end{equation}
The first and the second term in (\ref{eq:SymTFT_10}) are BF couplings of the $\mathbb{Z}_p$ valued 3-form background field and the 2-form background field respectively, and the last term is the 't Hooft anomaly~\cite{Bhardwaj:Generalized_Charge_II, Sakura:ICTP_Symmetries}. We take the NHC $\stackrel{\mathfrak{su}(3)}{3}$ as a concrete example, whose corresponding $S_{\mathcal{T}_{sym}}$ is
\begin{equation}\label{eq:SymTFT_NHC3}
    \frac{S_{\mathcal{T}_{sym}}}{2\pi} = \int_{\mathcal{M}^7} \frac{1}{6} a_3da_3 + \frac{1}{3} f_{2}^\rho df_{4}^\rho +  \frac{1}{3} a_3 f_{2}^\rho f_{2}^\rho\,.
\end{equation}
The result matches the result in~\cite{Apruzzi:6DSymTFT}. We note that the group theoretical coefficient $\alpha_G$ in~\cite{Apruzzi:6DSymTFT} is our $\frac{1}{2}L[u,t]_{ij}$.

Let us now explain why $\stackrel{\mathfrak{f}_4}{5}$ and $\stackrel{\mathfrak{e}_7}{7}$ were omitted from Table~\ref{tab:G_single_NHC}. The case $\stackrel{\mathfrak{e}_7}{7}$ simply falls outside the range of validity of~(\ref{10-SymTFT-p1}) since there exists an $I_1$ locus intersecting both the $(-7)$-curve and the boundary, i.e. there exists a flavor brane. For $\stackrel{\mathfrak{f}_4}{5}$ the boundary is $S^3/\mathbb{Z}_5$ where $\mathbb{Z}_5\simeq \frac{1}{5}(1,1)$ (c.f.~(\ref{eq:pq_action})). Therefore there is a background $\mathbb{Z}_5$ 3-form field in $S_{\mathcal{T}_{sym}}$. 

In this case for $\rho$ we have~\cite{Grassi:2018wfy}
\begin{equation}
    \rho = \begin{pmatrix}
        0 & 1 \\
        -1 & 1
    \end{pmatrix}\,.
\end{equation}
with which it is easy to check that:
\begin{equation}
    H^1(L,(\mathbb{Z}\oplus \mathbb{Z})_\rho) \cong \text{coker}(\rho - 1) = \emptyset\,.
\end{equation}
Therefore there are no background 2-form fields in $S_{\mathcal{T}_{sym}}$, which is consistent with the fact that the center of $F_4$ is trivial~\cite{Bhardwaj:Higher_form_5D6D}.

Due to the lack of 1-form and 0-form global symmetries, the SymTFT action for $\stackrel{\mathfrak{f}_4}{5}$ is simply
\be
\frac{S_{\mathcal{T}_{sym}}}{2\pi} = \int_{\mathcal{M}^7} \frac{1}{5} a_3da_3
\ee
and the SymTFT action for $\stackrel{\mathfrak{e}_7}{7}$ is
\be
\frac{S_{\mathcal{T}_{sym}}}{2\pi} = \int_{\mathcal{M}^7} \frac{1}{7} a_3da_3\,.
\ee

\paragraph{Crosscheck against 5D KK theory}

One can also check the results via the uplift of the 6D SymTFT of the 5D KK theory, which can be computed using the resolved elliptic CY3 in M-theory.

As a concrete example, let us consider the case of $\stackrel{\mathfrak{su}(3)}{3}$, which was discussed in \cite{Hubner:GenSymm_Ftheory, Apruzzi:6DSymTFT}. In the smooth elliptic CY3 $\widetilde{X}$, there are three compact exceptional divisors $S_1$, $S_2$ and $S_3$~\cite{DelZotto:2017pti}, which have the topology of Hirzebruch surface $\mb{F}_1$. We denote the compact curves by $C_1$, $\dots$, $C_4$, where $C_4=S_1\cdot S_2=S_1\cdot S_3=S_2\cdot S_3$ is the intersection curve between the divisors, and $C_1$, $C_2$, $C_3$ are the $\mb{P}^1$ fibers of $S_1$, $S_2$ and $S_3$. The intersection matrix of $\widetilde{X}$ is
\be
\mc{M}=
\begin{array} {c|cccc}
& C_1 & C_2 & C_3 & C_4\\
\hline   
S_1 & -2 & 1 & 1 & -1\\
S_2 & 1 & -2 & 1 & -1\\
S_3 & 1 & 1 & -2 & -1
\end{array}
\ee

From the Smith normal decomposition of $\mc{M}$, the 1-form symmetry of the 5d KK theory is $\Gamma_{\rm 5d}=\mb{Z}_3\oplus\mb{Z}_3$. The generator corresponding to $\tilde{b}_2$ can be chosen as
\be
S=\frac{1}{3}(S_3-S_2)\,,
\ee
while the generator corresponding to 
 $\tilde{b}_2^{\prime }$ is 
\be
S'=\frac{1}{3}(S_1+ S_2+ S_3)\,.
\ee

Note that $S'$ is proportional to the $T^2$ fibration over the $(-3)$-curve, which naturally uplifts to the generator of the 6d $\mb{Z}_3$ 2-form symmetry. On the other hand, $S$ is a linear combination of exceptional divisors, which corresponds to the generator of the 6d $\mb{Z}_3$ 1-form symmetry.

We get the intersection numbers
\be
S^3=S^{\prime 3}=S\cdot S^{\prime 2}=0\ ,\ S^2\cdot S^\prime=\frac{2}{3}\,.
\ee
Thus the final SymTFT expression for the 5D KK theory is
\be
\frac{\widetilde{S}_{\mathcal{T}_{sym}}}{2\pi} = \frac{1}{3}\int_{\mc{M}^6}\tilde{b}_2' d\tilde{a}_3+\frac{1}{3}\int_{\mc{M}^6}\left(\tilde{b}_2^{\prime}\tilde{b}_2\tilde{b}_2+2\tilde{a}_3\tilde{a}_1\tilde{b}_2\right)+\frac{1}{3}\int_{\mathcal{M}^6}\left(\tilde{b}_2 d\tilde{c}_3+\tilde{a}_1d\tilde{b}_4\right)\,.
\ee
The uplifted SymTFT action for the 6D (1,0) theory is:
\be
\frac{S_{\mathcal{T}_{sym}}}{2\pi} = \frac{1}{6} \int_{\mathcal{M}^7}a_3 d a_3 + \frac{1}{3} \int_{\mathcal{M}^7} a_3 f_2 f_2 +\frac{1}{3}\int_{\mathcal{M}^7}f_2 d f_4\,
\ee
which matches~(\ref{eq:SymTFT_NHC3}) and the known results~\cite{Hubner:GenSymm_Ftheory,Apruzzi:6DSymTFT}.

In this case, due to the coefficient $\frac{1}{6}$ in front of the term $a_3da_3$, the SymTFT does not admit a polarization strictly speaking, and the 6D (1,0) SCFT is always a relative theory.

\subsection{$\mathfrak{so}(2n+8)$ on a $(-4)$-curve}\label{sec:SO8on4_eg}

In this section, we consider the example of $\mathfrak{so}(2n+8)$ gauge algebra on a $(-4)$-curve with non-trivial 1-form, 2-form symmetries, as well as flavor branes giving rise to 0-form symmetries.

The tensor branch of this example is
\be
\overset{\mk{so}(2n+8)}{(-4)}-[\mk{sp}(2n)]\,.
\ee
In the SymTFT, we have the background gauge fields $f_2$ and $f_4$ for $\mb{Z}_2$ 1-form/3-form symmetry~\cite{Bhardwaj:Higher_form_5D6D}, $a_3$ for $\mb{Z}_4$ 2-form symmetry and $F_j$ ($j=1,\dots,2n)$ as the field strength of Cartan $\mk{u}(1)$ of the flavor symmetry $G_F=\mk{sp}(2n)$. Note that we use the standard labels for the Cartan subalgebra (simple roots) of $\mk{sp}(2n)$ Lie algebra
\be
\dynkin[edge length=0.8cm,labels*={1,2,n-2,n-1,n}]B{}
\ee
The resolved geometry can be found in e.g.~\cite{Apruzzi:2019kgb}. Labeling the compact exceptional divisors for $so(2n+8)$ as
\be
\dynkin[edge length=0.8cm,extended,labels*={U_1,U,U_2,U_3,U_{n-2},U_{n-1},U_n}] D{*oo...ooo}
\ee
The compact representative for the $\mb{Z}_4$ 2-form symmetry is
\be
S'=\frac{1}{4}(U+U_1+2\sum_{i=2}^{n-2}U_i+U_{n-1}+U_n)
\ee
while the compact representative for the $\mb{Z}_2$ 1-form symmetry is
\be
S=\frac{1}{2}(U+U_1)\,.
\ee

For the parts in the 7D SymTFT action without contributions from flavor branes, we compute from the resolved geometry
\be\label{eq:SOSp_action}
\frac{S_{\mathcal{T}_{sym}}}{2\pi}\supset\int_{\mc{M}^7}\frac{1}{8}a_3da_3+\frac{1}{2} f_2df_4+\frac{1}{2}a_3 f_2 f_2\,.
\ee

To obtain the $\mathbb{Z}_2$ 1-form symmetry as suggest by the SymTFT action~(\ref{eq:SOSp_action}) from IIB monodromy perspective, we apply the method described in section~\ref{sec:6D10_flav}. In this case the lattice $\Lambda_G = \mathbb{Z}^2/\text{coker}(\rho_G - 1)$ is $\mathbb{Z}_2\times \mathbb{Z}_2$ for even $n$ and $\mathbb{Z}_4$ for odd $n$ while the lattice $\Lambda_F = \mathbb{Z}^2/\text{coker}(\rho_F - 1)$ is $\mathbb{Z}_{4n}$ where we have dropped the non-torsional part. To study the intersection of lattices we denote by $e_1$ and $e_2$ the two basis vectors generating $\mathbb{Z}^2$. When $n$ is even the torsional parts of $\Lambda_G$ and $\Lambda_F$ are
\begin{equation}
    \Lambda_G = \langle (e_1,e_2)|2e_1 = 2e_2 = 0 \rangle,\ \Lambda_F = \langle e_2 | 4ne_2 = 0 \rangle\,.
\end{equation}
Therefore $\Lambda_G\cap \Lambda_F \cong \mathbb{Z}_2 = \langle e_2 | 2e_2 = 0 \rangle$. While when $n$ is odd we have
\begin{equation}
    \Lambda_G = \langle (e_1,e_2)|4e_1 = 2e_2 = 0 \rangle,\ \Lambda_F = \langle e_2 | 4ne_2 = 0 \rangle\,.
\end{equation}
Therefore we again have $\Lambda_G \cap \Lambda_F \cong \mathbb{Z}_2 = \langle e_2 | 2e_2 = 0 \rangle$. We see that in either case we have a $\mathbb{Z}_2$ lattice for $f_2^\rho$ in order to match the monodromies from gauge branes and flavor branes. Hence we conclude that the 1-form symmetry of this theory is $\mathbb{Z}_2$, which matches the result in~\cite{Bhardwaj:Higher_form_5D6D} via a different approach, i.e. counting the charged matters and charged instanton strings.

For the flavor center symmetry, first note that due to the presence of Tate monodromy
\be
\rho_{\rm Tate}=\bp -1 & 0\\0 & -1\ep\,,
\ee
the possible flavor center symmetry is given by~\cite{Apruzzi:2021mlh}
\be
\Lambda_F\cap \mb{Z}^2/\mathrm{coker}(\rho_{\rm Tate}-1)=\mb{Z}_2\,.
\ee

Using~(\ref{LSDD}) and plugging in~(\ref{F2-w2}) one can compute $L_{\gamma_1}^{ij}=\frac{1}{2}$ and $L_{\gamma_1}^i=\frac{1}{4}$ and get the SymTFT action (we have omitted the superscript $\rho'$)
\be
\frac{S_{\mathcal{T}_{sym}}}{2\pi}=\int_{\mc{M}^7}\frac{1}{8}a_3da_3+\frac{1}{2} f_2df_4+\frac{1}{2}a_3 f_2 f_2+\frac{1}{2}\j_1 d\Upsilon_5-\frac{1}{4}\j_1 f_2 w(F_{2,\mathfrak{sp}(2n)})^2-\frac{1}{4}a_3 w(F_{2,\mk{sp}(2n)})^2\,.
\ee
Here $a_3$ is the $\mb{Z}_4$-valued background gauge field for the 2-form symmetry, $(f_2,f_4)$ are the background gauge fields for the dual 1-form/3-form $\mb{Z}_2$ symmetries. $(\j_1,\Upsilon_5)$ are the background gauge fields for the dual 0-form/4-form $\mb{Z}_2$ symmetries, where $\j_1$ corresponds to the flavor center $\mb{Z}_2$. $F_{2,\mathfrak{sp}(2n)}$ is the field strength for the $\mathfrak{sp}(2n)$ non-abelian flavor symmetry. $w(F_{2,\mathfrak{sp}(2n)})$ is the second (generalized) Stiefel-Whitney class describing the obstruction of lifting an $Sp(2n)/\mb{Z}_2$ bundle to an $Sp(2n)$ bundle.

Now we discuss the topological boundary conditions for the gauge fields corresponding to different polarizations of the SymTFT. First, for the $a_3$ with BF term $\frac{1}{8}a_3da_3$, the only possible topological boundary condition is that it is restricted within a $\mb{Z}_2\subset \mb{Z}_4$ subgroup, i.e. we can set $a_3|_{M^6}=2A_3$ on the boundary $M^6$ where $A_3\in H^3(M^6,\mb{Z}_2)$, and the actual 6D 2-form symmetry is $\mb{Z}_2^{(2)}$. After choosing the polarization, we see that there are still a 't Hooft anomaly term 
\be
\frac{S_{\rm anomaly}}{2\pi}\supset\int_{\mc{M}^7}\frac{1}{2}A_3dA_3-\frac{1}{2}A_3 w(F_{2,\mk{sp}(2n)})^2
\ee
characterizing the obstruction to gauge this $\mb{Z}_2$ 2-form symmetry.

With such a restriction of $a_3$, the term $\frac{1}{2}a_3 f_2 f_2\sim A_3 f_2 f_2$ is trivialized, and there is no mixed 't Hooft anomaly between $a_3$ and $f_2$ any more. One can choose one of the following two topological boundary conditions for $(f_2,f_4)$ and for $(\j_1,\Upsilon_5)$:

\begin{enumerate}
\item Dirichlet b.c. on $f_2$ and Dirichlet b.c. on $\j_1$, leading to a $\mb{Z}_2^{(1)}$ 1-form symmetry and a $\mb{Z}_2^{(0)}$ flavor center 0-form symmetry in 6D.

In this case there is presence of mixed 't Hooft anomaly
\be
\frac{S_{\rm anomaly}}{2\pi}\supset\int_{\mc{M}^7}\frac{1}{4}\j_1 f_2 w(F_{2,\mathfrak{sp}(2n)})^2\,.
\ee

\item Dirichlet b.c. on $f_2$ and Dirichlet b.c. on $\Upsilon_5$, leading to a $\mb{Z}_2^{(1)}$ 1-form symmetry and a $\mb{Z}_2^{(4)}$ 4-form symmetry in 6D. 

From the e.o.m. of $\j_1$, there is the relation
\be
\pi(d\Upsilon_5-\frac{1}{2}f_2w(F_{2,\mathfrak{sp}(2n)})^2)=0\,.
\ee

\item Dirichlet b.c. on $f_4$ and Dirichlet b.c. on $\j_1$, leading to a $\mb{Z}_2^{(3)}$ 3-form symmetry and a $\mb{Z}_2^{(0)}$ flavor center 0-form symmetry in 6D.

From the e.o.m. of $f_2$, there is the relation
\be
\pi(df_4-\frac{1}{2}\j_1 w(F_{2,\mathfrak{sp}(2n)})^2)=0\,.
\ee

\item Dirichlet b.c. on $f_4$ and Dirichlet b.c. on $\Upsilon_5$, leading to a $\mb{Z}_2^{(3)}$ 3-form symmetry and a $\mb{Z}_2^{(4)}$ 4-form symmetry in 6D.

This choice is only possible when $w(F_{2,\mathfrak{sp}(2n)})^2=0$.

\end{enumerate}

\subsection{Cases with trivial 1-form symmetry but non-trivial flavor center}\label{sec:trivial1_nontri0_egs}

There is a large number of 6D (1,0) models with no 1-form symmetry, but with a non-trivial center in the non-abelian 0-form flavor symmetry. Computing the SymTFT of these theories provide us important information about the WZ-term of the flavor 7-branes.

Let us consider the example of the 6d (1,0) $(A_{n-1},A_{n-1})$ conformal matter~\cite{DelZotto:2014hpa}.
\be
[\mk{su}(n)]-\overset{\mk{su}(n)}{(-2)}-\overset{\mk{su}(n)}{(-2)}-[\mk{su}(n)]\,.
\ee
It is known that the global form of non-abelian 0-form symmetry group is $G_F=(SU(n)\times SU(n))/\mb{Z}_n$\footnote{Here we only count the non-abelian flavor symmetry group realized from a stack of D7-branes or equivalently Kodaira $I_n$ singularities.}, and the theory has a $\mb{Z}_n$ flavor center symmetry~\cite{Heckman:2022suy}.

In this example we have:
\begin{equation}
    L = S^3/\Gamma,\ \Gamma = \frac{1}{3}(1,2)
\end{equation}
hence $\Gamma =\mathbb{Z}_3$. This leads to a $\mathbb{Z}_3$ valued 3-form gauge field $a_3$ with the BF-coupling:
\begin{equation}
    \frac{S_{\mathcal{T}_{sym}}}{2\pi} \supset \int_{\mathcal{M}^7} \frac{1}{6} a_3 d a_3\,.
\end{equation}
In the resolved CY3, the generator of such $\mathbb{Z}_3$ corresponds to the linear combination of compact divisors
\be
S=\frac{1}{3}\sum_{i=1}^n (U_i-V_i)\,,
\ee
where $U_i$'s give the $I_n$ fibration over the first compact $(-2)$-curve and $V_i$'s give the $I_n$ fibration over the second compact $(-2)$-curve.

The total monodromy $\rho$ of this configuration is generated by:
\begin{equation}
    \rho' := \rho_1 = \rho_2 = \begin{pmatrix}
        1 & n \\
        0 & 1
    \end{pmatrix}.
\end{equation}
Here $\rho_{i}$ is the monodromy generated by the $i^{\text{th}}$ $SU(n)$ flavor brane. Since the two stacks of flavor branes must link each other on the $L$, a D3-brane probe looping around the 1-cycle that supports one of the flavor branes will experience the monodromy generated by the other. Thus it is clear that $\rho$ is trivial hence $\mathbb{Z}^2/\text{Im}(\rho-1) \cong \mathbb{Z}^2$. An argument similar to that based on~(\ref{eq:trivial_BF_20}) leads to vanishing BF-coupling involving any 2-form background gauge field for 1-form symmetry, which is expected since the reduction of $\breve{F}_3^\rho$ on $\breve{t}_1$ in the $\mathbb{Z}^2$ lattice leads to trivial 1-form symmetry.

The 0-form symmetry and its center are nevertheless non-trivial in this example. Using~(\ref{10-SymTFT}) we have
\begin{equation}
    \frac{S_{\mathcal{T}_{sym}}}{2\pi} \supset \int_{\mathcal{M}^7} L_{\gamma_1}^{ij} \j_{1i}^{\rho'} d \Upsilon_{5j}^{\rho'}
\end{equation}
where $\j_{1}^{\rho'}$ is the background 1-form field of the $\mathbb{Z}_n$ center 0-form symmetry. In this case we also have $L_{\gamma_1}^{ij} = \frac{1}{n}$~\cite{Hubner:GenSymm_Ftheory} which leads to the correct quantization condition. 

We also have a cubic term
\be
 \frac{S_{\mathcal{T}_{sym}}}{2\pi} \supset \sum_{\gamma_1}\frac{L_{\gamma_1}}{2}a_3\text{tr}F_{2,\gamma_1}^2
\ee
where the sum is over the two flavor $\mk{su}(n)$s on $\gamma_1$ and $\gamma_2$.

The coefficient $L_{\gamma_1}$ and $L_{\gamma_2}$ can be computed from the resolved CY3 from (\ref{LSDD}), and the result is
\be
L_{\gamma_1}=\frac{1}{3}\ ,\ L_{\gamma_2}=-\frac{1}{3}\,.
\ee
The final total SymTFT action is (we omited the superscript $\rho'$)
\be
\frac{S_{\mathcal{T}_{sym}}}{2\pi}=\int_{\mathcal{M}^7} \frac{1}{6} a_3 d a_3+\frac{1}{n}\j_{1} d \Upsilon_{5}+\frac{1}{6}a_3(\text{tr}F_{2,\mk{su}(n)_1}^2-\text{tr}F_{2,\mk{su}(n)_2}^2)\,.
\ee
$a_3$ is the background gauge field for potential $\mb{Z}_3$ 2-form symmetry. $(\j_{1},\Upsilon_5)$ are the background gauge fields of the dual $\mb{Z}_n$ 0-form/4-form symmetries, where $\j_1$ corresponds to the $\mb{Z}_n$ flavor center. 
$F_{2,\mk{su}(n)_i}$ denotes the field strength for the $i$-th $\mk{su}(n)$ flavor symmetry. 

Plug in (\ref{F2-w2}), we get
\be
\frac{S_{\mathcal{T}_{sym}}}{2\pi}=\int_{\mathcal{M}^7} \frac{1}{6} a_3 d a_3+\frac{1}{n}\j_{1} d \Upsilon_{5}-\frac{n-1}{3n}a_3(w(F_{2,\mk{su}(n)_1})^2-w(F_{2,\mk{su}(n)_2})^2)\,.
\ee

Note that there are no mixing between the flavor center symmetry gauge field $\j_1$ and the 2-form symmetry gauge field $a_3$.

\section{Conclusion}

In this work we studied the 7D SymTFT associated to 6D SCFT via dimensional reduction of topological IIB action in the presence of 7-branes on the boundary link of the non-compact base geometry of a non-compact elliptic Calabi-Yau threefold. In contrast to the approaches in literature that substantially rely on the topological data of the compact part of the base geometry (or equivalently the tensor branch of the corresponding 6D SCFT) in the derivation of the action of the 7D SymTFT, we make use of only the topological data of the boundary link (which is the philosophy of~\cite{Sakura:SymTFT_String, Sakura:3DSymTFT, Heckman:Brane_GenSymm, Apruzzi:2023uma} etc.) and obtain the SymTFT action without referring to the compact part of the geometry at all.

Another advantage of our approach is that since 6D SCFTs (and LSTs) have already been largely classified from F-theory geometric engineering in~\cite{Heckman:6DBaseClassification, Heckman:AtomicClassification, Bhardwaj:classification_LST}, it is natural to look for a classification of the corresponding 7D SymTFTs from F-theory as well. We show by examples that this can indeed be done by analyzing the untwisted and twisted (differential) cohomology of the boundary links of the base geometries in the classification program of 6D SCFTs and LSTs.

We also shortly comment on the cases of 6D (1,0) little string theories~\cite{Bhardwaj:classification_LST,DelZotto:2020sop,DelZotto:2022ohj,DelZotto:2022xrh,DelZotto:2023ahf}, which are realized as F-theory on 2D bases with a degenerate intersection matrix. In these cases, it is known that there is exists a $U(1)$ 1-form symmetry which forms a 2-group structure with the 0-form Poincar\'{e} symmetry and the R-symmetry. Nonetheless, in our geometric setup we cannot capture the information of Poincar\'{e} symmetry and the R-symmetry, and we do not have a non-trivial SymTFT action for the other higher-form symmetries.

In the paper we have not yet considered the cases with a non-trivial $U(1)$ flavor symmetries from Mordell-Weil groups~\cite{Lee:2018ihr,Apruzzi:2020eqi}, as their realizations in IIB is more subtle.

Another aspect that we have excluded is the terms involving $(-1)$-form symmetry. In other words, we have set the background gauge field of these $(-1)$-form symmetry to be trivial, which is a valid choice in the geometric engineering setups. They could arise from the dimensional reduction of various form fields in the IIB action and connect different 6D SCFTs. The detailed study of such terms would be subject to future work.

\vspace{.5cm}
\noindent \textbf{Acknowledgments.}

We would like to thank Christopher Beasley, Lakshya Bhardwaj, Andreas Braun, James Halverson, Max H\"ubner, Marwan Najjar, Washington Taylor and Yi Zhang for helpful discussions. J.T. is supported by National Natural Science Foundation of China under Grant No. 12405085 and by the Natural Science Foundation of Shanghai (Grant No. 24ZR1419300). Y-N.W. is supported by National Natural Science Foundation of China under Grant No. 12175004 and by Young Elite Scientists Sponsorship Program by CAST (2022QNRC001, 2023QNRC001, 2024QNRC001).

\appendix

\section{Validity of SymTFT computations}
\label{sec:CY3SymTFT}

In the section, we comment on the validity of the identification between differential cohomology classes and divisors in Section 3.3.1 of \cite{Sakura:SymTFT_String}.

For a non-compact elliptic CY3 $X_3$ with boundary space $L_5$, we have the long exact sequence 
\be
\dots\rightarrow H_4(X_3;\mb{Z})\overset{A}{\rightarrow}H_4(X_3,L_5;\mb{Z})\overset{f}{\rightarrow}H_3(L_5;\mb{Z})\overset{B}{\rightarrow}H_3(X_3;\mb{Z})\rightarrow\dots
\ee
In \cite{Sakura:SymTFT_String}, it was explained that when $H_3(X_3;\mb{Z})$ is trivial, then one can identify each free class of $H_3(L_5;\mb{Z})$ with a non-compact divisor in $H_4(X_3,L_5;\mb{Z})$, and each torsional class of $H_3(L_5;\mb{Z})$ with a compact divisor in $H_4(X_3;\mb{Z})$.

Here we first extend the discussion to the cases where
\be
H_3(X_3;\mb{Z})=\mb{Z}^{b_3}
\ee
is non-trivial but torsion-free. This is the case for many elliptic CY3 and isolated canonical threefold singularities~\cite{Closset:2020scj,Closset:2020afy,Closset:2021lwy,Mu:2023uws}.

Now we consider a torsional class $a\in H_3(L_5;\mb{Z})\cong H^2(L_5;\mb{Z})$ with torsional degree $n$, $na=0$. Although $f$ is not surjective, since $H_3(X_3;\mb{Z})$ is torsion-free, one can still pull $a$ back to an element $\kappa\in H_4(X_3,L_5;\mb{Z})$ such that $f(\kappa)=a$. We have
\be
0=na=nf(\kappa)=f(n\kappa)\,,
\ee
hence $n\kappa\in $ker$(f)=$im$(A)$. Thus there exists an element $Z\in H_4(X_3;\mb{Z})$ such that $A(Z)=n\kappa$. In conclusion, a torsional class $a\in H_3(L_5;\mb{Z})\cong H^2(L_5;\mb{Z})$ still maps to a compact divisor of $X_3$ in this case.

For a free class $b\in H^2(L_5;\mb{Z})$, there are two possibilities. 

\begin{enumerate}
\item If $b\in$im$(f)$, then one can pull it back to an element $\xi\in H_4(X_3,L_5;\mb{Z})$, such that $f(\xi)=b$. In this case, $b$ still corresponds to a non-compact divisor of $X_3$.

\item If $b\notin$ im$(f)$, then one can map it to an element $B(b)\in H_3(X_3;\mb{Z})$, which is a free compact 3-cycle of $X_3$. In this case $b$ is not interpreted as a non-compact (relative) cycle of $X_3$, and it does not participate in the global symmetry discussions.
\end{enumerate}

In conclusion, the correspondence between non-compact/compact divisors with differential cohomology classes are the same as in \cite{Sakura:SymTFT_String}. The intersection number calculations in \cite{Sakura:SymTFT_String} can also be applied here.

\section{Homology and cohomology with twisted coefficients}\label{app:cohomo_twistcoeff}

In this appendix, we summarize relevant facts about homology and cohomology with twisted coefficients that are useful in this work~\footnote{We thank Benjamin Sung for helpful discussions on certain details of the calculations in this Appendix.}. We will follow Chapter 5 of~\cite{KirkDavis_lecture_AT} and Chapter 3.H of~\cite{hatcher2002algebraic}.

The \emph{group ring} $\mathbb{Z}[\pi]$ of a group $\pi$ is a ring whose elements are $\sum_im_i g_i$, $m_i\in\mathbb{Z}$, $g_i\in\pi$. We will consider modules over $\mathbb{Z}[\pi]$. It is a basic fact that a representation of $\pi$ on an abelian group $A$ is the same thing as a $\mathbb{Z}[\pi]$-module.

Let $\pi = \pi_1(X)$ for a path connected and locally path-connected space $X$ which admits a universal cover $\widetilde{X}$ with group of covering transformation $\pi$. The singular complex $C_*(\widetilde{X})$ of $\widetilde{X}$ with integer coefficients is a right $\mathbb{Z}[\pi]$-module. We have the following~\cite{KirkDavis_lecture_AT}:
\begin{defn}\label{eq:twist_homo}
    Given a $\mathbb{Z}[\pi]$-module $A$, form the tensor product
    \begin{align*}
        C_*(X,A) = C_*(\widetilde{X}) \otimes_{\mathbb{Z}[\pi]} A
    \end{align*}
    where $\mathbb{Z}[\pi]$ acts on $C_*(\widetilde{X})$ from the right as the deck transformation and on $A$ from the left as a linear combination of $g_i\in \pi$ acting on elements of $A$. This is a chain complex whose homology is called the homology of $X$ with local coefficients in $A$ and is denoted by $H_*(X,A)$.
\end{defn}
\noindent When the $\mathbb{Z}[\pi]$-module $A$ is specified by a representation $\rho: \pi\rightarrow Aut(A)$, we decorate $A$ with a subscript $\rho$ and call $H_*(X,A_\rho)$ the \emph{homology of $X$ twisted by $\rho$}.

The corresponding cohomology is defined as~\cite{KirkDavis_lecture_AT}:
\begin{defn}\label{defn:twist_coho}
    Given a left $\mathbb{Z}[\pi]$-module $A$, form the tensor product:
    \begin{align*}
        C^*(X,A) = \text{Hom}_{\mathbb{Z}[\pi]}(C_*(\widetilde{X}), A).
    \end{align*}
    The cohomology of this chain complex is called the cohomology of $X$ with local coefficients in $A$ and is denoted by $H^*(X,A)$.
\end{defn}
\noindent Similarly, when $A$ is specified by a representation $\rho: \pi\rightarrow Aut(A)$ we denote the cohomology group by $H^*(X,A_\rho)$ and call it the \emph{cohomology of $X$ twisted by $\rho$}.

As a simple example, when $\rho: \pi \rightarrow Aut(A)$ is trivial, we have:
\begin{equation}
    C_*(X,A) = C_*(\widetilde{X}) \otimes_{\mathbb{Z}[\pi]} A_\rho \cong C_*(X)\otimes_{\mathbb{Z}} A.
\end{equation}
The first equality is nothing but Definition~\ref{eq:twist_homo}. The second equality is due the fact that $A_\rho = A$ since $\rho$ is trivial representation of $\pi$ and the fact that a deck transformation $g\in \pi$ acting from the right on $C_*(\widetilde{X})$ descends to identity on $C_*(X)$. Therefore, we have $H_*(X,A_\rho) = H_*(X,A)$, the homology of $X$ with coefficients in $A$ as a $\mathbb{Z}$-module. When $\widetilde{X} = S^3$ with $\pi: \widetilde{X} \rightarrow X = \widetilde{X}/\Gamma$ and $A = \mathbb{Z}\oplus \mathbb{Z}$ as a $\mathbb{Z}$-module, we have $H_*(S^3/\Gamma,A_\rho) = H_*(S^3/\Gamma,A) = H_*(S^3/\Gamma,\mathbb{Z})\oplus H_*(S^3/\Gamma,\mathbb{Z})$ whose cohomological version leads to~(\ref{eq:Z2_20}). Note that here the fact that the quotient $\pi$ does not lead to any non-trivial $\rho(\pi)$ acting on $A$ implies that there is no 7-brane in the corresponding system, hence $\Gamma\subset SU(2)$.

More generally, we consider the space $S^3/\Gamma = \partial (\mathbb{C}^2/\Gamma)$ where $\Gamma$ is a subgroup of $U(2)$. The action of $\Gamma$ is a $\mathbb{Z}_p$-action $(z_1,z_2)\rightarrow (\omega z_1, \omega z_2)$ defined in~(\ref{eq:pq_action}). In the language used in this appendix we have $X = S^3/\Gamma$ and $\widetilde{X} = S^3$. The local coefficients of interest live in a $\mathbb{Z}_p$-module $A\cong \mathbb{Z}\oplus\mathbb{Z}$ and a twist is given by $\rho: \Gamma = \mathbb{Z}_p\rightarrow Aut(A)\subset GL(2,\mathbb{Z})$. In other words, the $\mathbb{Z}_p$-action on $A$ is given by $\rho(g)\in GL(2,\mathbb{Z})$ acting on $A$ where $g\in\Gamma$ and $\rho^p = 1$. For simplicity from now on we will not distinguish between the map $\rho$ and its image in $Aut(A)$ which is a $2\times 2$ matrix. To calculate the twisted cohomology $H^*(X,A_\rho)$, we need to study the cohomology of the chain complex $\text{Hom}_{\mathbb{Z}[\mathbb{Z}_p]}(C_*(\widetilde{X}), A)$ as given by Definition~\ref{defn:twist_coho}.

To calculate the cohomology of the chain complex $\text{Hom}_{\mathbb{Z}_p}(C_*(S^3), A)$, we note that the cell decomposition of $S^3$ is given by the following good cover:
\begin{equation}
    S^3 = U_1 \cup U_2 \cup U_3 \cup U_4 \cup U_5
\end{equation}
with $U_i$ being the $i^{\text{th}}$ facet of the triangulation of a 4-ball as a 4-simplex (i.e. a 4D tetrahedron with vertices $(1,0,0,0)$, $(0,1,0,0)$, $(0,0,1,0)$, $(0,0,0,1)$ and $(0,0,0,0)$). The degree-0 \v{C}ech cohomology can be calculated as follows:
\begin{equation}
    \delta_0 f(U_i\cap U_j) = f(U_i) - f(U_j) = (1-\rho) f(U_i)
\end{equation}
for $f \in \text{Hom}_{\mathbb{Z}[\mathbb{Z}_p]}(C_0(\widetilde{X}), A)\cong A^{\oplus 5}$. The coboundary map $\delta_0: A^{\oplus 5}\rightarrow A^{\oplus 10}$ can be written as a $10\times 5$ matrix each row of which has two non-zero elements $1$ and $-\rho$ placed in cyclic order. More precisely, the map is given by:
\begin{equation}
    \delta_0 = \begin{pmatrix}
        1 & -\rho & 0 & 0 & 0 \\
        1 & 0 & -\rho & 0 & 0 \\
        1 & 0 & 0 & -\rho & 0 \\
        1 & 0 & 0 & 0 & -\rho \\
        0 & 1 & -\rho & 0 & 0 \\
        0 & 1 & 0 & -\rho & 0 \\
        0 & 1 & 0 & 0 & -\rho \\
        0 & 0 & 1 & -\rho & 0\\
        0 & 0 & 1 & 0 & -\rho \\
        0 & 0 & 0 & 1 & -\rho
    \end{pmatrix}.
\end{equation}
The map $\delta_0$ acting on $A^{\oplus 5}$ is thus equivalent to $1-\rho$ acting on $A$. In other words, the degree-0 coboundary map can be written as $ 1-\rho: A\rightarrow A$. The higher degree coboundary maps are then fixed by the condition $\delta_{i+1}\circ\delta_i = 0$ and the fact $\rho^p = 1$. Therefore, $H^*(X, A_\rho)$ is given by the cohomology of the following chain complex:
\begin{equation}
\label{Arho-chain}
    A \xrightarrow{1 - \rho} A \xrightarrow{1 + \rho + \cdots + \rho^{p-1}} A \xrightarrow{1 - \rho} A.
\end{equation}
In general $1 + \rho + \cdots + \rho^{p-1} = 0$ since $1 - \rho \in GL(2,\mathbb{Z})$, hence we have:
\begin{equation}
    H^0(X,A_\rho) = H^2(X,A_\rho) = 0,\ H^1(X,A_\rho) = H^3(X,A_\rho) = (\mathbb{Z}\oplus \mathbb{Z})/\text{Im}(1-\rho)
\end{equation}

On the other hand, for the cases of type $I_n$ and $I_n^*$, where there does not exist $p\in\mb{Z}$ such that $\rho^p=1$, the middle map in the chain complex (\ref{Arho-chain}) should be modified. For instance in the type $I_n$ case when
\be
\rho=\bp 1 & n\\0 & 1\ep\,,
\ee
we can choose 
\begin{equation}
    A \xrightarrow{\bp 0 & -n\\0 & 0\ep} A \xrightarrow{\bp 0 & k\\0 & 0\ep} A \xrightarrow{\bp 0 & -n\\0 & 0\ep} A.
\end{equation}
If $k=0$, we can simply compute
\begin{equation}
    H^0(X,A_\rho) =H^2(X,A_\rho)= \mb{Z}\ ,\ H^1(X,A_\rho)= H^3(X,A_\rho) = \mb{Z}\oplus\mb{Z}_n\,.
\end{equation}
The $\mb{Z}$ factors above would not have physical effects as the torsional flavor center symmetry. 

For type $I_n^*$, 
\be
\rho=\bp -1 & n\\0 & -1\ep\,,
\ee
and the chain complex is 
\begin{equation}
    A \xrightarrow{\bp 2 & -n\\0 & 2\ep} A \xrightarrow{\bp 0 & 0\\0 & 0\ep} A \xrightarrow{\bp 2 & -n\\0 & 2\ep} A\,,
\end{equation}
leading to
\begin{equation}
    H^0(X,A_\rho) =H^2(X,A_\rho)=0\ ,\ H^1(X,A_\rho)= H^3(X,A_\rho) = \left\{\begin{array}{rl} \mb{Z}_4 & (n\text{\ odd})\\
    \mb{Z}_2\oplus\mb{Z}_2 & (n\text{\ even})\end{array}\right.\,.
\end{equation}

\bibliographystyle{JHEP}

\bibliography{refs}

\providecommand{\href}[2]{#2}\begingroup\raggedright\begin{thebibliography}{100}

\bibitem{Witten:AdSCFT&TFT}
E.~Witten, \emph{{AdS / CFT correspondence and topological field theory}}, \href{http://dx.doi.org/10.1088/1126-6708/1998/12/012}{\emph{JHEP} {\bf 12} (1998) 012}, [\href{http://arxiv.org/abs/hep-th/9812012}{{\tt hep-th/9812012}}].

\bibitem{Kong:2020cie}
L.~Kong, T.~Lan, X.-G. Wen, Z.-H. Zhang and H.~Zheng, \emph{{Algebraic higher symmetry and categorical symmetry -- a holographic and entanglement view of symmetry}}, \href{http://dx.doi.org/10.1103/PhysRevResearch.2.043086}{\emph{Phys. Rev. Res.} {\bf 2} (2020) 043086}, [\href{http://arxiv.org/abs/2005.14178}{{\tt 2005.14178}}].

\bibitem{Gaiotto:2020iye}
D.~Gaiotto and J.~Kulp, \emph{{Orbifold groupoids}}, \href{http://dx.doi.org/10.1007/JHEP02(2021)132}{\emph{JHEP} {\bf 02} (2021) 132}, [\href{http://arxiv.org/abs/2008.05960}{{\tt 2008.05960}}].

\bibitem{Sakura:SymTFT_String}
F.~Apruzzi, F.~Bonetti, I.~Garc\'ia~Etxebarria, S.~S. Hosseini and S.~Schafer-Nameki, \emph{{Symmetry TFTs from String Theory}}, \href{http://dx.doi.org/10.1007/s00220-023-04737-2}{\emph{Commun. Math. Phys.} {\bf 402} (2023) 895--949}, [\href{http://arxiv.org/abs/2112.02092}{{\tt 2112.02092}}].

\bibitem{Apruzzi:6DSymTFT}
F.~Apruzzi, \emph{{Higher form symmetries TFT in 6d}}, \href{http://dx.doi.org/10.1007/JHEP11(2022)050}{\emph{JHEP} {\bf 11} (2022) 050}, [\href{http://arxiv.org/abs/2203.10063}{{\tt 2203.10063}}].

\bibitem{Moradi:2022lqp}
H.~Moradi, S.~F. Moosavian and A.~Tiwari, \emph{{Topological holography: Towards a unification of Landau and beyond-Landau physics}}, \href{http://dx.doi.org/10.21468/SciPostPhysCore.6.4.066}{\emph{SciPost Phys. Core} {\bf 6} (2023) 066}, [\href{http://arxiv.org/abs/2207.10712}{{\tt 2207.10712}}].

\bibitem{Hubner:GenSymm_Ftheory}
M.~Hubner, D.~R. Morrison, S.~Schafer-Nameki and Y.-N. Wang, \emph{{Generalized Symmetries in F-theory and the Topology of Elliptic Fibrations}}, \href{http://dx.doi.org/10.21468/SciPostPhys.13.2.030}{\emph{SciPost Phys.} {\bf 13} (2022) 030}, [\href{http://arxiv.org/abs/2203.10022}{{\tt 2203.10022}}].

\bibitem{Freed:2022qnc}
D.~S. Freed, G.~W. Moore and C.~Teleman, \emph{{Topological symmetry in quantum field theory}},  \href{http://arxiv.org/abs/2209.07471}{{\tt 2209.07471}}.

\bibitem{Sakura:NoninvertibleHolography}
F.~Apruzzi, I.~Bah, F.~Bonetti and S.~Schafer-Nameki, \emph{{Noninvertible Symmetries from Holography and Branes}}, \href{http://dx.doi.org/10.1103/PhysRevLett.130.121601}{\emph{Phys. Rev. Lett.} {\bf 130} (2023) 121601}, [\href{http://arxiv.org/abs/2208.07373}{{\tt 2208.07373}}].

\bibitem{Kaidi:2022cpf}
J.~Kaidi, K.~Ohmori and Y.~Zheng, \emph{{Symmetry TFTs for Non-invertible Defects}}, \href{http://dx.doi.org/10.1007/s00220-023-04859-7}{\emph{Commun. Math. Phys.} {\bf 404} (2023) 1021--1124}, [\href{http://arxiv.org/abs/2209.11062}{{\tt 2209.11062}}].

\bibitem{Sakura:3DSymTFT}
M.~van Beest, D.~S.~W. Gould, S.~Schafer-Nameki and Y.-N. Wang, \emph{{Symmetry TFTs for 3d QFTs from M-theory}}, \href{http://dx.doi.org/10.1007/JHEP02(2023)226}{\emph{JHEP} {\bf 02} (2023) 226}, [\href{http://arxiv.org/abs/2210.03703}{{\tt 2210.03703}}].

\bibitem{Freed:2022iao}
D.~S. Freed, \emph{{Introduction to topological symmetry in QFT}}, {\emph{Proc. Symp. Pure Math.} {\bf 107} (2024) 93--106}, [\href{http://arxiv.org/abs/2212.00195}{{\tt 2212.00195}}].

\bibitem{Kaidi:2023maf}
J.~Kaidi, E.~Nardoni, G.~Zafrir and Y.~Zheng, \emph{{Symmetry TFTs and anomalies of non-invertible symmetries}}, \href{http://dx.doi.org/10.1007/JHEP10(2023)053}{\emph{JHEP} {\bf 10} (2023) 053}, [\href{http://arxiv.org/abs/2301.07112}{{\tt 2301.07112}}].

\bibitem{Chen:2023qnv}
J.~Chen, W.~Cui, B.~Haghighat and Y.-N. Wang, \emph{{SymTFTs and duality defects from 6d SCFTs on 4-manifolds}}, \href{http://dx.doi.org/10.1007/JHEP11(2023)208}{\emph{JHEP} {\bf 11} (2023) 208}, [\href{http://arxiv.org/abs/2305.09734}{{\tt 2305.09734}}].

\bibitem{Bhardwaj:Generalized_Charge_II}
L.~Bhardwaj and S.~Schafer-Nameki, \emph{{Generalized Charges, Part II: Non-Invertible Symmetries and the Symmetry TFT}},  \href{http://arxiv.org/abs/2305.17159}{{\tt 2305.17159}}.

\bibitem{Cao:2023rrb}
W.~Cao and Q.~Jia, \emph{{Symmetry TFT for subsystem symmetry}}, \href{http://dx.doi.org/10.1007/JHEP05(2024)225}{\emph{JHEP} {\bf 05} (2024) 225}, [\href{http://arxiv.org/abs/2310.01474}{{\tt 2310.01474}}].

\bibitem{Apruzzi:2023uma}
F.~Apruzzi, F.~Bonetti, D.~S.~W. Gould and S.~Schafer-Nameki, \emph{{Aspects of Categorical Symmetries from Branes: SymTFTs and Generalized Charges}},  \href{http://arxiv.org/abs/2306.16405}{{\tt 2306.16405}}.

\bibitem{Bhardwaj:2023idu}
L.~Bhardwaj, L.~E. Bottini, D.~Pajer and S.~Sch\"afer-Nameki, \emph{{Gapped Phases with Non-Invertible Symmetries: (1+1)d}},  \href{http://arxiv.org/abs/2310.03784}{{\tt 2310.03784}}.

\bibitem{Bhardwaj:2023fca}
L.~Bhardwaj, L.~E. Bottini, D.~Pajer and S.~Schafer-Nameki, \emph{{Categorical Landau Paradigm for Gapped Phases}},  \href{http://arxiv.org/abs/2310.03786}{{\tt 2310.03786}}.

\bibitem{Baume:2023kkf}
F.~Baume, J.~J. Heckman, M.~H\"ubner, E.~Torres, A.~P. Turner and X.~Yu, \emph{{SymTrees and Multi-Sector QFTs}}, \href{http://dx.doi.org/10.1103/PhysRevD.109.106013}{\emph{Phys. Rev. D} {\bf 109} (2024) 106013}, [\href{http://arxiv.org/abs/2310.12980}{{\tt 2310.12980}}].

\bibitem{Wen:2023otf}
R.~Wen and A.~C. Potter, \emph{{Classification of 1+1D gapless symmetry protected phases via topological holography}},  \href{http://arxiv.org/abs/2311.00050}{{\tt 2311.00050}}.

\bibitem{Lan:2023uuq}
T.~Lan, G.~Yue and L.~Wang, \emph{{Category of SET orders}},  \href{http://arxiv.org/abs/2312.15958}{{\tt 2312.15958}}.

\bibitem{Bhardwaj:2023bbf}
L.~Bhardwaj, L.~E. Bottini, D.~Pajer and S.~Schafer-Nameki, \emph{{The Club Sandwich: Gapless Phases and Phase Transitions with Non-Invertible Symmetries}},  \href{http://arxiv.org/abs/2312.17322}{{\tt 2312.17322}}.

\bibitem{Brennan:2024fgj}
T.~D. Brennan and Z.~Sun, \emph{{A SymTFT for Continuous Symmetries}},  \href{http://arxiv.org/abs/2401.06128}{{\tt 2401.06128}}.

\bibitem{Antinucci:2024zjp}
A.~Antinucci and F.~Benini, \emph{{Anomalies and gauging of U(1) symmetries}},  \href{http://arxiv.org/abs/2401.10165}{{\tt 2401.10165}}.

\bibitem{Bonetti:2024cjk}
F.~Bonetti, M.~Del~Zotto and R.~Minasian, \emph{{SymTFTs for Continuous non-Abelian Symmetries}},  \href{http://arxiv.org/abs/2402.12347}{{\tt 2402.12347}}.

\bibitem{Apruzzi:2024htg}
F.~Apruzzi, F.~Bedogna and N.~Dondi, \emph{{SymTh for non-finite symmetries}},  \href{http://arxiv.org/abs/2402.14813}{{\tt 2402.14813}}.

\bibitem{DelZotto:2024tae}
M.~Del~Zotto, S.~N. Meynet and R.~Moscrop, \emph{{Remarks on geometric engineering, symmetry TFTs and anomalies}}, \href{http://dx.doi.org/10.1007/JHEP07(2024)220}{\emph{JHEP} {\bf 07} (2024) 220}, [\href{http://arxiv.org/abs/2402.18646}{{\tt 2402.18646}}].

\bibitem{Bhardwaj:2024qrf}
L.~Bhardwaj, D.~Pajer, S.~Schafer-Nameki and A.~Warman, \emph{{Hasse Diagrams for Gapless SPT and SSB Phases with Non-Invertible Symmetries}},  \href{http://arxiv.org/abs/2403.00905}{{\tt 2403.00905}}.

\bibitem{Wen:2024udn}
R.~Wen, W.~Ye and A.~C. Potter, \emph{{Topological holography for fermions}},  \href{http://arxiv.org/abs/2404.19004}{{\tt 2404.19004}}.

\bibitem{Franco:2024mxa}
S.~Franco and X.~Yu, \emph{{Generalized Symmetries in 2D from String Theory: SymTFTs, Intrinsic Relativeness, and Anomalies of Non-invertible Symmetries}},  \href{http://arxiv.org/abs/2404.19761}{{\tt 2404.19761}}.

\bibitem{Putrov:2024uor}
P.~Putrov and R.~Radhakrishnan, \emph{{Non-anomalous non-invertible symmetries in 1+1D from gapped boundaries of SymTFTs}},  \href{http://arxiv.org/abs/2405.04619}{{\tt 2405.04619}}.

\bibitem{Bhardwaj:2024kvy}
L.~Bhardwaj, L.~E. Bottini, S.~Schafer-Nameki and A.~Tiwari, \emph{{Lattice Models for Phases and Transitions with Non-Invertible Symmetries}},  \href{http://arxiv.org/abs/2405.05964}{{\tt 2405.05964}}.

\bibitem{Bhardwaj:2024ydc}
L.~Bhardwaj, K.~Inamura and A.~Tiwari, \emph{{Fermionic Non-Invertible Symmetries in (1+1)d: Gapped and Gapless Phases, Transitions, and Symmetry TFTs}},  \href{http://arxiv.org/abs/2405.09754}{{\tt 2405.09754}}.

\bibitem{Copetti:2024onh}
C.~Copetti, \emph{{Defect Charges, Gapped Boundary Conditions, and the Symmetry TFT}},  \href{http://arxiv.org/abs/2408.01490}{{\tt 2408.01490}}.

\bibitem{Antinucci:2024bcm}
A.~Antinucci, F.~Benini and G.~Rizi, \emph{{Holographic duals of symmetry broken phases}},  \href{http://arxiv.org/abs/2408.01418}{{\tt 2408.01418}}.

\bibitem{Antinucci:2024ltv}
A.~Antinucci, C.~Copetti and S.~Schafer-Nameki, \emph{{SymTFT for (3+1)d Gapless SPTs and Obstructions to Confinement}},  \href{http://arxiv.org/abs/2408.05585}{{\tt 2408.05585}}.

\bibitem{Cvetic:2024dzu}
M.~Cveti\v{c}, R.~Donagi, J.~J. Heckman, M.~H\"ubner and E.~Torres, \emph{{Cornering Relative Symmetry Theories}},  \href{http://arxiv.org/abs/2408.12600}{{\tt 2408.12600}}.

\bibitem{Bhardwaj:2024xcx}
L.~Bhardwaj, T.~D\'ecoppet, S.~Schafer-Nameki and M.~Yu, \emph{{Fusion 3-Categories for Duality Defects}},  \href{http://arxiv.org/abs/2408.13302}{{\tt 2408.13302}}.

\bibitem{GarciaEtxebarria:2024jfv}
I.~n. Garc\'\i{}a~Etxebarria, J.~Huertas and A.~M. Uranga, \emph{{SymTFT Fans: The Symmetry Theory of 4d N=4 Super Yang-Mills on spaces with boundaries}},  \href{http://arxiv.org/abs/2409.02156}{{\tt 2409.02156}}.

\bibitem{Bhardwaj:2024igy}
L.~Bhardwaj, C.~Copetti, D.~Pajer and S.~Schafer-Nameki, \emph{{Boundary SymTFT}},  \href{http://arxiv.org/abs/2409.02166}{{\tt 2409.02166}}.

\bibitem{Argurio:2024ewp}
R.~Argurio, A.~Collinucci, G.~Galati, O.~Hulik and E.~Paznokas, \emph{{Non-Invertible T-duality at Any Radius via Non-Compact SymTFT}},  \href{http://arxiv.org/abs/2409.11822}{{\tt 2409.11822}}.

\bibitem{Chen:2024fno}
J.~Chen, W.~Cui, B.~Haghighat and Y.~Sun, \emph{{Modularity of Vafa-Witten Partition Functions from SymTFT}},  \href{http://arxiv.org/abs/2409.19397}{{\tt 2409.19397}}.

\bibitem{Gaiotto_2015}
D.~Gaiotto, A.~Kapustin, N.~Seiberg and B.~Willett, \emph{{Generalized Global Symmetries}}, \href{http://dx.doi.org/10.1007/JHEP02(2015)172}{\emph{JHEP} {\bf 02} (2015) 172}, [\href{http://arxiv.org/abs/1412.5148}{{\tt 1412.5148}}].

\bibitem{McGreevy:2022oyu}
J.~McGreevy, \emph{{Generalized Symmetries in Condensed Matter}}, \href{http://dx.doi.org/10.1146/annurev-conmatphys-040721-021029}{\emph{Ann. Rev. Condensed Matter Phys.} {\bf 14} (2023) 57--82}, [\href{http://arxiv.org/abs/2204.03045}{{\tt 2204.03045}}].

\bibitem{Sakura:ICTP_Symmetries}
S.~Schafer-Nameki, \emph{{ICTP lectures on (non-)invertible generalized symmetries}}, \href{http://dx.doi.org/10.1016/j.physrep.2024.01.007}{\emph{Phys. Rept.} {\bf 1063} (2024) 1--55}, [\href{http://arxiv.org/abs/2305.18296}{{\tt 2305.18296}}].

\bibitem{Brennan:2023mmt}
T.~D. Brennan and S.~Hong, \emph{{Introduction to Generalized Global Symmetries in QFT and Particle Physics}},  \href{http://arxiv.org/abs/2306.00912}{{\tt 2306.00912}}.

\bibitem{Bhardwaj:2023kri}
L.~Bhardwaj, L.~E. Bottini, L.~Fraser-Taliente, L.~Gladden, D.~S.~W. Gould, A.~Platschorre et~al., \emph{{Lectures on generalized symmetries}}, \href{http://dx.doi.org/10.1016/j.physrep.2023.11.002}{\emph{Phys. Rept.} {\bf 1051} (2024) 1--87}, [\href{http://arxiv.org/abs/2307.07547}{{\tt 2307.07547}}].

\bibitem{Luo:2023ive}
R.~Luo, Q.-R. Wang and Y.-N. Wang, \emph{{Lecture notes on generalized symmetries and applications}}, \href{http://dx.doi.org/10.1016/j.physrep.2024.02.002}{\emph{Phys. Rept.} {\bf 1065} (2024) 1--43}, [\href{http://arxiv.org/abs/2307.09215}{{\tt 2307.09215}}].

\bibitem{Vafa:Ftheory}
C.~Vafa, \emph{{Evidence for F theory}}, \href{http://dx.doi.org/10.1016/0550-3213(96)00172-1}{\emph{Nucl. Phys. B} {\bf 469} (1996) 403--418}, [\href{http://arxiv.org/abs/hep-th/9602022}{{\tt hep-th/9602022}}].

\bibitem{Morrison:1996na}
D.~R. Morrison and C.~Vafa, \emph{{Compactifications of F theory on Calabi-Yau threefolds. 1}}, \href{http://dx.doi.org/10.1016/0550-3213(96)00242-8}{\emph{Nucl. Phys.} {\bf B473} (1996) 74--92}, [\href{http://arxiv.org/abs/hep-th/9602114}{{\tt hep-th/9602114}}].

\bibitem{Morrison:1996pp}
D.~R. Morrison and C.~Vafa, \emph{{Compactifications of F theory on Calabi-Yau threefolds. 2.}}, \href{http://dx.doi.org/10.1016/0550-3213(96)00369-0}{\emph{Nucl. Phys. B} {\bf 476} (1996) 437--469}, [\href{http://arxiv.org/abs/hep-th/9603161}{{\tt hep-th/9603161}}].

\bibitem{Weigand:Ftheory}
T.~Weigand, \emph{{F-theory}}, {\emph{PoS} {\bf TASI2017} (2018) 016}, [\href{http://arxiv.org/abs/1806.01854}{{\tt 1806.01854}}].

\bibitem{Heckman:6DBaseClassification}
J.~J. Heckman, D.~R. Morrison and C.~Vafa, \emph{{On the Classification of 6D SCFTs and Generalized ADE Orbifolds}}, \href{http://dx.doi.org/10.1007/JHEP05(2014)028}{\emph{JHEP} {\bf 05} (2014) 028}, [\href{http://arxiv.org/abs/1312.5746}{{\tt 1312.5746}}].

\bibitem{Heckman:AtomicClassification}
J.~J. Heckman, D.~R. Morrison, T.~Rudelius and C.~Vafa, \emph{{Atomic Classification of 6D SCFTs}}, \href{http://dx.doi.org/10.1002/prop.201500024}{\emph{Fortsch. Phys.} {\bf 63} (2015) 468--530}, [\href{http://arxiv.org/abs/1502.05405}{{\tt 1502.05405}}].

\bibitem{Dierigl:2020myk}
M.~Dierigl, P.-K. Oehlmann and F.~Ruehle, \emph{{Non-Simply-Connected Symmetries in 6D SCFTs}}, \href{http://dx.doi.org/10.1007/JHEP10(2020)173}{\emph{JHEP} {\bf 10} (2020) 173}, [\href{http://arxiv.org/abs/2005.12929}{{\tt 2005.12929}}].

\bibitem{Bhardwaj:Higher_form_5D6D}
L.~Bhardwaj and S.~Sch\"afer-Nameki, \emph{{Higher-form symmetries of 6d and 5d theories}}, \href{http://dx.doi.org/10.1007/JHEP02(2021)159}{\emph{JHEP} {\bf 02} (2021) 159}, [\href{http://arxiv.org/abs/2008.09600}{{\tt 2008.09600}}].

\bibitem{Apruzzi:2020zot}
F.~Apruzzi, M.~Dierigl and L.~Lin, \emph{{The fate of discrete 1-form symmetries in 6d}}, \href{http://dx.doi.org/10.21468/SciPostPhys.12.2.047}{\emph{SciPost Phys.} {\bf 12} (2022) 047}, [\href{http://arxiv.org/abs/2008.09117}{{\tt 2008.09117}}].

\bibitem{Apruzzi:2021mlh}
F.~Apruzzi, L.~Bhardwaj, D.~S.~W. Gould and S.~Schafer-Nameki, \emph{{2-Group symmetries and their classification in 6d}}, \href{http://dx.doi.org/10.21468/SciPostPhys.12.3.098}{\emph{SciPost Phys.} {\bf 12} (2022) 098}, [\href{http://arxiv.org/abs/2110.14647}{{\tt 2110.14647}}].

\bibitem{Heckman:2022suy}
J.~J. Heckman, C.~Lawrie, L.~Lin, H.~Y. Zhang and G.~Zoccarato, \emph{{6D SCFTs, center-flavor symmetries, and Stiefel-Whitney compactifications}}, \href{http://dx.doi.org/10.1103/PhysRevD.106.066003}{\emph{Phys. Rev. D} {\bf 106} (2022) 066003}, [\href{http://arxiv.org/abs/2205.03411}{{\tt 2205.03411}}].

\bibitem{DelZotto:2014hpa}
M.~Del~Zotto, J.~J. Heckman, A.~Tomasiello and C.~Vafa, \emph{{6d Conformal Matter}}, \href{http://dx.doi.org/10.1007/JHEP02(2015)054}{\emph{JHEP} {\bf 02} (2015) 054}, [\href{http://arxiv.org/abs/1407.6359}{{\tt 1407.6359}}].

\bibitem{Bhardwaj:Generalized_Charge_I}
L.~Bhardwaj and S.~Schafer-Nameki, \emph{{Generalized charges, part I: Invertible symmetries and higher representations}}, \href{http://dx.doi.org/10.21468/SciPostPhys.16.4.093}{\emph{SciPost Phys.} {\bf 16} (2024) 093}, [\href{http://arxiv.org/abs/2304.02660}{{\tt 2304.02660}}].

\bibitem{Polchinski:String_vol2}
J.~Polchinski, \emph{{String theory. Vol. 2: Superstring theory and beyond}}.
\newblock Cambridge Monographs on Mathematical Physics. Cambridge University Press, 12, 2007, \href{http://dx.doi.org/10.1017/CBO9780511618123}{10.1017/CBO9780511618123}.

\bibitem{Belov:2006xj}
D.~M. Belov and G.~W. Moore, \emph{{Type II Actions from 11-Dimensional Chern-Simons Theories}},  \href{http://arxiv.org/abs/hep-th/0611020}{{\tt hep-th/0611020}}.

\bibitem{Lawrie:2023tdz}
C.~Lawrie, X.~Yu and H.~Y. Zhang, \emph{{Intermediate defect groups, polarization pairs, and noninvertible duality defects}}, \href{http://dx.doi.org/10.1103/PhysRevD.109.026005}{\emph{Phys. Rev. D} {\bf 109} (2024) 026005}, [\href{http://arxiv.org/abs/2306.11783}{{\tt 2306.11783}}].

\bibitem{Yu:2023nyn}
X.~Yu, \emph{{Noninvertible symmetries in 2D from type IIB string theory}}, \href{http://dx.doi.org/10.1103/PhysRevD.110.065008}{\emph{Phys. Rev. D} {\bf 110} (2024) 065008}, [\href{http://arxiv.org/abs/2310.15339}{{\tt 2310.15339}}].

\bibitem{Bergshoeff:2001pv}
E.~Bergshoeff, R.~Kallosh, T.~Ortin, D.~Roest and A.~Van~Proeyen, \emph{{New formulations of D = 10 supersymmetry and D8 - O8 domain walls}}, \href{http://dx.doi.org/10.1088/0264-9381/18/17/303}{\emph{Class. Quant. Grav.} {\bf 18} (2001) 3359--3382}, [\href{http://arxiv.org/abs/hep-th/0103233}{{\tt hep-th/0103233}}].

\bibitem{HopkinsSinger}
M.~J. Hopkins and I.~M. Singer, \emph{{Quadratic functions in geometry, topology, and M theory}}, {\emph{J. Diff. Geom.} {\bf 70} (2005) 329--452}, [\href{http://arxiv.org/abs/math/0211216}{{\tt math/0211216}}].

\bibitem{Hsieh:2020jpj}
C.-T. Hsieh, Y.~Tachikawa and K.~Yonekura, \emph{{Anomaly Inflow and p-Form Gauge Theories}}, \href{http://dx.doi.org/10.1007/s00220-022-04333-w}{\emph{Commun. Math. Phys.} {\bf 391} (2022) 495--608}, [\href{http://arxiv.org/abs/2003.11550}{{\tt 2003.11550}}].

\bibitem{Douglas:Cremmer-Scherk}
M.~R. Douglas, D.~S. Park and C.~Schnell, \emph{{The Cremmer-Scherk Mechanism in F-theory Compactifications on K3 Manifolds}}, \href{http://dx.doi.org/10.1007/JHEP05(2014)135}{\emph{JHEP} {\bf 05} (2014) 135}, [\href{http://arxiv.org/abs/1403.1595}{{\tt 1403.1595}}].

\bibitem{Katz:DimRed_B-field}
S.~Katz and W.~Taylor, \emph{{Dimensional reduction of B-fields in F-theory}}, \href{http://dx.doi.org/10.4310/PAMQ.2022.v18.n4.a10}{\emph{Pure Appl. Math. Quart.} {\bf 18} (2022) 1621--1660}, [\href{http://arxiv.org/abs/2204.13146}{{\tt 2204.13146}}].

\bibitem{Heckman:Brane_GenSymm}
J.~J. Heckman, M.~H\"ubner, E.~Torres and H.~Y. Zhang, \emph{{The Branes Behind Generalized Symmetry Operators}}, \href{http://dx.doi.org/10.1002/prop.202200180}{\emph{Fortsch. Phys.} {\bf 71} (2023) 2200180}, [\href{http://arxiv.org/abs/2209.03343}{{\tt 2209.03343}}].

\bibitem{GarciaEtxebarria:IIB_flux_noncomm}
I.~Garc\'\i{}a~Etxebarria, B.~Heidenreich and D.~Regalado, \emph{{IIB flux non-commutativity and the global structure of field theories}}, \href{http://dx.doi.org/10.1007/JHEP10(2019)169}{\emph{JHEP} {\bf 10} (2019) 169}, [\href{http://arxiv.org/abs/1908.08027}{{\tt 1908.08027}}].

\bibitem{Witten:Geometric_Langlands_6D}
E.~Witten, \emph{{Geometric Langlands From Six Dimensions}},  \href{http://arxiv.org/abs/0905.2720}{{\tt 0905.2720}}.

\bibitem{Morrison:HigherForm_5D}
D.~R. Morrison, S.~Schafer-Nameki and B.~Willett, \emph{{Higher-Form Symmetries in 5d}}, \href{http://dx.doi.org/10.1007/JHEP09(2020)024}{\emph{JHEP} {\bf 09} (2020) 024}, [\href{http://arxiv.org/abs/2005.12296}{{\tt 2005.12296}}].

\bibitem{Vandermeulen:2022edk}
T.~Vandermeulen, \emph{{Lower-Form Symmetries}},  \href{http://arxiv.org/abs/2211.04461}{{\tt 2211.04461}}.

\bibitem{Aloni:2024jpb}
D.~Aloni, E.~Garc\'\i{}a-Valdecasas, M.~Reece and M.~Suzuki, \emph{{Spontaneously Broken $(-1)$-Form U(1) Symmetries}},  \href{http://arxiv.org/abs/2402.00117}{{\tt 2402.00117}}.

\bibitem{Santilli:2024dyz}
L.~Santilli and R.~J. Szabo, \emph{{Higher form symmetries and orbifolds of two-dimensional Yang-Mills theory}},  \href{http://arxiv.org/abs/2403.03119}{{\tt 2403.03119}}.

\bibitem{ToricVarieties:CoxLittleSchenck}
D.~Cox, J.~Little and H.~Schenck, \emph{Toric varieties}, vol.~124 of \emph{Graduate Studies in Mathematics}.
\newblock American Mathematical Society, United States, 2011, \href{http://dx.doi.org/10.1090/gsm/124}{10.1090/gsm/124}.

\bibitem{Cvetic:Cut&Glue}
M.~Cveti\v{c}, J.~J. Heckman, M.~H\"ubner and E.~Torres, \emph{{0-form, 1-form, and 2-group symmetries via cutting and gluing of orbifolds}}, \href{http://dx.doi.org/10.1103/PhysRevD.106.106003}{\emph{Phys. Rev. D} {\bf 106} (2022) 106003}, [\href{http://arxiv.org/abs/2203.10102}{{\tt 2203.10102}}].

\bibitem{Morrison:NHC}
D.~R. Morrison and W.~Taylor, \emph{{Classifying bases for 6D F-theory models}}, \href{http://dx.doi.org/10.2478/s11534-012-0065-4}{\emph{Central Eur. J. Phys.} {\bf 10} (2012) 1072--1088}, [\href{http://arxiv.org/abs/1201.1943}{{\tt 1201.1943}}].

\bibitem{Halverson:Strong_Coupling}
J.~Halverson, \emph{{Strong Coupling in F-theory and Geometrically Non-Higgsable Seven-branes}}, \href{http://dx.doi.org/10.1016/j.nuclphysb.2017.02.014}{\emph{Nucl. Phys. B} {\bf 919} (2017) 267--296}, [\href{http://arxiv.org/abs/1603.01639}{{\tt 1603.01639}}].

\bibitem{Grassi:D3brane}
A.~Grassi, J.~Halverson, F.~Ruehle and J.~L. Shaneson, \emph{{Dualities of deformed $ \mathcal{N}=2 $ SCFTs from link monodromy on D3-brane states}}, \href{http://dx.doi.org/10.1007/JHEP09(2017)135}{\emph{JHEP} {\bf 09} (2017) 135}, [\href{http://arxiv.org/abs/1611.01154}{{\tt 1611.01154}}].

\bibitem{Grassi:2018wfy}
A.~Grassi, J.~Halverson, C.~Long, J.~L. Shaneson and J.~Tian, \emph{{Non-simply-laced Symmetry Algebras in F-theory on Singular Spaces}}, \href{http://dx.doi.org/10.1007/JHEP09(2018)129}{\emph{JHEP} {\bf 09} (2018) 129}, [\href{http://arxiv.org/abs/1805.06949}{{\tt 1805.06949}}].

\bibitem{Grassi:2021ptc}
A.~Grassi, J.~Halverson, C.~Long, J.~L. Shaneson, B.~Sung and J.~Tian, \emph{{6D anomaly-free matter spectrum in F-theory on singular spaces}}, \href{http://dx.doi.org/10.1007/JHEP08(2022)182}{\emph{JHEP} {\bf 08} (2022) 182}, [\href{http://arxiv.org/abs/2110.06943}{{\tt 2110.06943}}].

\bibitem{Aharony:Sfold_N=3SCFT}
O.~Aharony, Y.~Tachikawa and K.~Gomi, \emph{{S-folds and 4d N=3 superconformal field theories}}, \href{http://dx.doi.org/10.1007/JHEP06(2016)044}{\emph{JHEP} {\bf 06} (2016) 044}, [\href{http://arxiv.org/abs/1602.08638}{{\tt 1602.08638}}].

\bibitem{Cvetic:HigherForm_Anomalies}
M.~Cvetic, M.~Dierigl, L.~Lin and H.~Y. Zhang, \emph{{Higher-form symmetries and their anomalies in M-/F-theory duality}}, \href{http://dx.doi.org/10.1103/PhysRevD.104.126019}{\emph{Phys. Rev. D} {\bf 104} (2021) 126019}, [\href{http://arxiv.org/abs/2106.07654}{{\tt 2106.07654}}].

\bibitem{Heckman:TopDown_TopologicalDefects}
J.~J. Heckman, M.~Hubner, E.~Torres, X.~Yu and H.~Y. Zhang, \emph{{Top down approach to topological duality defects}}, \href{http://dx.doi.org/10.1103/PhysRevD.108.046015}{\emph{Phys. Rev. D} {\bf 108} (2023) 046015}, [\href{http://arxiv.org/abs/2212.09743}{{\tt 2212.09743}}].

\bibitem{Douglas:brane_within_branes}
M.~R. Douglas, \emph{{Branes within branes}}, {\emph{NATO Sci. Ser. C} {\bf 520} (1999) 267--275}, [\href{http://arxiv.org/abs/hep-th/9512077}{{\tt hep-th/9512077}}].

\bibitem{Sen:tachyon_condensation}
A.~Sen, \emph{{Tachyon condensation on the brane anti-brane system}}, \href{http://dx.doi.org/10.1088/1126-6708/1998/08/012}{\emph{JHEP} {\bf 08} (1998) 012}, [\href{http://arxiv.org/abs/hep-th/9805170}{{\tt hep-th/9805170}}].

\bibitem{Witten:DbraneKtheory}
E.~Witten, \emph{{D-branes and K-theory}}, \href{http://dx.doi.org/10.1088/1126-6708/1998/12/019}{\emph{JHEP} {\bf 12} (1998) 019}, [\href{http://arxiv.org/abs/hep-th/9810188}{{\tt hep-th/9810188}}].

\bibitem{Cordova:2019uob}
C.~C\'ordova, D.~S. Freed, H.~T. Lam and N.~Seiberg, \emph{{Anomalies in the Space of Coupling Constants and Their Dynamical Applications II}}, \href{http://dx.doi.org/10.21468/SciPostPhys.8.1.002}{\emph{SciPost Phys.} {\bf 8} (2020) 002}, [\href{http://arxiv.org/abs/1905.13361}{{\tt 1905.13361}}].

\bibitem{Apruzzi:Fate_1form_6D}
F.~Apruzzi, M.~Dierigl and L.~Lin, \emph{The fate of discrete 1-form symmetries in 6d}, \href{http://dx.doi.org/10.21468/scipostphys.12.2.047}{\emph{{SciPost} Physics} {\bf 12} (feb, 2022) }.

\bibitem{Aharony:2013hda}
O.~Aharony, N.~Seiberg and Y.~Tachikawa, \emph{{Reading between the lines of four-dimensional gauge theories}}, \href{http://dx.doi.org/10.1007/JHEP08(2013)115}{\emph{JHEP} {\bf 08} (2013) 115}, [\href{http://arxiv.org/abs/1305.0318}{{\tt 1305.0318}}].

\bibitem{Banks:probing}
T.~Banks, M.~R. Douglas and N.~Seiberg, \emph{{Probing F theory with branes}}, \href{http://dx.doi.org/10.1016/0370-2693(96)00808-8}{\emph{Phys. Lett. B} {\bf 387} (1996) 278--281}, [\href{http://arxiv.org/abs/hep-th/9605199}{{\tt hep-th/9605199}}].

\bibitem{DeWolfe:1998bi}
O.~DeWolfe, T.~Hauer, A.~Iqbal and B.~Zwiebach, \emph{{Constraints on the BPS spectrum of N=2, D = 4 theories with A-D-E flavor symmetry}}, \href{http://dx.doi.org/10.1016/S0550-3213(98)00652-X}{\emph{Nucl. Phys. B} {\bf 534} (1998) 261--274}, [\href{http://arxiv.org/abs/hep-th/9805220}{{\tt hep-th/9805220}}].

\bibitem{Nardoni:2024sos}
E.~Nardoni, M.~Sacchi, O.~Sela, G.~Zafrir and Y.~Zheng, \emph{{Dimensionally reducing generalized symmetries from (3+1)-dimensions}}, \href{http://dx.doi.org/10.1007/JHEP07(2024)110}{\emph{JHEP} {\bf 07} (2024) 110}, [\href{http://arxiv.org/abs/2403.15995}{{\tt 2403.15995}}].

\bibitem{Park:2011ji}
D.~S. Park, \emph{{Anomaly Equations and Intersection Theory}}, \href{http://dx.doi.org/10.1007/JHEP01(2012)093}{\emph{JHEP} {\bf 01} (2012) 093}, [\href{http://arxiv.org/abs/1111.2351}{{\tt 1111.2351}}].

\bibitem{Apruzzi:2019opn}
F.~Apruzzi, C.~Lawrie, L.~Lin, S.~Sch\"afer-Nameki and Y.-N. Wang, \emph{{Fibers add Flavor, Part I: Classification of 5d SCFTs, Flavor Symmetries and BPS States}}, \href{http://dx.doi.org/10.1007/JHEP11(2019)068}{\emph{JHEP} {\bf 11} (2019) 068}, [\href{http://arxiv.org/abs/1907.05404}{{\tt 1907.05404}}].

\bibitem{DelZotto:2017pti}
M.~Del~Zotto, J.~J. Heckman and D.~R. Morrison, \emph{{6D SCFTs and Phases of 5D Theories}}, \href{http://dx.doi.org/10.1007/JHEP09(2017)147}{\emph{JHEP} {\bf 09} (2017) 147}, [\href{http://arxiv.org/abs/1703.02981}{{\tt 1703.02981}}].

\bibitem{Apruzzi:2019kgb}
F.~Apruzzi, S.~Schafer-Nameki and Y.-N. Wang, \emph{{5d SCFTs from Decoupling and Gluing}}, \href{http://dx.doi.org/10.1007/JHEP08(2020)153}{\emph{JHEP} {\bf 08} (2020) 153}, [\href{http://arxiv.org/abs/1912.04264}{{\tt 1912.04264}}].

\bibitem{Bhardwaj:classification_LST}
L.~Bhardwaj, M.~Del~Zotto, J.~J. Heckman, D.~R. Morrison, T.~Rudelius and C.~Vafa, \emph{{F-theory and the Classification of Little Strings}}, \href{http://dx.doi.org/10.1103/PhysRevD.93.086002}{\emph{Phys. Rev. D} {\bf 93} (2016) 086002}, [\href{http://arxiv.org/abs/1511.05565}{{\tt 1511.05565}}].

\bibitem{DelZotto:2020sop}
M.~Del~Zotto and K.~Ohmori, \emph{{2-Group Symmetries of 6D Little String Theories and T-Duality}}, \href{http://dx.doi.org/10.1007/s00023-021-01018-3}{\emph{Annales Henri Poincare} {\bf 22} (2021) 2451--2474}, [\href{http://arxiv.org/abs/2009.03489}{{\tt 2009.03489}}].

\bibitem{DelZotto:2022ohj}
M.~Del~Zotto, M.~Liu and P.-K. Oehlmann, \emph{{Back to heterotic strings on ALE spaces. Part I. Instantons, 2-groups and T-duality}}, \href{http://dx.doi.org/10.1007/JHEP01(2023)176}{\emph{JHEP} {\bf 01} (2023) 176}, [\href{http://arxiv.org/abs/2209.10551}{{\tt 2209.10551}}].

\bibitem{DelZotto:2022xrh}
M.~Del~Zotto, M.~Liu and P.-K. Oehlmann, \emph{{Back to heterotic strings on ALE spaces. Part II. Geometry of T-dual little strings}}, \href{http://dx.doi.org/10.1007/JHEP01(2024)109}{\emph{JHEP} {\bf 01} (2024) 109}, [\href{http://arxiv.org/abs/2212.05311}{{\tt 2212.05311}}].

\bibitem{DelZotto:2023ahf}
M.~Del~Zotto, M.~Liu and P.-K. Oehlmann, \emph{{6D heterotic little string theories and F-theory geometry: An introduction}}, {\emph{Proc. Symp. Pure Math.} {\bf 107} (2024) 179--200}, [\href{http://arxiv.org/abs/2303.13502}{{\tt 2303.13502}}].

\bibitem{Lee:2018ihr}
S.-J. Lee, D.~Regalado and T.~Weigand, \emph{{6d SCFTs and U(1) Flavour Symmetries}}, \href{http://dx.doi.org/10.1007/JHEP11(2018)147}{\emph{JHEP} {\bf 11} (2018) 147}, [\href{http://arxiv.org/abs/1803.07998}{{\tt 1803.07998}}].

\bibitem{Apruzzi:2020eqi}
F.~Apruzzi, M.~Fazzi, J.~J. Heckman, T.~Rudelius and H.~Y. Zhang, \emph{{General prescription for global $U(1)$\textquoteright{}s in 6D SCFTs}}, \href{http://dx.doi.org/10.1103/PhysRevD.101.086023}{\emph{Phys. Rev. D} {\bf 101} (2020) 086023}, [\href{http://arxiv.org/abs/2001.10549}{{\tt 2001.10549}}].

\bibitem{Closset:2020scj}
C.~Closset, S.~Schafer-Nameki and Y.-N. Wang, \emph{{Coulomb and Higgs Branches from Canonical Singularities: Part 0}}, \href{http://dx.doi.org/10.1007/JHEP02(2021)003}{\emph{JHEP} {\bf 02} (2021) 003}, [\href{http://arxiv.org/abs/2007.15600}{{\tt 2007.15600}}].

\bibitem{Closset:2020afy}
C.~Closset, S.~Giacomelli, S.~Schafer-Nameki and Y.-N. Wang, \emph{{5d and 4d SCFTs: Canonical Singularities, Trinions and S-Dualities}}, \href{http://dx.doi.org/10.1007/JHEP05(2021)274}{\emph{JHEP} {\bf 05} (2021) 274}, [\href{http://arxiv.org/abs/2012.12827}{{\tt 2012.12827}}].

\bibitem{Closset:2021lwy}
C.~Closset, S.~Sch\"afer-Nameki and Y.-N. Wang, \emph{{Coulomb and Higgs branches from canonical singularities. Part I. Hypersurfaces with smooth Calabi-Yau resolutions}}, \href{http://dx.doi.org/10.1007/JHEP04(2022)061}{\emph{JHEP} {\bf 04} (2022) 061}, [\href{http://arxiv.org/abs/2111.13564}{{\tt 2111.13564}}].

\bibitem{Mu:2023uws}
J.~Mu, Y.-N. Wang and H.~N. Zhang, \emph{{5d SCFTs from isolated complete intersection singularities}}, \href{http://dx.doi.org/10.1007/JHEP02(2024)155}{\emph{JHEP} {\bf 02} (2024) 155}, [\href{http://arxiv.org/abs/2311.05441}{{\tt 2311.05441}}].

\bibitem{KirkDavis_lecture_AT}
J.~Davis and P.~Kirk, \emph{{Lecture Notes in Algebraic Topology}}, vol.~35 of \emph{Graduate Studies in Mathematics}.
\newblock American Mathematical Society, Providence, R.I., 2001, \href{http://dx.doi.org/10.1090/gsm/035}{10.1090/gsm/035}.

\bibitem{hatcher2002algebraic}
A.~Hatcher, \emph{Algebraic Topology}.
\newblock Algebraic Topology. Cambridge University Press, 2002, \href{http://dx.doi.org/https://www.cambridge.org/9780521795401}{https://www.cambridge.org/9780521795401}.

\end{thebibliography}\endgroup

\end{document}